%% file: ms.tex
\documentclass[twocolumn]{aastex631}

\newcommand{\HeI}{\hbox{[{\rm He}\kern 0.1em{\sc i}]}}
\newcommand{\Ha}{\hbox{{\rm H}\kern 0.1em$\alpha$}}
\newcommand{\Hb}{\hbox{{\rm H}\kern 0.1em$\beta$}}
\newcommand{\Hd}{\hbox{{\rm H}\kern 0.1em$\delta$}}
\newcommand{\Hg}{\hbox{{\rm H}\kern 0.1em$\gamma$}}

\newcommand{\MgII}{\hbox{{\rm Mg}\kern 0.1em{\sc ii}}}
\newcommand{\CIV}{\hbox{{\rm C}\kern 0.1em{\sc iv}}}
\newcommand{\CIII}{\hbox{{\rm C}\kern 0.1em{\sc iii}]}}

\newcommand{\NeV}{\hbox{[{\rm Ne}\kern 0.1em{\sc v}]}}
\newcommand{\OII}{\hbox{[{\rm O}\kern 0.1em{\sc ii}]}}
\newcommand{\NeIII}{\hbox{[{\rm Ne}\kern 0.1em{\sc iii}]}}
\newcommand{\OIII}{\hbox{[{\rm O}\kern 0.1em{\sc iii}]$\lambda$5007}}
\newcommand{\OIIIs}{\hbox{[{\rm O}\kern 0.1em{\sc iii}]$\lambda$4364}}

\newcommand{\NII}{\hbox{[{\rm N}\kern 0.1em{\sc ii}]}}
\newcommand{\SII}{\hbox{[{\rm S}\kern 0.1em{\sc ii}]}}

\newcommand{\lmass}{$\log\mathrm{M\!_\star/M}_{\odot}$}

\newcommand{\uvnir}{[0.3$-$0.9 $\mu$m]~}
\newcommand{\nearmidir}{[1$-$3 $\mu$m]~}

\usepackage{amsmath,amstext,natbib,longtable,verbatim,float,graphicx,color, longtable, soul, enumitem} 

\definecolor{citeRGB}{rgb}{0,0.1,0.7}
\usepackage{hyperref}
\begin{document}

\title{From ``The Cliff" to ``Virgil": Mapping the Spectral Diversity of Little Red Dots with JWST/NIRSpec}


\author[0000-0001-6813-875X]{Guillermo Barro}
\affiliation{University of the Pacific, Stockton, CA 90340 USA}

\author[0000-0003-4528-5639]{Pablo G. P\'erez-Gonz\'alez}
\affiliation{Centro de Astrobiolog\'{\i}a (CAB), CSIC-INTA, Ctra. de Ajalvir km 4, Torrej\'on de Ardoz, E-28850, Madrid, Spain}

\author[0000-0002-8360-3880]{Dale Kocevski}
\affiliation{Department of Physics and Astronomy, Colby College, Waterville, ME 04901, USA}

\author[0000-0002-1410-0470]{Jonathan R. Trump}
\affil{Department of Physics, 196A Auditorium Road, Unit 3046, University of Connecticut, Storrs, CT 06269, USA}

\author[0000-0001-5414-5131]{Mark Dickinson}
\affiliation{NSF National Optical-Infrared Astronomy Research Laboratory, 950 N. Cherry Ave., Tucson, AZ 85719, USA}

\author[0000-0002-7959-8783]{Pablo Arrabal Haro}
\affiliation{Center for Space Sciences and Technology, UMBC, 5523 Research Park Dr, Baltimore, MD 21228 USA}
\affiliation{Astrophysics Science Division, NASA Goddard Space Flight Center, 8800 Greenbelt Rd, Greenbelt, MD 20771, USA}

\author[0000-0001-5384-3616]{Madisyn Brooks}
\affiliation{Department of Physics, 196A Auditorium Road, Unit 3046, University of Connecticut, Storrs, CT 06269, USA}

\author[0000-0002-7622-0208]{Callum T. Donnan}
\affiliation{NSF National Optical-Infrared Astronomy Research Laboratory, 950 N. Cherry Ave., Tucson, AZ 85719, USA}

\author[0000-0002-1404-5950]{James S. Dunlop}
\affiliation{Institute for Astronomy, University of Edinburgh, Royal Observatory, Edinburgh, EH9 3HJ, UK}

\author[0000-0001-8519-1130]{Steven L. Finkelstein}
\affiliation{Department of Astronomy, The University of Texas at Austin, Austin, TX, USA}
\affiliation{Cosmic Frontier Center, The University of Texas at Austin, Austin, TX, USA}

\author[0000-0002-3560-8599]{Maximilien Franco}
\affiliation{Université Paris-Saclay, Université Paris Cité, CEA, CNRS, AIM, 91191 Gif-sur-Yvette, France}.

\author[0000-0003-3248-5666]{Giovanni Gandolfi}
\affiliation{INAF – Osservatorio Astronomico di Roma, via Frascati 33, 00078, Monteporzio Catone, Italy}

\author[0000-0002-7831-8751]{Mauro Giavalisco}
\affiliation{University of Massachusetts Amherst, 710 North Pleasant Street, Amherst, MA 01003-9305, USA}

\author[0000-0001-9440-8872]{Norman A. Grogin}
\affiliation{Space Telescope Science Institute, 3700 San Martin Drive, Baltimore, MD 21218, USA}

\author[0000-0002-3301-3321]{Michaela Hirschmann}
\affiliation{Institute of Physics, Laboratory of Galaxy Evolution, Ecole Polytechnique Federale de Lausanne (EPFL), Observatoire de Sauverny, 1290 Versoix, Switzerland}

\author[0000-0001-9187-3605]{Jeyhan S. Kartaltepe}
\affiliation{Laboratory for Multiwavelength Astrophysics, School of Physics and Astronomy, Rochester Institute of Technology, 84 Lomb Memorial Drive, Rochester, NY 14623, USA}

\author[0000-0002-6610-2048]{Anton M. Koekemoer}
\affiliation{Space Telescope Science Institute, 3700 San Martin Drive, Baltimore, MD 21218, USA}

\author[0000-0003-2366-8858]{Rebecca L. Larson}
\altaffiliation{Giacconi Postdoctoral Fellow}
\affil{Space Telescope Science Institute, 3700 San Martin Drive, Baltimore, MD 21218, USA}

\author[0000-0002-9393-6507]{Gene C. K. Leung}
\affiliation{MIT Kavli Institute for Astrophysics and Space Research, 77 Massachusetts Ave., Cambridge, MA 02139, USA}

\author[0000-0003-1581-7825]{Ray A. Lucas}
\affiliation{Space Telescope Science Institute, 3700 San Martin Drive, Baltimore, MD 21218, USA}

\author[0000-0001-8688-2443]{Elizabeth J.\ McGrath}
\affiliation{Department of Physics and Astronomy, Colby College, Waterville, ME 04901, USA}

\author[0000-0001-7503-8482]{Casey Papovich}
\affiliation{Department of Physics and Astronomy, Texas A\&M University, College Station, TX, 77843-4242 USA}
\affiliation{George P.\ and Cynthia Woods Mitchell Institute for Fundamental Physics and Astronomy, Texas A\&M University, College Station, TX, 77843-4242 USA}

\author[0000-0002-0939-9156]{Borja P\'erez-D\'iaz}
\affiliation{INAF – Osservatorio Astronomico di Roma, via Frascati 33, 00078, Monteporzio Catone, Italy}

\author[0000-0002-6748-6821]{Rachel S. Somerville}
\affiliation{Center for Computational Astrophysics, Flatiron Institute, 162 5th Avenue, New York, NY 10010, USA}

\author[0000-0001-8728-2984]{Elizabeth Taylor}
\affiliation{Institute for Astronomy, University of Edinburgh, Royal Observatory, Edinburgh EH9 3HJ, UK}

\author[0000-0003-1282-7454]{Anthony J. Taylor}
\affiliation{Department of Astronomy, The University of Texas at Austin, Austin, TX, USA}
\affiliation{Cosmic Frontier Center, The University of Texas at Austin, Austin, TX, USA}

\author[0000-0002-9909-3491]{Roberta Tripodi}
\affiliation{INAF – Osservatorio Astronomico di Roma, via Frascati 33, 00078, Monteporzio Catone, Italy}

\author[0000-0003-3466-035X]{{L. Y. Aaron} {Yung}}
\altaffiliation{Giacconi Postdoctoral Fellow}
\affiliation{Space Telescope Science Institute, 3700 San Martin Dr., Baltimore, MD 21218, USA}

\author[0000-0002-9373-3865]{Xin Wang}
\affiliation{School of Astronomy and Space Science, University of Chinese Academy of Sciences (UCAS), Beijing 100049, China}
\affiliation{National Astronomical Observatories, Chinese Academy of Sciences, Beijing 100101, China}
\affiliation{Institute for Frontiers in Astronomy and Astrophysics, Beijing Normal University, Beijing 102206, China}
\shorttitle{The spectral types of LRDs}

\begin{abstract} 

One of JWST’s most unexpected discoveries is the emergence of “Little Red Dots” (LRDs): compact sources at $z \gtrsim 3$ with blue rest-frame UV continua, red optical slopes, and broad Balmer emission lines that challenge standard models and suggest a population of early, unusual active galactic nuclei (AGNs). Using a comprehensive photometric selection and public NIRSpec/PRISM spectroscopy across six JWST deep fields, we identify a large sample of 118 LRDs with high-S/N spectra, enabling a population-wide analysis of their UV–optical continuum and emission lines. We find clear correlations between rest-frame UV–optical color ([0.3$-$0.9 $\mu$m]) and slopes: bluer LRDs have blue UV slopes ($\beta_{\nu,\mathrm{UV}} \sim 0.3$) and red optical slopes, while redder LRDs exhibit redder UV slopes ($\beta_{\nu,\mathrm{UV}} \sim 1.1$). The continuum shape shows a similar trend: redder LRDs display prominent Balmer breaks and curvature, while bluer LRDs follow power-law–like optical SEDs. From literature compilations, $\sim$60\% of known broad-line AGNs satisfy our LRD criteria, and up to 90\% of LRDs show broad Balmer lines. Emission-line diagnostics reveal a shift from high \Ha/\Hb\ and low \OIII/\Hb\ in redder LRDs to the opposite in bluer ones, along with stronger narrow-line equivalent widths, suggesting a transition from AGN- to host-dominated emission. We fit the spectra with a two-component model combining a gas-enshrouded black hole (BH) and a galaxy host. Redder LRDs require higher-luminosity, unreddened BHs and modestly reddened hosts; bluer LRDs require lower-luminosity, reddened BHs and dust-free galaxies. This framework reproduces the diversity in colors and spectral shape by varying BH luminosity, obscuration, and BH-to-host luminosity ratio.

\end{abstract}


\section{Introduction}
\label{s:intro}

One of the most surprising and actively debated discoveries from the first years of JWST observations is the emergence of a populous class of compact, red sources at $z > 4$, now widely known as Little Red Dots (LRDs). These objects exhibit: (1) remarkably small, point-like morphologies; (2) unusual spectral energy distributions (SEDs) combining blue rest-frame ultraviolet (UV) continua with red rest-frame optical colors; and (3) a high incidence of broad permitted emission lines, with FWHM values of 2000–4000 km~s$^{-1}$ \citep{labbe23, barro24, greene23, akins24, kokorev24, kocevski24, pgp23b, taylor24}.

Early interpretations of these properties suggest they could be dusty, compact, massive starburst galaxies \citep{labbe23, baggen23, pgp23a, barro24}, faint, obscured active galactic nuclei (AGNs) \citep{kocevski23, greene23, labbe24}, or a combination of both \citep{killi23, wang24_lrd, barro24, pg24a, leung24}. Despite the growing body of photometric and spectroscopic data, the physical nature of LRDs remains unresolved, as their properties differ significantly from those of typical high-redshift galaxies or AGNs, and are subject to strong modeling degeneracies that make interpretation difficult.

From the photometric perspective, large-area JWST surveys now routinely identify hundreds of LRDs across multiple deep fields using single- or multi-color selections designed to capture their distinctive “V-shaped” SEDs over a wide redshift range \citep{kokorev24, kocevski24, akins24, barro24}. However, the inferred surface densities span a wide range, from $\sim$0.25 to 1 arcmin$^{-2}$, depending on the adopted color thresholds, and increase systematically from the reddest to the bluest selections (e.g., F277W$-$F444W $>$ 1.5 versus F200W$-$F444W $>$ 1; \citealt{akins24, barro25}). This dependence on selection criteria indicates that LRDs constitute a far more diverse population than initially recognized, extending beyond the most extreme red objects and spanning a broad range of rest-frame UV-to-NIR colors of up to four magnitudes \citep{pg24a, barro25}.

Mid- and far-infrared photometry has played a key role in constraining the dust content and emission properties of LRDs, placing important limits on both star-forming and AGN-driven models. The lack of detections in individual ALMA observations and deep submillimeter stacks suggests that LRDs do not exhibit the high infrared luminosities typical of dusty star-forming galaxies \citep{labbe24, akins24, casey24}. If LRDs are dusty starbursts, their IR emission must arise from extremely compact, hot regions with unusually short-wavelength dust peaks \citep{pgp23a, pg24a}. Conversely, near-to-mid IR colors from MIRI observations \citep{williams24, pg24a, akins24} are inconsistent with the warm dust torus emission expected from classical obscured AGNs. If LRDs are AGNs, they may instead host hot-dust-deficient tori \citep[e.g.,][]{lyu17}, with emission peaking at much longer wavelengths ($\sim$10–100~$\mu$m) that remain largely inaccessible to current infrared facilities \citep{casey24, leung24, barro24}.

On the spectroscopic front, JWST/NIRSpec MSA \citep{ferruit22} has delivered over 200 spectra of LRDs, revealing emission lines in nearly all cases and firmly establishing the widespread presence of broad-line emission across the population \citep{greene23, kocevski24, taylor24, juodzbalis25, hviding25}. Just as critical, continuum detections have significantly shaped physical interpretations of LRDs. Initial reports of Balmer breaks in several sources pointed to stellar-dominated continua \citep{wang24_lrd, wang24b, labbe24b}. However, further observations revealed larger breaks increasingly inconsistent with stellar populations \citep{labbe24b, ma25, ji25}, culminating in extreme cases with break strengths above 5 magnitudes \citep{degraaff25, naidu25, taylor25} that challenged the stellar origin.

This has led to alternative AGN models that invoke dense gas envelopes around accreting black holes similar to “quasi-stars” \citep{begelman25} or “black hole stars” \citep{inayoshi24, ji25} that can naturally produce Balmer breaks as strong or stronger than those of evolved stars. These dense gas cocoons may also account for the narrow absorption features seen in some broad Balmer lines \citep{matthee24, kocevski24, deugenio25b, torralba25b}, and their intrinsically red optical continua reduce the need for high dust reddening, easing tensions with mid- and far-infrared constraints. These advantages make variations of the dense-gas scenario a promising framework for explaining the nature of LRDs.

Still, while dense gas envelopes appear important in some LRDs, they do not yet offer a population-wide explanation. Can this model account for the full range of observed colors? What about LRDs without strong Balmer breaks or absorption features? Answering these questions requires determining what fraction of the population exhibits these traits. However, most current studies rely on small or heterogeneous samples, often based on photometric selections without spectroscopic confirmation or spectroscopic datasets that do not consistently apply photometric LRD criteria. This mismatch makes it difficult to assess whether features like broad lines, steep optical slopes, or extreme Balmer breaks are common or limited to specific subsets.

In this paper, we address these issues by constructing a large, uniformly selected sample of photometric LRDs across six JWST deep fields. In \S~\ref{s:sample}, we apply an inclusive, color-based selection that recovers the full photometric diversity reported in previous studies, including sources missed by more restrictive criteria. We then cross-match this sample with all available NIRSpec spectroscopy compiled in the Dawn JWST Archive (DJA)\footnote{\url{https://dawn-cph.github.io/dja}}, enabling a systematic comparison between photometric and spectroscopic properties. In \S~\ref{s:restSED}, we use continuum spectroscopy to measure rest-frame UV and optical slopes and colors, free from emission-line contamination, and to characterize population-wide correlations among these quantities. Finally, in \S~\ref{s:model}, we interpret these trends within a semi-empirical two-component model combining a galaxy and a BH$^{\star}$ component, demonstrating that it naturally reproduces the observed range of LRD continuum and emission-line properties.

Throughout the paper, we assume a flat cosmology with $\mathrm{\Omega_M\, =\, 0.3,\, \Omega_{\Lambda}\, =\, 0.7}$, and a Hubble constant $\mathrm{H_0\, =\, 70\, km\,s^{-1} Mpc^{-1}}$. We use AB magnitudes \citep{oke83}.

\section{Data}
\label{s:data}

\subsection{JWST/NIRCam Photometry}

For source identification and photometry in the CEERS-EGS, PRIMER-UDS, PRIMER-COSMOS, and UNCOVER fields, we use version v7.2 of the publicly available photometric catalogs from the DJA. These catalogs are derived from JWST/NIRCam Level-2 imaging processed with the grizli pipeline \citep{brammer_grizli}, including PSF-matching across filters. Source detection is performed using SEP \citep{SEP}, a Python implementation of SExtractor \citep{sex}, applied to a detection image typically based on F444W. Photometry is measured in 0\farcs5 diameter circular apertures and corrected to total flux using Kron scaling \citep{kron80}, with filter-dependent aperture corrections applied based on empirical PSFs (see \citealt{valentino23} and Brammer et al., in prep., for details). For the JADES fields (GOODS-N and GOODS-S), we use the official Data Release 3 (DR3) multi-band photometric catalogs \citep{eisenstein24,deugeniojades25}, which include consistent photometry across NIRCam bands based on the JADES pipeline.

\subsection{JWST/NIRSpec Spectroscopy}

For NIRSpec spectroscopy, we focus primarily on low-resolution (R~$\sim$~100) PRISM observations, as our primary goal is to characterize the continuum shapes of LRDs. Some results based on higher-resolution grating data are also discussed where relevant. Spectra are drawn from version 4.4 of the DJA, which compiles uniformly reduced NIRSpec data across all major extragalactic surveys. Reductions were performed with the latest version of the open-source pipeline \texttt{msaexp} \citep{brammer23}, as described in \citet{degraaff24} and \citet{heintz25}. The DJA processing includes background subtraction, flat-fielding, 2D rectification, and optimal 1D extraction, along with masking of contaminated regions and quality flags for each pixel. All spectra include associated error, contamination flags, and metadata for reproducibility.

Specifically, we cross-match source with robust spectroscopic redshift (grade=3) in the DJA spectroscopic table with photometric catalogs in each field using a 0\farcs5 matching radius to identify spectra corresponding to photometrically selected LRDs (see next section). Approximately three-quarters of the spectra come from three large programs: the Red Unknowns: Bright Infrared Extragalactic Survey (RUBIES; GO-4233; \citealt{degraaff24}), and the CANDELS-Area Prism Epoch of Reionization Survey (CAPERS; GO-6838; PI: Dickinson), which target the CEERS-EGS, PRIMER-UDS, and PRIMER-COSMOS (CAPERS only) fields, and the JWST Advanced Deep Extragalactic Survey (JADES; GTO-1180, 1181, 1286; \citealt{curtislake25, scholtz25}) covering GOODS-N and GOODS-S. These contribute 30, 28, and 30 spectra to our sample, respectively (25\%, 23\%, and 25\%). An additional 11 LRDs ($\sim$10\%) come from spectroscopic follow-up in the UNCOVER field (GO-2561; \citealt{bezanson24}). The remaining 20 sources ($\sim$17\%) are drawn from a range of smaller programs, including GTO-1215 \citep{maseda24}, GO-5224 \citep{naidu25}, CEERS (GO-1345; \citealt{finkelstein25}), DDT-2750 (PI: Arrabal Haro), GO-4106 (PIs: Nelson \& Labbe), GO-6585 (PI: Coulter), GO-3073 (PI: Castellano), DDT-6541 (PI: Egami), and GO-2198 (PI: Barrufet).

\section{Photometric selection of LRDs}
\label{s:sample}

\subsection{Selection criteria and completeness}
\label{ss:selection}

We identify LRDs using the photometric selection method introduced by \citet{barro25}, applied uniformly across six JWST deep fields. Briefly, we use an observed color–color criterion (F200W–F444W vs. F115W–F200W) designed to identify the characteristic blue UV and red optical SEDs of LRDs. The selection region boundaries are defined to encompass the redshift–color tracks of representative LRDs spanning a broad range of rest-frame colors at $z \gtrsim 3$. We use F200W, rather than the previously adopted F277W, as the pivot filter to better sample the rest-frame UV at lower redshifts and avoid contamination from emission lines in both color axes. As in prior work, we also apply a compactness cut based on the flux ratio in two small apertures (F444W[0\farcs5]/F444W[0\farcs2]~$<$~1.5).

\begin{figure}
\centering
\includegraphics[width=8.2cm,angle=0]{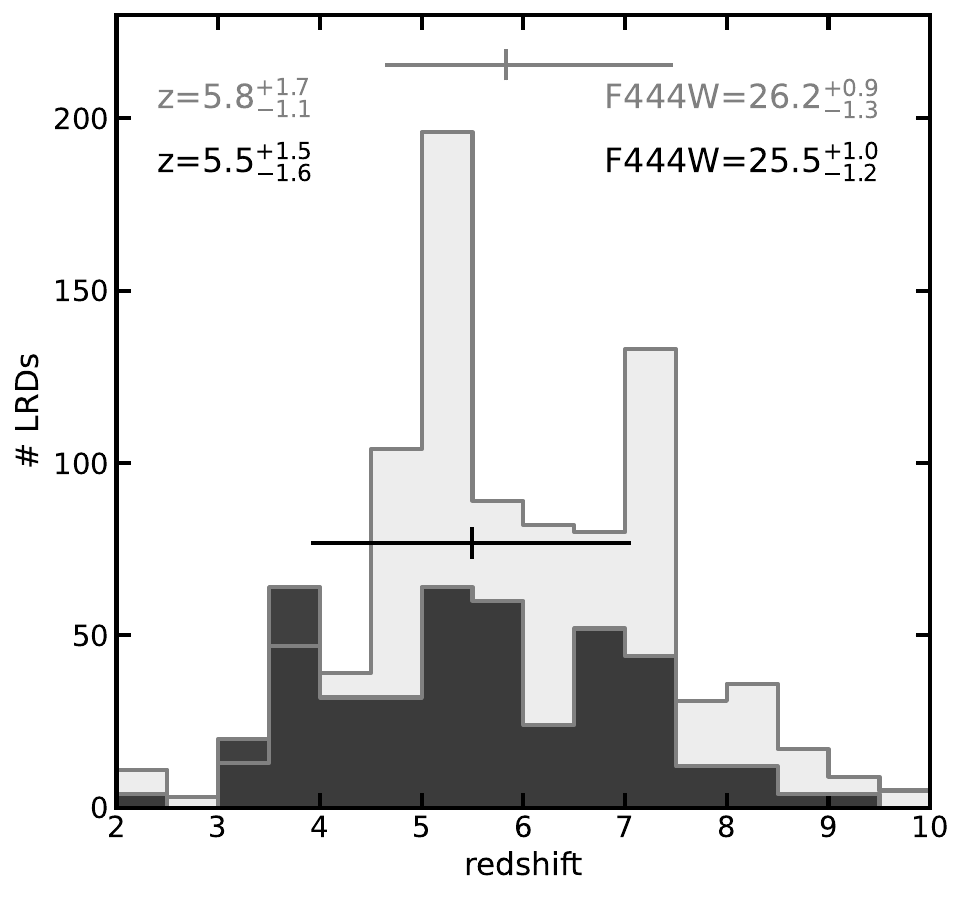}
\caption{ Redshift distribution of the photometric LRD sample (gray) and the NIRSpec/PRISM spectroscopic subset (black, scaled by a factor of 3). Horizontal lines show the median redshift and percentiles for each sample. The full sample spans $z\sim3–9$ with peaks at $z\sim5$ and $z\sim7$; the spectroscopic subset is skewed slightly lower, likely due to selection biases in spectroscopic follow-up favoring brighter sources.}
\label{fig:redshift_hist}
\end{figure}

This approach captures nearly all candidates identified by the selection criteria of \citet{kocevski24} and \citet{kokorev24}, while recovering an additional $\sim$20\% of sources with bluer UV-to-NIR colors that fall outside those earlier definitions. Although optimized for completeness at $z \sim 3$–7, the selection becomes increasingly incomplete at higher redshifts ($z \gtrsim 8$), where the F150W–F200W color begins to trace the Lyman break, shifting sources out of the selection region. Nonetheless, this method probes a broader region of color–color space, providing the most inclusive and comprehensive sampling of the photometric diversity within the LRD population.

\begin{figure*}
\centering
\includegraphics[width=18cm,angle=0]{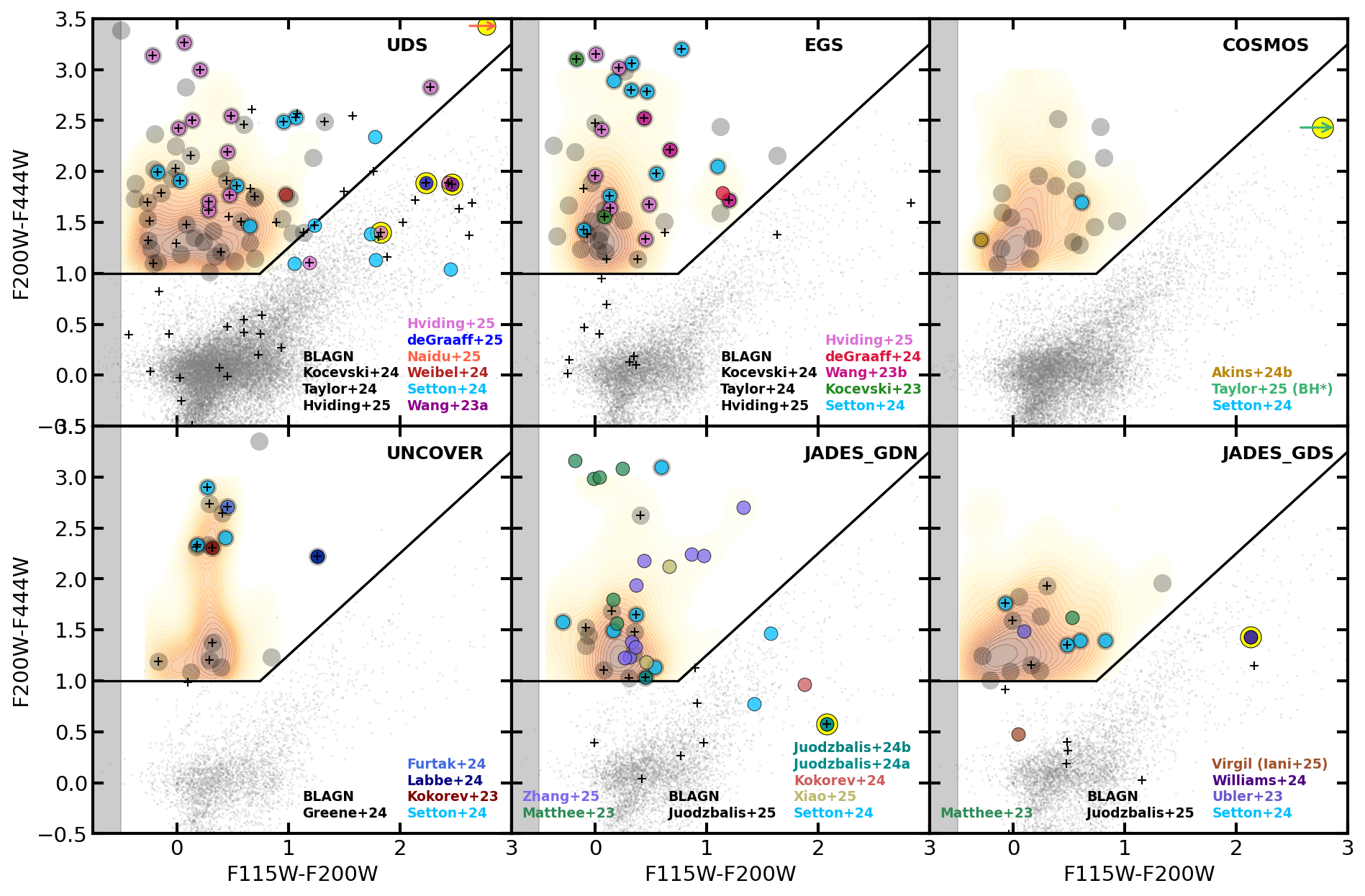}
\caption{Color–color selection diagram of LRDs based on the \citet{barro25} criterion (solid lines; gray area to reject brown dwarf contaminants) for the six JWST deep fields analyzed in this work. Small gray dots show the full photometric catalog, while the orange density map marks the distribution of sources selected as photometric LRDs. Gray circles indicate LRDs with high-quality NIRSpec/PRISM spectra from DJA. Colored symbols highlight literature LRDs within this sample; when an object appears in multiple studies, only the original reference is shown. Black crosses denote sources classified as BLAGNs in previous studies.  Yellow circles indicate compact spectroscopic LRDs that were missed by the photometric selection and were added manually for completeness. The diagram demonstrates that the vast majority of previously identified spectroscopic LRDs fall within the photometric selection region.}
\label{fig:selection_criteria}
\end{figure*}

Overall, we identify $\sim$1000 LRDs down to a limiting magnitude of F444W$=27.5$ (28 in the JADES fields), including 191 LRDs with NIRSpec/PRISM spectra in DJA over an area of 632.4 arcmin$^{2}$. Approximately 70\% of the spectra have sufficient signal-to-noise in the rest-frame UV and optical continuum for SED fitting (see \S~\ref{s:restSED} and \ref{s:model}). Sample sizes by field are summarized in Table~\ref{tab:agn_lrd_fractions}. Figure~\ref{fig:redshift_hist} shows the redshift distribution for the full photometric LRD sample (gray) and the NIRSpec/PRISM spectroscopic subset (black, scaled by a factor of 3). The photometric sample spans a wide range from $z\sim3$ to 9, with two prominent peaks at $z\sim5$ and $z\sim7$, and a median redshift of $z=5.8^{+1.7}_{-1.1}$. This is consistent with \citet{barro25}, but our analysis expands the sample from 248 to nearly 1000 LRDs and extends coverage to all six JWST deep fields. As in \citet{barro25}, the lower-redshift peak is partly driven by LRDs with bluer optical colors and strong \OIII\ emission that boosts F277W; these are recovered by our selection and by \citet{kokorev24}, but are largely missed by redder cuts such as those in \citet{kocevski24}, which yield a higher median redshift of $z\sim6.4$. In contrast, our selection becomes increasingly incomplete at $z\gtrsim8$, where the F150W-F200W color begins tracing the Lyman break. The spectroscopic subset shows a slightly lower median redshift of $z=5.5^{+1.5}_{-1.6}$, likely reflecting selection biases toward brighter targets more easily observed in follow-up. This is consistent with their brighter median F444W$=25.5$, compared to 26.2 for the full photometric sample.

Figure~\ref{fig:selection_criteria} shows the LRD color–color selection diagram from \citet{barro25} for all six fields, with the photometric LRD population shown as an orange density map and spectroscopic LRDs as gray circles. Most spectroscopically identified LRDs from the literature fall within our photometric selection region, demonstrating excellent agreement between photometric and spectroscopic classifications. This includes all NIRCam/WFSS BLAGNs from \citet{matthee24} and $\sim$76\% of the red, high–equivalent-width (EW$>400$,\AA) sources from \citet{setton24}. The remaining $\sim$20\% are predominantly extended sources \citep{setton24} with red rest-frame UV–optical SEDs characteristic of dusty, massive galaxies (e.g., 6585-COS-60504, RUBIES-UDS-41598, RUBIES-UDS-152282, RUBIES-UDS-19735, WIDE-GDN-6406). Excluding these extended sources increases the selection completeness to 87\%.

There are, however, a small number of compact, spectroscopic LRDs with broad emission lines that are missed by our photometric selection, as well as by most other criteria. These sources typically have relatively low redshifts ($z\lesssim3$) and/or unusually red rest-frame UV slopes, which shift their F115W$-$F200W colors out of the LRD selection region and into the locus of massive dusty galaxies (see Figure~1 of \citealt{barro25}). We identify seven such objects in the literature and include them manually in the photometric sample for completeness; they are marked with yellow circles in Figure~\ref{fig:selection_criteria}. Notably, RUBIES-BLAGN-1 \citep{wang24b}, JADES-1180-12402 \citep{setton24,williams23}, the ``Rosetta'' AGN \citep{juodzbalis24a}, and RUBIES-UDS-144195 \citep{hviding25} are among the lowest-redshift LRDs, at $z=3.11$, 3.19, 2.26, and 3.404, respectively. All but the ``Rosetta'' source show significantly redder UV slopes (see discussion in \S~\ref{s:restSED}). We additionally include the three reported extreme Balmer break sources (``black hole stars'', BH$^{\star}$) from \citet{degraaff25}, \citet{naidu25}, and \citet{taylor25}, which exhibit weak UV emission and are identified solely by their extremely red optical colors.


While the addition of these six spectroscopic LRDs contributes negligibly to the overall photometric sample, and less than 1\% to the spectroscopic one, they highlight the need for a more complete diagnostic criterion based on rest-frame continuum properties. Classification using rest-frame UV and optical slopes \citep{kocevski24, hainline24} or UV-to-optical colors \citep{barro24} offers a more robust way to organize and distinguish LRD subtypes. As we show in \S~\ref{s:restSED}, these six sources fall at the red end of the rest-frame color diagram but are otherwise consistent with the broader LRD population.

Rest-frame colors or slopes provide a redshift-independent and physically motivated method for identifying LRDs based on their distinctive SED shapes, similar to how the UVJ diagram \citep{wuyts07, williams10} separates quiescent, dusty, and star-forming galaxies. However, such classifications depend critically on how rest-frame properties are derived. The most reliable approach would use low-resolution, broad-wavelength spectroscopy (e.g., NIRSpec/PRISM), which clearly separates continuum and line emission. However, while low-resolution spectra are only available for limited samples, broad photometric coverage extending into the mid-IR (e.g., NIRCam + MIRI F560W/F770W) can effectively trace the rising optical SED even at $z \gtrsim 6$ and mitigate contamination from strong emission lines in NIRCam bands (especially F277W, F356W, and F444W).

A key limitation of color-based selections in discrete redshift bins, such as that used by \citet{kocevski24}, is that individual bands can be biased by line contamination (e.g., F277W at $z \sim 5$; see \citealt{barro24}), resulting in incompleteness. By contrast, fitting the full photometric SED at the best available redshift allows for interpolation across uncontaminated bands or SED modeling with templates that include emission lines, yielding more reliable estimates of the intrinsic continuum slope or color.

\subsection{Other compact, non-LRD sources}
\label{s:other_nonlrd}

Compact quiescent galaxies represent another population that can overlap with and contaminate LRD selections. These sources share key observable features, red colors, compact morphologies, and, in some cases, broad emission lines \citep{kocevski23b}, but their intrinsic SEDs differ markedly from those of LRDs. Specifically, quiescent galaxies exhibit red UV slopes, and while their optical continua are steep near the Balmer break, they flatten rapidly at longer wavelengths, deviating from the approximate power-law behavior typical of LRDs.

Figure~\ref{fig:selection_criteria} highlights the locations of three recently reported quiescent galaxies at high redshift from \citet{kokorev24}, \citet{weibel24}, and \citet{degraaff24} in the photometric selection diagram. The latter two fall within the LRD selection region near its red boundary, making them potential contaminants. This occurs because, at $z>5$, the F200W band samples wavelengths blueward of the Balmer break, causing F200W–F444W colors to appear redder and F115W–F200W colors bluer, thereby pushing quiescent galaxies into the LRD selection box. In the next section, we demonstrate that a rest-frame color–color diagram can effectively distinguish these contaminants based on their distinct SED shapes.

Another interesting source is ``Virgil", a spectroscopically confirmed Lyman-$\alpha$ emitter at z$=$6.6 in JADES-GDS (\citealt{iani24}; \citealt{rinaldi25}), which exhibits a rest-frame UV–optical SED typical of LAEs at similar redshifts. In NIRCam bands, its SED is entirely blue and would place it well outside the standard LRD selection region, in the lower-left corner of the photometric diagram. However, MIRI photometry reveals a steeply rising red continuum beyond rest-frame $\sim$0.7$\mu$m, producing an optical-to-NIR slope similar to that seen in LRDs. If rest-frame colors incorporating these longer-wavelength data were used, ``Virgil" would fall within the LRD color space. This example highlights the possibility of a hidden population of LRD-like sources that are missed by traditional NIRCam-based selections. It suggests a possible scenario in which the component dominating the rest-frame optical emission, likely an exotic type of AGN or compact dusty star formation,  can vary in luminosity relative to a relatively blue, UV-dominated host, producing a continuum of UV–NIR colors. We explore this connection further in the next section using a rest-frame color–color diagram.

\begin{table*}
\centering
\small
\begin{tabular}{llcccccc}
\hline
\textbf{Field} & \textbf{Area}   & \textbf{Phot. LRDs} & \textbf{Spec. LRDs} & \textbf{S/N$_{\rm UV}>2$} & \textbf{BLAGNs LRDs} & \textbf{LRDs BL} \\
 & (arcmin$^{2}$)       &        &     &    &    &                        \\
\hline
CEERS-EGS   & 88.6        &      127            &     45                &    25                        & 61\%                               & 89\%                                \\
PRIMER-UDS   & 243.0  &      353            &     68                &    33                        & 59\%                               & 89\%                                \\
JADES-GDS    & 62.1  &      92             &     21                &    17                        & 50\%                               & 70\%                                \\
JADES-GDN    & 56.0  &      112            &     19                &    16                        & 67\%                               & 54\%                                \\
UNCOVER       & 43.8 &      45             &     14                &    12                        & 92\%                               & 65\%                                \\
PRIMER-COSMOS  & 138.9 &      207            &     23                &    15                         & N/A                                & N/A                                 \\
\hline
\hline
Total & 632.4 & 968 & 191 & 118 & - & - \\
\hline
\end{tabular}
\caption{Summary of LRD and AGN statistics by field, including photometric and spectroscopic sample sizes, and the fraction of BLAGNs classified as LRDs and LRDs with broad emission lines.}
\label{tab:agn_lrd_fractions}
\end{table*}

\subsection{Spectroscopic samples of BLAGNs}

Figure~\ref{fig:selection_criteria} shows the distribution of a large compilation of broad-line AGNs (BLAGNs) from the literature in the photometric selection diagrams for each field (black crosses). These samples are primarily based on NIRSpec grating observations \citep{kocevski23, harikane23, kocevski24, maiolino23, taylor24, hviding25}, prism spectra \citep{greene23}, and NIRCam/WFSS data \citep{matthee24}. The goal is to quantify (i) what fraction of BLAGNs are identified as LRDs by photometry, and (ii) what fraction of photometric LRDs exhibit broad lines.

In CEERS-EGS, we compile BLAGN measurements from \citet{kocevski23}, \citet{harikane23}, \citet{kocevski24}, \citet{taylor24}, and \citet{hviding25}. The first two studies identified BLAGNs using data from the CEERS spectroscopic program. \citet{kocevski24} extended this work using early observations from the RUBIES program. \citet{hviding25} and \citet{taylor24} further expanded the sample, conducting more comprehensive searches for broad H$\alpha$ emission using the full RUBIES and combined RUBIES+CEERS datasets, respectively. Consequently, the \citet{hviding25} and \citet{taylor24} samples are the largest. We identify 25 of the 28 BLAGNs from \citet{taylor24} within the area covered by our photometric catalog, and all 28 from \citet{hviding25}. Combined, these yield 40 unique BLAGNs, with only $\sim$30\% (13) in common between the two samples. All 8 sources from \citet{kocevski24}, and all but one from \citet{harikane23}, are also included in this combined set. Including the additional \citet{harikane23} source brings the total to 41 BLAGNs. Of these, 25 (61\%) satisfy our photometric LRD selection.

A fraction of 61\% (25/41) of BLAGNs classified as photometric LRDs is notably higher than the $\sim$34\% reported in \citet{taylor24} and the $\sim$54\% in \citet{hviding25}. The larger discrepancy with \citet{taylor24} arises because their analysis identified photometric LRDs using only the less complete \citet{kocevski24} sample. The difference with \citet{hviding25} likely stems from the inclusion of bluer LRDs in our selection. Their classification relies on a combination of compact morphology, broad-line emission, and a “V-shaped” SED. The latter requires a spectroscopic optical slope of $\beta_{\lambda,\rm OPT}>0$ (i.e., $\beta_{\nu,\rm OPT}>2$), similar to the threshold in \citet{kocevski24}, whereas our method extends to bluer slopes (see next section). In turn, all of the spectroscopic LRDs (15) in \citet{hviding25} are photometric LRDs by our selection.

Estimating the fraction of LRDs that exhibit broad lines is more complex, as it depends on knowing which photometric LRDs were analyzed in BLAGN surveys with sufficiently deep spectra for BLAGN identification. To compare with the more complete samples from \citet{hviding25} and \citet{taylor24}, we identify photometric LRDs with reliable (grade=3 in DJA) G395M grating spectra from RUBIES and redshifts in the range $3.5 < z < 6.8$, ensuring that H$\alpha$ falls within the spectral coverage. We find 19 such LRDs, of which 17 (89\%) exhibit broad lines. The two exceptions, RUBIES-EGS-69475 and -57375, have very low S/N in the optical continuum, which may prevent reliable linewidth measurements. One additional LRD (RUBIES-EGS-9809) lacks a broad-line classification because H$\alpha$ falls in the chip gap.

In UDS, we again use BLAGN samples from \citet{kocevski24}, \citet{taylor24}, and \citet{hviding25}, all based on spectra from the RUBIES survey. We identify all 35 sources in \citet{taylor24}, 50 of 51 from \citet{hviding25}, and 6 from \citet{kocevski24}. The combined \citet{taylor24} and \citet{hviding25} samples yield a total of 69 unique BLAGNs, with only $\sim$25\% (17) overlapping between them. Adding two additional sources found only in \citet{kocevski24} brings the total to 71. Of these, 42 (59\%) satisfy our photometric LRD criterion. This fraction is again higher than those reported in \citet{taylor24} and \citet{hviding25}, which are $34\%$ and $40\%$, respectively.

We also find that 19 of the 20 spectroscopic LRDs from \citet{hviding25} within our photometric catalog are classified as photometric LRDs. However, three of these were manually added due to their combination of lower redshift and/or red UV slopes that make them fall outside our photometric selection (RUBIES-BLAGN1,``The Cliff", and RUBIES-144195; see previous section). The one remaining exception, RUBIES-33938, is relatively faint with a low S/N spectrum, has a moderate color (F277W$-$F444W $\sim$ 1~mag), and is not identified as a photometric LRD in any prior studies.

Among LRDs with good-quality G395M grating spectra from RUBIES and spectroscopic redshifts between $3.5 < z < 6.8$, we identify 36 sources. Of these, 32 (89\%) are classified as BLAGNs in the literature. The remaining four: RUBIES-166271, 40800, 59492, and 125917, have low optical S/N, which complicates reliable line width measurements and does not rule out the presence of faint broad lines.

In JADES-GDS and GDN, we use BLAGN identifications from \citet{juodzbalis25}, based on multiple GTO spectroscopic programs in the JADES fields. Most of these measurements rely on R1000 grating data, primarily G395M, with G235M used at lower redshift. The sample includes 12 BLAGNs from \citet{maiolino23} and one from \citet{matthee24}. In GDS, we identify 14 of the 16 BLAGNs in \citet{juodzbalis25}; the other two are either outside the photometric mosaic or blended with a foreground galaxy. Of the 14, seven (50\%) satisfy the photometric LRD selection. To estimate the LRD fraction with broad lines, we identify ten LRDs with good-quality grating spectra ($z < 6.8$), of which seven (70\%) show broad lines in \citet{juodzbalis25}. In GDN, we identify all 15 BLAGNs in \citet{juodzbalis25}, of which 10 (67\%) are photometric LRDs. We also find 13 LRDs with grating spectra at $z < 6.8$, of which seven (54\%) have broad lines. For completeness, we note that all seven BLAGNs in JADES-GDS and GDN identified by \citet{matthee24} also satisfy the LRD photometric selection.

In UNCOVER, we use the BLAGN sample from \citet{greene23}. This field differs from the others in that the broad-line measurements are based on lower-resolution, but typically higher-S/N, NIRSpec prism data rather than gratings. We identify all 12 sources in the sample, of which 11 (92\%) satisfy the photometric LRD selection (ID 35488 lies just outside the selection boundary). To estimate the spectroscopic overlap, we identify 17 photometric LRDs that were targeted by the UNCOVER spectroscopic program; of these, 11 (65\%) are classified as BLAGNs in \citet{greene23}.

In PRIMER-COSMOS, no large-scale spectroscopic programs for BLAGN identification are currently available. The ongoing CAPERS program (PI: Dickinson) may provide this information in the future, but only from PRISM data.

Overall, we find that the fraction of LRDs among samples of BLAGNs is relatively high, typically ranging from ~60\% up to 90\% in UNCOVER. These values are higher than those reported in earlier studies using the same BLAGN samples, primarily due to (1) the adoption of a more flexible photometric selection that extends to bluer UV-to-optical colors (i.e., shallower optical slopes), and (2) the inclusion of a small number of spectroscopic LRDs at lower redshift that fall outside traditional color cuts. For example, \citet{taylor24} report an LRD fraction of $\sim$34\% among BLAGN in CEERS-EGS and PRIMER-UDS, while \citet{hainline24} find a value of $\sim$30\% in JADES after excluding three LRD-targeted sources. Indeed, the spectroscopic follow-up strategy plays a significant role in these differences. The 90\% fraction in UNCOVER reflects a strong targeting bias, as many sources were selected specifically to follow up the LRDs identified in \citet{labbe23}. Similarly, the RUBIES survey prioritized red NIRCam sources, which naturally favors LRD-like objects. As noted by \citet{greene23}, both the equivalent width and FWHM of broad lines tend to be larger in the reddest sources (e.g., F277W$-$F444W $>$ 1.6), making BLAGNs that are also LRDs easier to detect. \citet{hainline24} and \citet{zhang25} reach a similar conclusion, showing that LRDs dominate the high-luminosity end of the AGN population. Consequently, it is clear that larger, unbiased spectroscopic samples are needed to reliably assess the intrinsic fraction of LRDs among the full BLAGN population.

The fraction of LRDs with broad emission lines is also relatively high, typically around 50–90\%. Importantly, these values represent lower limits, as nearly all LRDs show clear emission lines, and the identification of broad components is often limited by the depth and sensitivity of the observations.


\section{Rest-frame UV Slopes and Optical Colors}
\label{s:restSED}

This section analyzes the distribution of spectroscopic LRDs in a rest-frame color versus slope diagram that characterizes their UV and optical continuum shapes. In the following subsections, we use this diagram as a framework to classify the LRD population and investigate trends in their SEDs, including the shape of the continuum and the properties of their emission lines such as FWHM, flux ratios, and equivalent widths.

\subsection{Rest-frame color - slope diagram classification}

Here, we analyze a subset of 118 LRDs (out of 191) with NIRSpec/PRISM spectroscopy and high S/N continuum in the UV. We require an average S/N per pixel $>2$ in the range $\lambda_{\rm UV}=1200-3500$~\AA. Since LRDs are typically red, this threshold also implies higher S/N in the optical, ensuring sufficient signal for reliable SED modeling. We compute UV slopes by fitting a power law to the continuum between the Lyman and Balmer limits, using the convention $f_{\nu} \propto \lambda^{\beta_{\nu}}$. The slope is measured in $f_{\nu}$ rather than the traditional $f_{\lambda}$, where $\beta_{\nu} = \beta_{\lambda} + 2$, to enable direct comparison with the rest-frame UV-to-optical color, \uvnir, defined using synthetic tophat filters of width 100~\AA.

\begin{figure}
\includegraphics[width=9cm,angle=0]{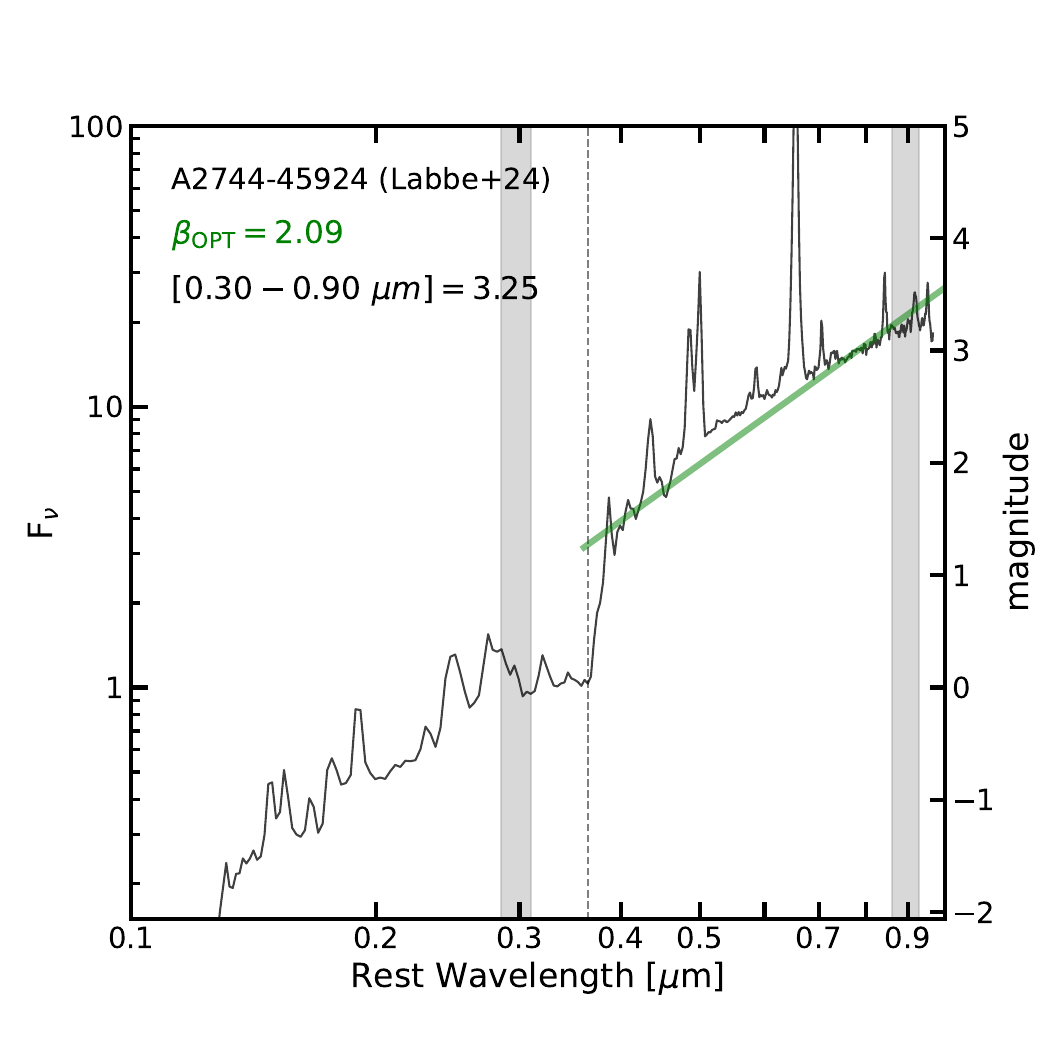}
\caption{Example of an LRD with a prominent Balmer break illustrating the difference between the optical slope and the rest-frame UV-to-optical color. The black line shows the NIRSpec/PRISM continuum spectrum, with a green line indicating the best-fit power law to the optical continuum redward of the Balmer limit (H$_\infty$). The shaded gray regions mark the wavelength ranges used to compute the \uvnir color. In LRDs with prominent continuum breaks, the rest-frame color is often significantly larger than the optical slope, as the latter fails to capture the presence of a sharp discontinuity.}
\label{fig:beta_vs_color}
\end{figure}

To characterize the optical continuum, we adopt a rest-frame color rather than a formal slope. As noted in previous studies, many LRDs show significant continuum curvature and strong Balmer breaks \citep[e.g.,][]{wang24_lrd, furtak23, labbe24b, setton24}, making simple power-law slopes inadequate. Figure~\ref{fig:beta_vs_color} illustrates one such case: A2744-45924 \citep{labbe24b}, which exhibits a prominent Balmer break. The optical slope measured redward of H$\infty$ (green line) underestimates the true curvature and fails to capture the discontinuity. In contrast, the \uvnir color spans a broader wavelength baseline across the Balmer break and reflects those features more robustly. As a result, LRDs with prominent breaks often have \uvnir colors that exceed what their optical slope would predict.

For comparison with earlier work, this rest-frame optical color can be interpreted approximately as a two-point slope, with a numerical relation similar to:

\begin{equation}
    \beta_{\nu, {\rm OPT}} = \frac{0.4(m_{0.3} - m_{0.9})}{\log(0.3\,\mu\mathrm{m} / 0.9\,\mu\mathrm{m})} \quad = 1.2\cdot[0.3{-}0.9\,\mu\mathrm{m}]
\end{equation}

\begin{figure*}
\centering
\includegraphics[width=9.2cm,angle=0]{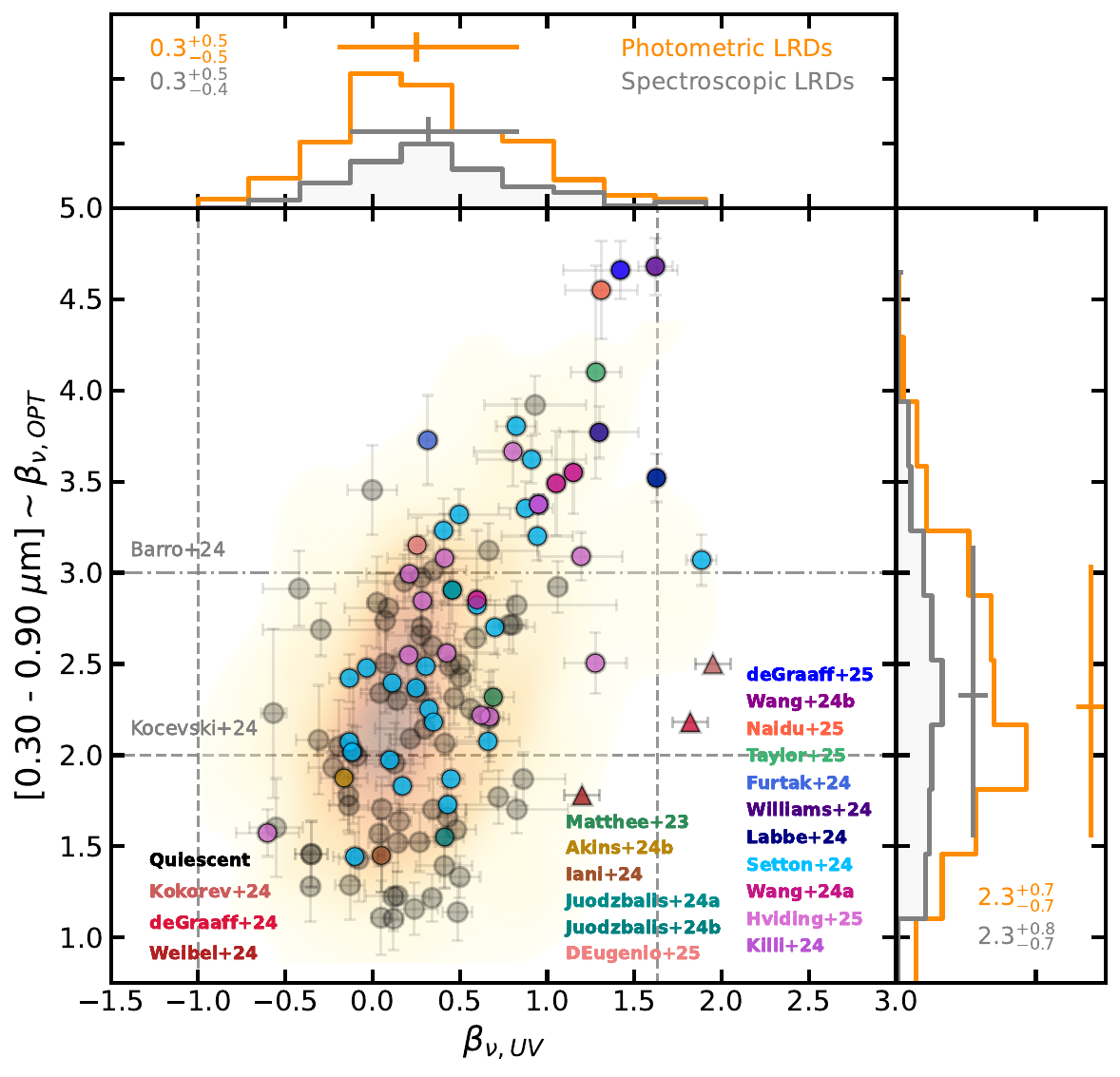}
\includegraphics[width=8cm,angle=0]{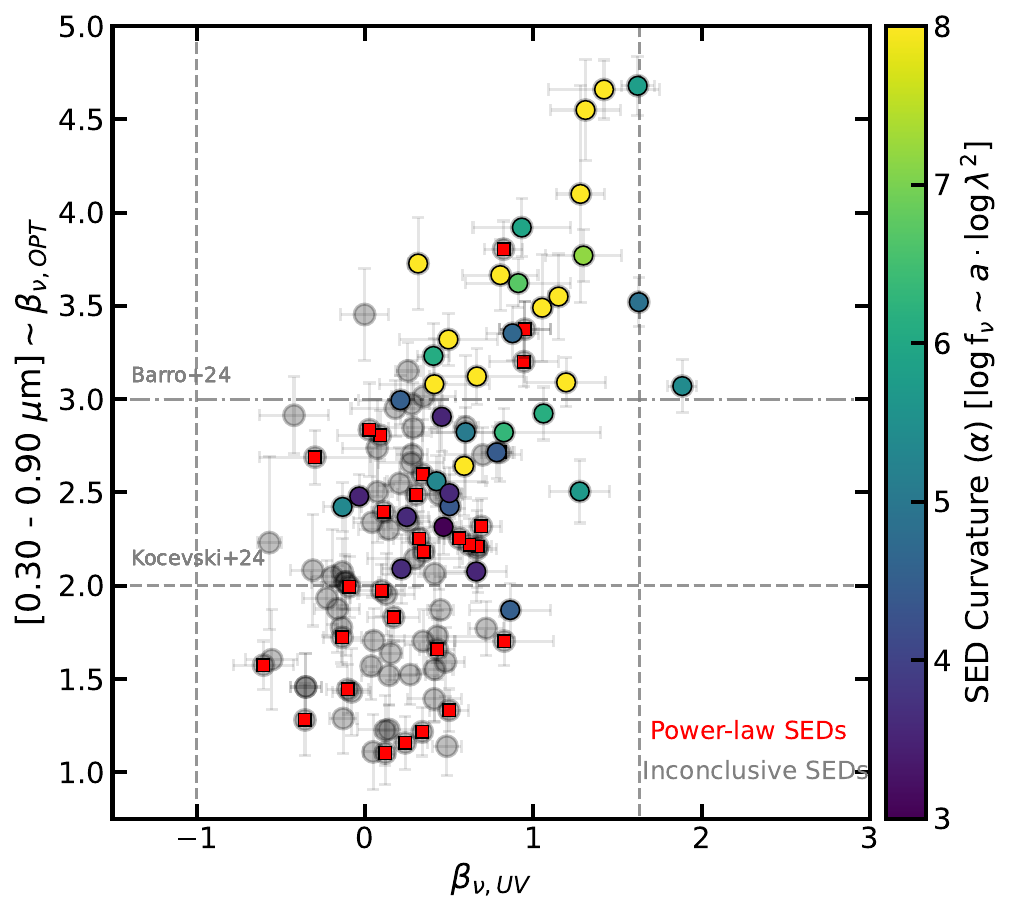}
\caption{Rest-frame UV slope versus optical color for LRDs with high-S/N (S/N$_{\rm UV}$~$>$2) NIRSpec/Prism continuum detections. \textit{Left:} Distribution for 118 spectroscopic LRDs. Colored markers highlight sources from the literature, including those reported to show strong continuum breaks near the Balmer limit \citep[e.g.,][]{wang24_lrd, furtak24, labbe24b, degraaff25, naidu25, taylor25}. Dashed and dashed-dotted lines indicate selection thresholds from \citet{kocevski24} and \citet{barro24}. Orange density contours and histograms show the full photometric LRD sample using SED-fit values; gray histograms show the same for the spectroscopic subset. Medians and 68\% percentiles for both samples are indicated. The LRD population spans nearly 4 magnitudes in \uvnir color, reflecting substantial diversity in their SEDs. Most LRDs have blue UV slopes ($\beta_{\nu, \mathrm{UV}} = 0.3^{+0.5}_{-0.5}$), but there is a clear trend toward redder UV slopes at redder colors, reaching $\beta_{\nu, \mathrm{UV}} \sim 1$ for the reddest 7\% with \uvnir$>$3.5. \textit{Right:} Same color–slope diagram, now color-coded by the curvature of the rest-optical continuum, computed from second-degree polynomial fits to the log-log SEDs. Red squares mark LRDs best fit by power laws. Strongly curved SEDs (yellow) are concentrated among the reddest LRDs, while bluer LRDs typically exhibit power-law–like continua. Some overlap remains around the median \uvnir color (see discussion in \S~\ref{ss:breakers_vs_powerlaws}).}
\label{fig:rest_slopes}
\end{figure*}

Figure~\ref{fig:rest_slopes} shows the distribution of UV slope versus rest-frame optical color (\uvnir) for the LRD sample. Spectroscopic LRDs are shown as grey circles, while orange density contours represent the broader photometric sample, with slopes and colors derived from best-fit SEDs. This diagram conveys similar information to the UV and optical slope classifications proposed by \citet{kocevski24} and \citet{hainline25}, but based here on continuum spectroscopy. The top and right histograms show the UV slope and optical color distributions for the photometric (orange) and spectroscopic (black) samples. Vertical and horizontal dashed lines indicate the \citet{kocevski24} slope thresholds ($\beta_{\nu, \mathrm{OPT}} > 2$; $\beta_{\nu, \mathrm{UV}} > 1.27$), while the dashed-dotted line marks the approximate limit used by \citet{barro24} (F277W–F444W $>$ 1.5; $\beta_{\nu, \mathrm{OPT}} \gtrsim 3$). Colored symbols mark prominent LRDs from previous studies.

The median optical color of the photometric LRDs is \uvnir$=2.3^{+0.9}_{-0.7}$ mag, but the population spans nearly 4 magnitudes in color, showing that LRDs comprise a diverse set of sources with likely varied intrinsic properties. The selection used by \citet{barro25} includes many bluer LRDs with optical slopes below the typical threshold of \citet{kocevski24}, comprising the bottom $\sim$30\% of the sample. These include well-studied sources like those in \citet{akins24b}, \citet{juodzbalis24a}, and ``Virgil," which was confirmed as an LRD only after incorporating MIRI data. These bluer LRDs are critical for probing the boundary between LRDs and normal galaxies, where the AGN becomes subdominant.
At the red end, LRDs selected with a stricter observed color cut (F277W–F444W~$>$~1.5; \uvnir$\gtrsim 3$), as in \citet{barro24} and \citet{akins24}, make up only the top 18\% of the sample and do not include the median LRD. Beyond this threshold, the number density drops quickly. The most extreme cases, with \uvnir$\gtrsim 3.5$, represent just 7\% of the population, yet include most of the prominent LRDs in the literature. While these have been critical to shaped our understanding of the population, they are not representative but rather occupy the red extreme of the distribution.

\begin{figure*}
\centering
\includegraphics[width=18cm,angle=0]{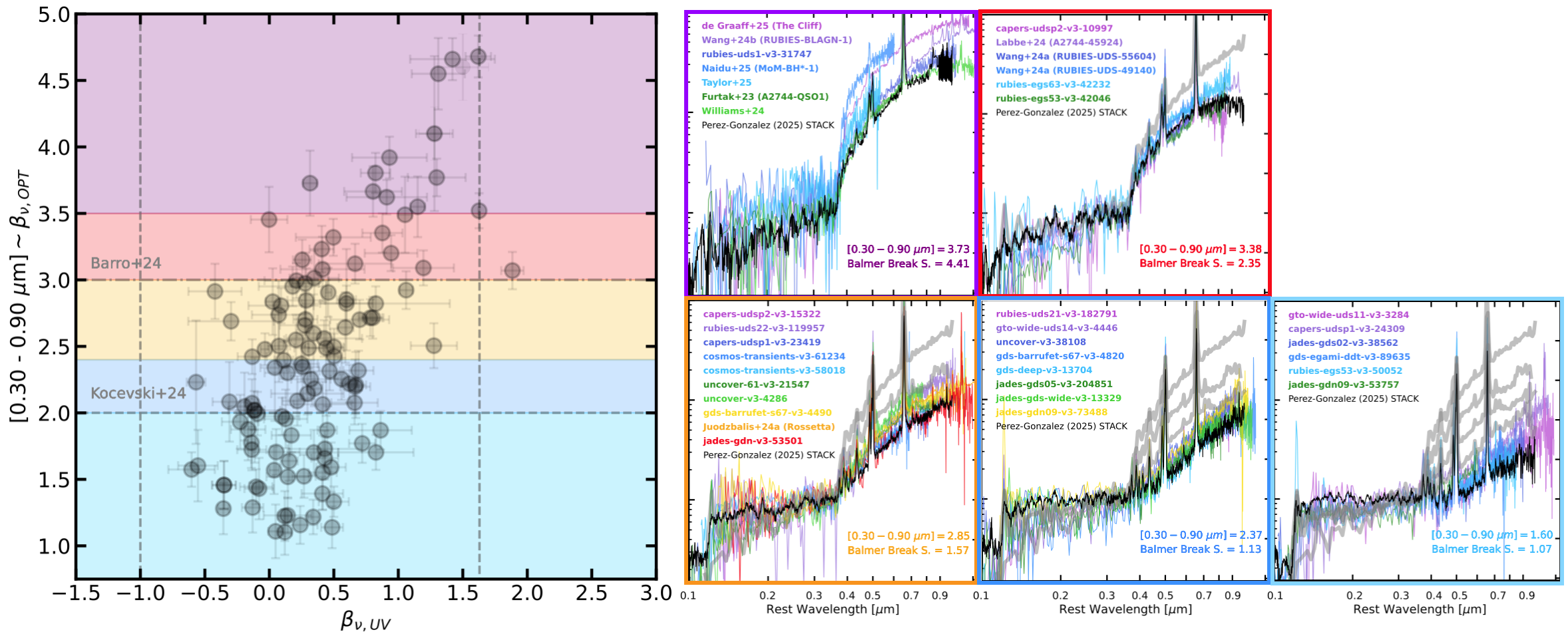}
\caption{\textit{Left:} Same color–slope diagram as in Figure~\ref{fig:rest_slopes}, now divided into five bins of \uvnir color from reddest to bluest. \textit{Right:} Top to bottom, the five panels show individual spectra of representative LRDs with high continuum S/N in each color bin, highlighting the diversity in SED shapes. Black lines show stacked LRD spectra from Perez-Gonzalez (in prep.) for the same bins. Gray lines show the stacked spectra from redder bins, carried forward to each panel to highlight differences across the sequence. The reddest LRDs (top two panels; \uvnir$>$~3, $\sim$20\% of the sample) exhibit strong Balmer breaks, pronounced curvature, and redder UV slopes. Toward bluer LRDs (bottom three panels), the breaks and curvature diminish, giving way to smoother, power-law–like SEDs. The UV slope also becomes markedly bluer, with median values near $\beta \sim 0.3$ across all three bluest bins.}
\label{fig:LRDtypes}
\end{figure*}

Figure~\ref{fig:rest_slopes} also shows a clear trend toward redder UV slopes at higher \uvnir. Most LRDs follow the canonical ``V-shaped" SED with blue UV slopes ($\beta_{\nu, \mathrm{UV}} = 0.3^{+0.5}_{-0.5}$) and red optical colors, but among redder sources, the UV slope steepens. For photometric LRDs with \uvnir$>$3 and $>$3.5, the median UV slope increases to $\beta_{\nu, \mathrm{UV}} = 0.6^{+0.4}_{-0.6}$ and $0.8^{+0.4}_{-0.3}$, respectively. Among spectroscopic LRDs, the trend is even stronger, with medians of $0.9^{+0.4}_{-0.4}$ and $1.0^{+0.6}_{-0.2}$ in the same bins. These include several well-studied LRDs known for their strong continuum breaks near the Balmer limit \citep{wang24_lrd, furtak23, labbe24b, degraaff25, naidu25, taylor25}, all with UV slopes $\beta_{\nu, \mathrm{UV}} \gtrsim 1$.

This trend poses a challenge for photometric selections, which are tuned to detect the combination of blue UV and red optical slopes. Instead, the reddest LRDs begin to resemble quiescent or dusty galaxies with uniformly red continua, making them harder to identify in standard color–color diagrams (as noted in Section~\ref{ss:selection}). Several LRDs with red UV slopes were thus identified only a posteriori from spectroscopy \citep{wang24_lrd, juodzbalis24a, williams24}. Yet once placed on the color–slope diagram, they fall well within the main LRD distribution, supporting their classification as true LRDs.

In contrast, candidate quiescent galaxies from \citet{weibel24}, \citet{kokorev24b}, and \citet{degraaff24b} lie significantly outside the LRD locus. Although some have red optical colors (\uvnir$\sim$2–2.5 mag), they show much redder UV slopes ($\beta_{\nu, \mathrm{UV}} \gtrsim 1$) than LRDs with similar colors. This separation highlights the utility of the color–slope diagram for distinguishing LRDs from quiescent interlopers in photometric samples.

\subsection{Curvature vs. power-law in the rest-optical SED}

We use the detailed spectral coverage of the optical SED provided by NIRSpec/PRISM to examine the continuum shape of LRDs. Our goal is to quantify what fraction of the population shows significant curvature or breaks in the optical, which may point to stellar-dominated SEDs \citep[e.g.,][]{williams24, pg24a}, or more exotic BH$^{\star}$ or quasi-stellar components \citep[e.g.,][]{inayoshi24, naidu25}. These cases differ from the smoother power-law continua often associated with canonical obscured AGNs in broadband photometry \citep[e.g.,][]{polletta06}.

To quantify these differences without relying on specific physical models, we fit the rest-frame SEDs in log($f_{\nu}$)–log($\lambda$) space using two simple functions: (1) a linear model representing a power-law ($a \cdot \log \lambda$), and (2) a second-degree polynomial ($a \cdot \log \lambda^2$ + $b \cdot \log \lambda$) that captures curvature. We compare the fits using both $\chi^2$ and the Akaike Information Criterion (AIC). For spectra better fit by the quadratic model, we use the coefficient $a$ as a proxy for curvature strength. Sources with $a \sim 0$ are classified as power-law, while those with $a > 0$ but no strong AIC preference are labeled as inconclusive.

The right panel of Figure~\ref{fig:rest_slopes} shows the same color–slope diagram as the left, now color-coded by curvature for the best-fit quadratic models. Red markers indicate power-law SEDs, and grey circles mark inconclusive cases. The curved SEDs cluster toward the red end of the distribution (\uvnir$\gtrsim 3$), overlapping with LRDs previously reported to exhibit Balmer breaks. These sources show steep optical slopes with a sharp flattening toward the NIR, distinct from power-law shapes. In contrast, LRDs with bluer optical colors follow the typical ``V-shaped'' SED with blue UV and red optical power-law slopes.

A potential complication from this trend is that some of the reddest LRDs, despite lacking blue UV slopes or power-law continuum shapes, still show large differences between their UV and optical slopes. As a result, ``V-shaped'' metrics like $\Delta k = \beta_{\rm OPT} - \beta_{\rm UV}$ \citep[e.g.,][]{setton24} can yield similar values for both curved and canonical blue plus red LRDs. This occurs because $\Delta k$ runs diagonally across the color–slope diagram, grouping together sources from the top-right (strongly curved) and bottom-left (classical power-law) that have fundamentally different SED shapes.

\subsection{The variety of SED types in LRDs}


\begin{figure*}
\centering
\includegraphics[width=8.5cm,angle=0]{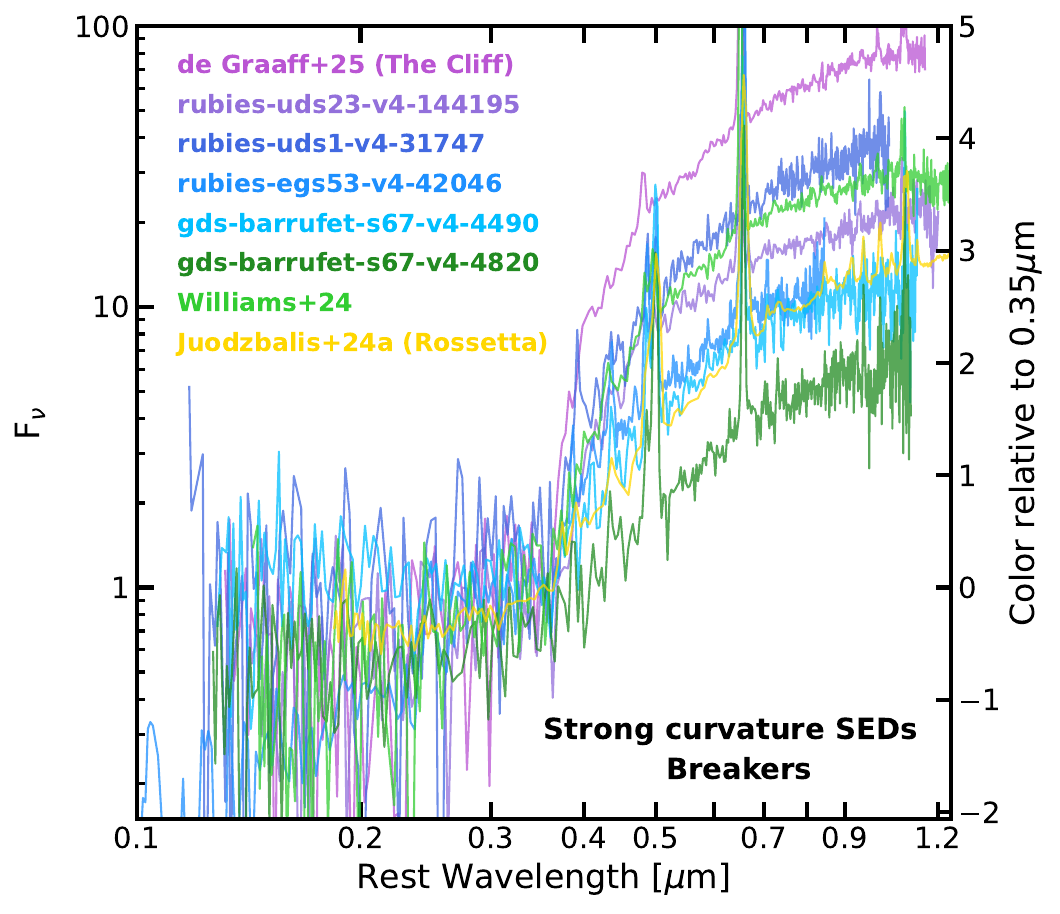}
\hfill
\includegraphics[width=8.5cm,angle=0]{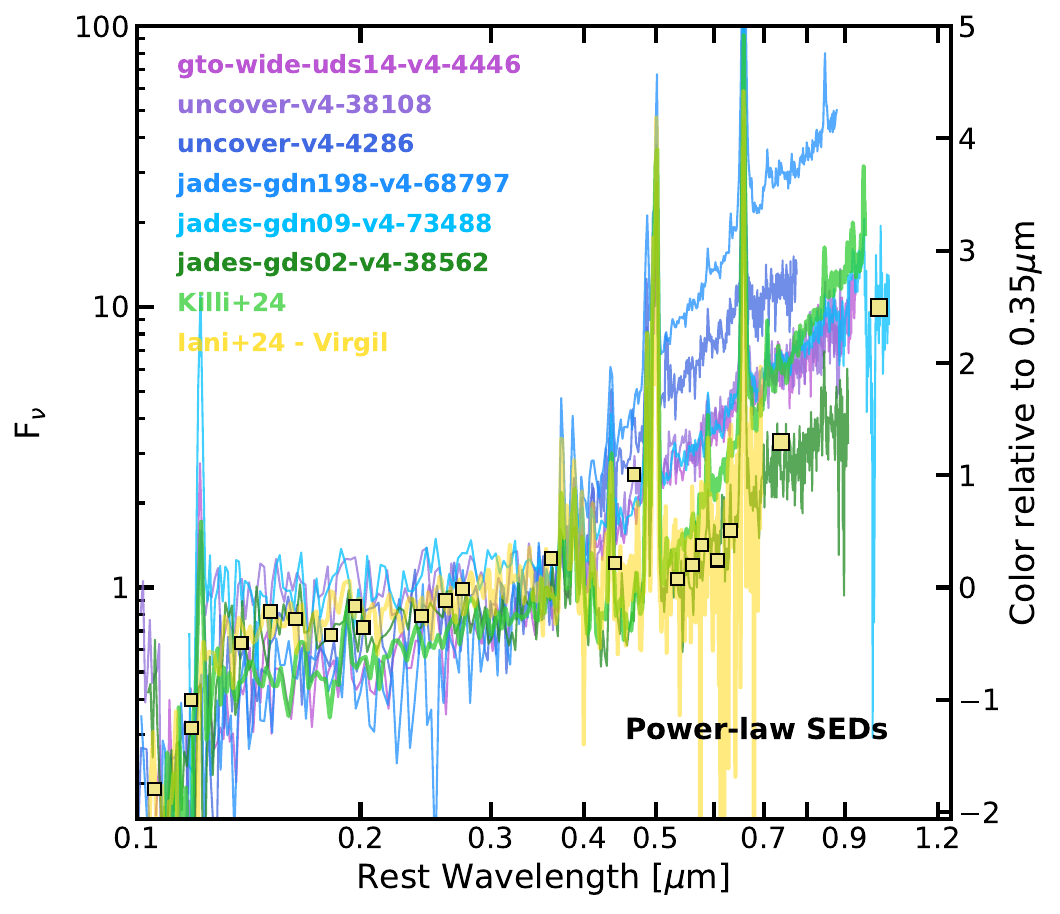}
\caption{Examples of rest-frame spectra illustrating the two dominant spectral shapes among LRDs, strong curvature (``Breakers") and power-laws, normalized at H$_{\infty}$. \textit{Left:} LRDs with strong curvatures show SEDs that rise steeply redward of the Balmer limit followed by a flattening toward the NIR. \textit{Right:} LRDs with power-law–like SEDs show smooth, constant slopes across the optical. While breakers typically have red \uvnir colors and power-laws are generally bluer, there is overlap at intermediate colors—similar rest-frame colors can result from distinct SED shapes. In breakers, the optical continuum rises sharply at H$_{\infty}$, whereas in bluer power-law LRDs the inflection occurs at longer wavelengths. This is illustrated by ``Virgil” (yellow), whose optical rise is delayed to $\sim$0.6~$\mu$m and is visible only with MIRI photometry (large squares).}
\label{fig:breakers_powerlaws}
\end{figure*}

We further investigate the diversity of LRD SEDs as a function of rest-frame UV-to-optical color by examining both individual and median spectra for subsamples of LRDs with high-S/N ($>$5) optical continuum. We divide the sample into five color bins spanning \uvnir$\sim$1 to 5 mag. The bin widths are roughly 0.5 mag, adjusted slightly to include a similar number of sources per bin. The left and right panels of Figure~\ref{fig:LRDtypes} show the bin definitions and corresponding SEDs, all normalized at the Balmer limit (H$_\infty$). The black line in each panel shows the median SED for each bin. The average color and Balmer break are also indicated.

The reddest bin (\uvnir$>3.5$~mag) contains LRDs with the strongest curvature, reddest UV slopes, and most pronounced continuum breaks, with an average Balmer break strength of 4.41, far beyond what standard stellar population models can reproduce. This bin includes the three extreme LRDs from \citet{degraaff25}, \citet{naidu25}, and \citet{taylor25} that motivated the black hole star (BH$^{\star}$) model, in which a dense gas cocoon surrounds an accreting black hole, causing the sharp decline. We further discuss this modeling in \S~\ref{s:model}. In addition to their breaks, these LRDs exhibit sharp flattening toward the NIR, consistent with the weak MIRI colors reported for the reddest LRDs \citep[e.g.,][]{degraaff25, setton25, ronayne25}. The adjacent bin ($3 <$~\uvnir$< 3.5$~mag) shows similar overall shapes, with somewhat weaker curvature, red UV slopes, and a lower average Balmer break amplitude of 2.35. This group includes some of the first spectroscopically confirmed LRDs with strong but less extreme Balmer breaks \citep{kocevski24, labbe24b, wang24b}. In both of the two reddest bins, LRDs tend to exhibit prominent, broad Balmer emission lines, with weaker or undetected metal lines. We discuss line ratios and equivalent widths in more detail in \S~\ref{ss:color_lineratios}.

The three bottom bins correspond to LRDs with \uvnir$< 3$, approaching the median color of the photometric sample (\uvnir$=2.3$~mag). These sources display a remarkably similar flat UV continuum, with $\beta_{\nu, \mathrm{UV}} \sim 0.3$, much bluer than in the redder bins. As \uvnir decreases from 3 to 1, the average Balmer break amplitude declines steadily to 1.57, 1.13, and 1.07, values more typical of normal blue star-forming galaxies. Similarly, SED curvature diminishes toward shallower, power-law–like optical slopes. In these bins, the composite SED closely resembles the canonical combination of blue plus red power laws, with an inflection near the Balmer limit. In parallel, the median spectra show a growing prominence of narrow metal lines, especially the coronal lines around the Balmer break, such as \OII\ and \NeIII, while the relative strength of \Ha\ appears to decline compared to the blended \Hb+\OIII\ complex.

Notably, the median SEDs in the two bluest bins suggest that the inflection may occur redward of H$_\infty$, near the \OIII\ emission line. This is consistent with ``Virgil" \citep{iani24, rinaldi25}, a blue LRD with a seemingly flat optical continuum that rises sharply beyond \Ha. A shift in the inflection is quite relevant. One of the main arguments against two-component models (e.g., galaxy+AGN) has been the consistency of the inflection at H$_\infty$, which would favor a single-component origin \citep{setton24}. If the inflection point varies with color, it would instead favor a multi-component scenario for the peculiar SED of LRDs. We revisit this scenario in detail in \S~\ref{s:model}.

\subsection{Breakers vs. Power-Laws: Diversity and Degeneracy in LRD SEDs}
\label{ss:breakers_vs_powerlaws}

Figure~\ref{fig:rest_slopes} shows that redder LRDs tend to have more pronounced curvature in their optical SEDs, while bluer LRDs are more consistent with power-law shapes. However, there is some clear overlap. I.e., similar \uvnir colors can result from either a steep Balmer break followed by a flattening or a continuously rising power-law. Figure~\ref{fig:breakers_powerlaws} illustrates this degeneracy, showing examples of both types of SEDs across a range of rest-frame \uvnir colors.

The left panel of Figure~\ref{fig:breakers_powerlaws} shows LRDs with pronounced curvature, following the trend between \uvnir color and curvature strength ($a \cdot \log\lambda^2$; see color-coding in Figure~\ref{fig:rest_slopes}). The reddest, highest-curvature sources exhibit the most extreme Balmer breaks: their SEDs rise steeply in the optical and then flatten sharply toward the NIR. Moving toward bluer LRDs, the break becomes less pronounced, the optical slope shallower, and the flattening shifts to longer wavelengths. A simple interpretation in the context of a two-component model, would be that the relative luminosity of the break-dominated component gradually decreases top-to-bottom, effectively hiding the steep slope and pushing the inflection point redward.

\begin{figure*}
\centering
\includegraphics[width=9.1cm,angle=0]{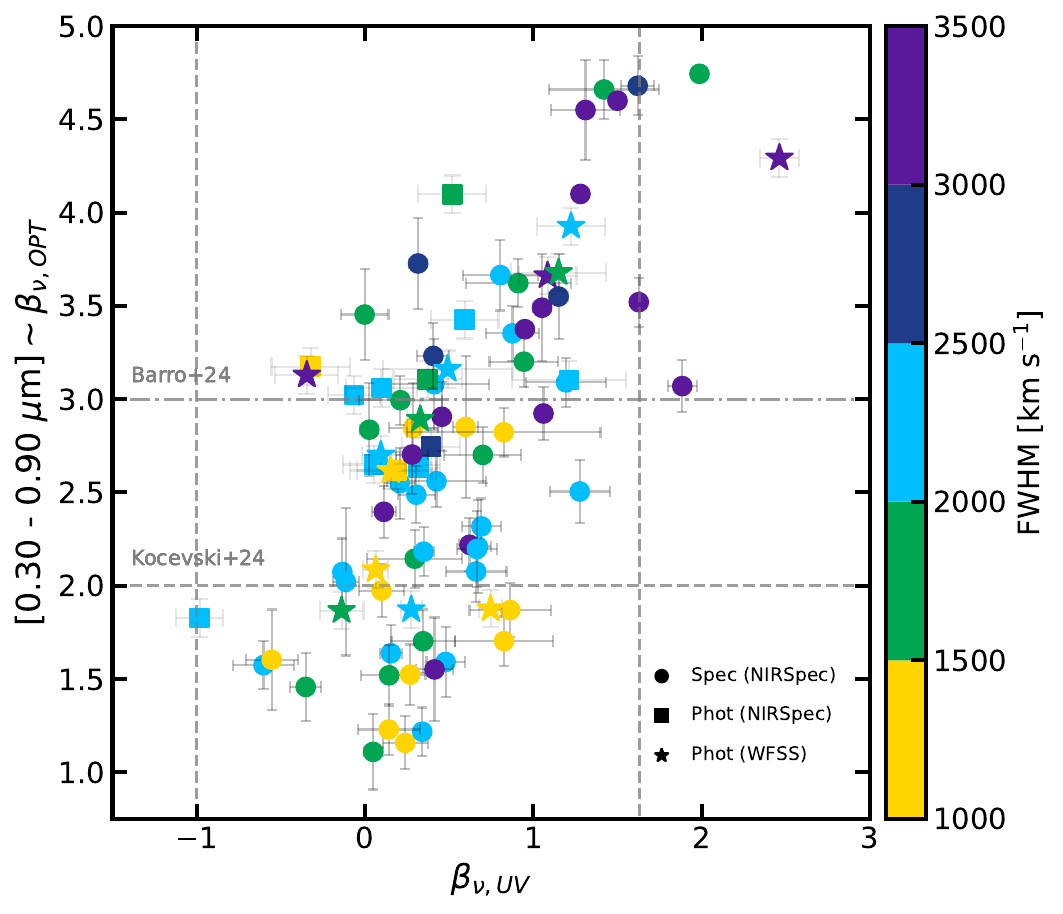}
\includegraphics[width=8.2cm,angle=0]{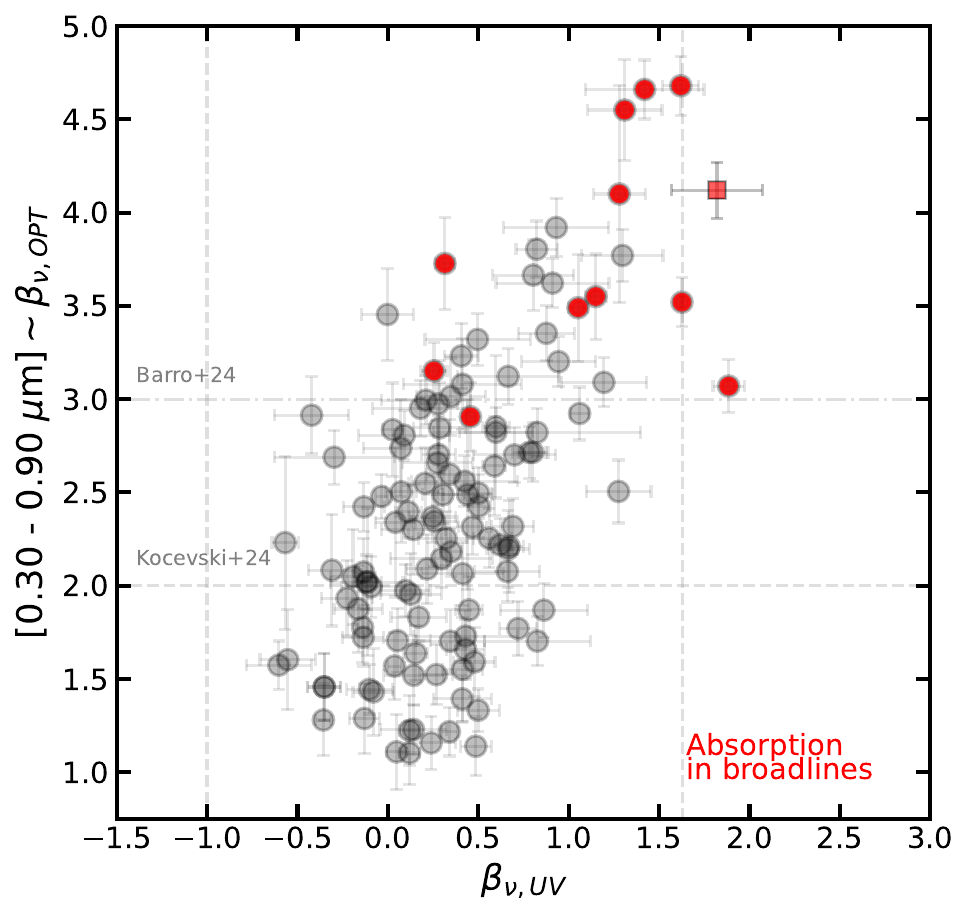}
\caption{{\it Left}: Same color-slope diagram as in Figure~\ref{fig:rest_slopes} but color-coded by the FWHM of broad emission lines compiled from the literature \citep{greene23, kocevski23, taylor24, juodzbalis25, akins24, degraaff25, naidu25, matthee24, zhang25}. Circles show LRDs with spectroscopic color measurements; squares and stars indicate colors derived from photometric SED fits for sources with lower S/N or no NIRSpec continuum. LRDs with the reddest colors (\uvnir$>$3.5 mag) and redder UV slopes tend to have the broadest emission lines. {\it Right}: Same diagram, now highlighting LRDs with narrow absorptions in their broad emission lines, as compiled from the literature (see text). Red circles mark the LRDs with reported absorption features; the open red square indicates the candidate from \citet{torralba25b}, for which no public spectrum is available, and the classification is based on photometry. Grey dots show other NIRSpec LRDs with no reported absorptions. Absorbers occupy the reddest end of the distribution and exhibit the characteristic strong breaks and optical curvature of this population.}
\label{fig:linewidths}
\end{figure*}

The right panel of Figure~\ref{fig:breakers_powerlaws} shows power-law LRDs. While most are blue, a small fraction ($\sim$10\%) exhibit very red colors (\uvnir$>3$ in Figure~\ref{fig:rest_slopes}). These red power-law sources show a range of optical slopes, suggesting variation in the physical conditions shaping the continuum, such as dust attenuation or gas density in BH$^{\star}$ models. Notably, the reddest of these LRDs, JADES-GDN-68797 and MACSJ0647.7+7015 (one of the first reported LRDs; \citealt{killi23}), exhibit remarkably similar optical slopes, despite the latter being about 1 mag bluer. This also agrees with the apparent shift in the inflection point from the flat UV to the rising optical continuum in this object, which occurs at longer wavelengths. A similar shift is seen in ``Virgil" (yellow) and other bluer LRDs, as discussed in the previous section. We return to this point in \S~\ref{s:model}.


\subsection{Color Dependence of the Emission Line FWHMs}


Next, we analyze how the widths of broad emission lines correlate with rest-frame colors. The ubiquitous broad lines observed in LRDs were among the key motivations for their initial interpretation as AGNs. Early black hole mass estimates based on these lines suggested surprisingly high black hole–to–stellar mass ratios \citep{kocevski23,harikane23,greene23, maiolino23}. While recent studies have identified exponential wings in the broad-line profiles, potentially indicating that standard virial mass estimators may not apply \citep{naidu25, rusakov25}, it remains important to test whether line width correlates with overall rest-frame color, which, as we have shown, is closely tied to the diversity of LRD SEDs.

The left panel of Figure~\ref{fig:linewidths} shows the same rest-frame color–slope diagram as Figure~\ref{fig:rest_slopes}, now color-coded by the FWHM of the broad lines, compiled from BLAGNs in the literature as described in \S~3.3. We use measurements from NIRSpec grating observations reported in \citet{kocevski23}, \citet{taylor24}, \citet{juodzbalis25}, and \citet{hviding25}, along with individual values from other studies \citep[e.g.,][]{akins24, degraaff25, naidu25}. We also include NIRCam/WFSS-based estimates from \citet{matthee24} and \citet{zhang25}. For LRDs with NIRSpec/Prism continuum and $S/N_{\rm UV} > 2$, rest-frame colors are computed directly from the spectra (circles). For lower-S/N sources or those without spectra, we use rest-frame colors derived from best-fit photometric SEDs (squares and stars).

The colormap suggests a positive correlation, with the reddest LRDs showing the highest FWHMs. A color vs. FWHM linear fit yields a statistically significant correlation, albeit with substantial scatter (Pearson $r = 0.33$, $p = 2.7 \times 10^{-3}$; Spearman $\rho = 0.36$, $p = 1.0 \times 10^{-3}$). The median FWHM, binned by rest-frame color as in Figure~\ref{fig:LRDtypes}, declines from 2327~$\pm$~760~km~s$^{-1}$ for \uvnir$>$~3, to 2067~$\pm$~877~km~s$^{-1}$ for $2<$~\uvnir~$<3$, and 1677~$\pm$~955~km~s$^{-1}$ for \uvnir$<$~2. This trend agrees with the results of \citet{greene23}, who found broader lines in redder LRDs (e.g., F277W$-$F444W~$>$~1.6) in the UNCOVER field.

Despite this general trend, the scatter is large, and there are notable exceptions. Some blue LRDs exhibit very broad lines, for example, GOODS-N-1001830 from \citet{juodzbalis24a} and UNCOVER-38018 from \citet{greene23} both show FWHM~$>$~4000~km~s$^{-1}$. Conversely, some of the reddest LRDs, such as RUBIES-UDS-16053 \citep{taylor25} and ``The Cliff” \citep{degraaff25}, exhibit narrower lines with FWHM~$\lesssim$~1500~km~s$^{-1}$.




\subsection{LRDs with narrow absorption in the broad emission lines}

We now examine the subset of LRDs that exhibit narrow absorption features superimposed on their broad Balmer lines.  These absorptions exhibit large EWs that exceed expectations from stellar photospheres, pointing instead to very dense gas near the broad-line region \citep{matthee24,deugenio25,deugenio25b,torralba25b}. Consequently, their presence has become a key prediction of the BH$^{\star}$ scenario, in which the optically thick gas cocoon produces both strong Balmer breaks and broad absorption features \citep{inayoshi24, ji25}.

We compile 12 such LRDs with reported absorptions across the six JWST fields, as identified in the literature by \citet{kocevski24}, \citet{wang24b} (RUBIES-EGS-55604, -49140, -42046), \citet{juodzbalis24a} (1181-28074; ``Rosetta"), \citet{furtak23}, \citet{ji25} (A2744-QSO1), \citet{labbe24b} (A2744-45924), \citet{wang24_lrd} (RUBIES-UDS-BLAGN1), \citet{degraaff25} (RUBIES-UDS-154183; ``The Cliff”), \citet{naidu25} (MoM-BH$^{\star}$-1), \citet{deugenio25b} (1286-159717), \citet{taylor25} (CAPERS-BH$^{\star}$-1), and \citet{torralba25b} (GDN-9771). The last object is included for completeness, though no public spectrum is available, and its properties are inferred from photometry.

\begin{figure*}
\centering
\includegraphics[width=18cm,angle=0]{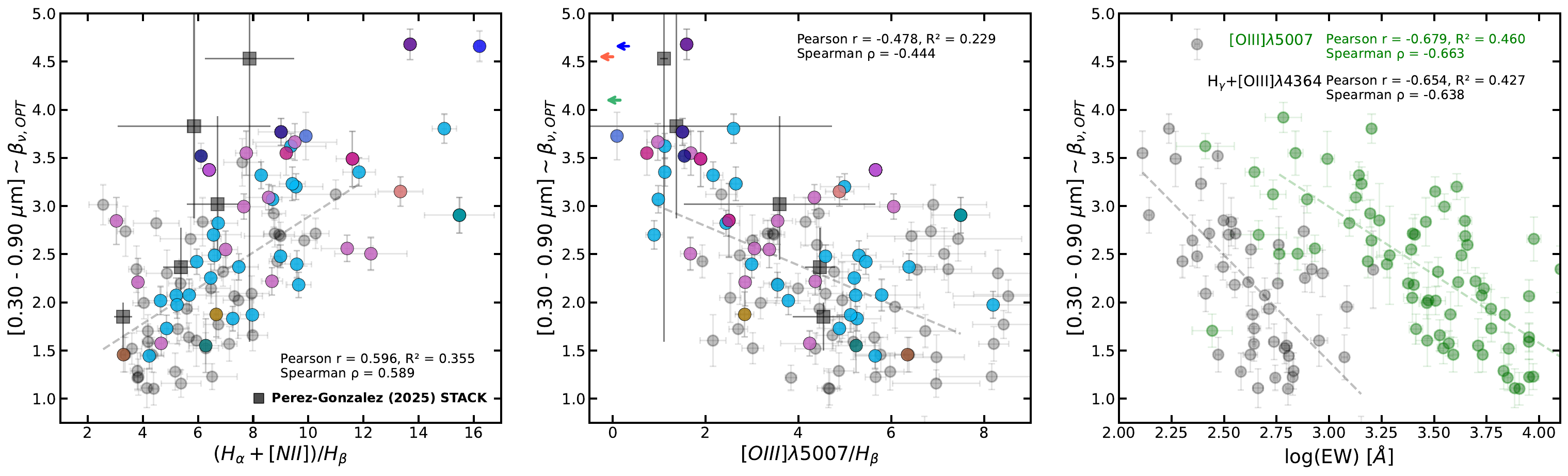}
\caption{Emission line ratios and equivalent widths for spectroscopic LRDs as a function of \uvnir color. Only lines with high S/N~$>$5 are included. Colored markers in the first two panels highlight LRDs from the literature, as in Figure~\ref{fig:rest_slopes}. Black squares show the same line ratios measured from the stacked spectra in each color bin by Pérez-González (in prep.), as shown in Figure~\ref{fig:LRDtypes}. \textit{Left:} The Balmer decrement (\Ha/\Hb) increases with \uvnir color, peaking in the reddest LRDs. \textit{Center:} The \OIII/\Hb\ ratio declines with \uvnir color. Bluest LRDs show the highest values, while the reddest, BH$^{\star}$-like LRDs exhibit weak or absent \OIII. \textit{Right:} The equivalent widths of \OIII\ (green) and \Hg+\OIII$\lambda$4364 (gray) both decline toward redder \uvnir values, suggesting that narrow emission lines become stronger in bluer LRDs, suggesting a more dominant role of the galaxy host.}
\label{fig:lineratios}
\end{figure*}

The right panel of Figure~\ref{fig:linewidths} shows the positions of these broad-line absorbers in the rest-frame color–slope diagram. Nearly all lie at the red end of the distribution, with a mean optical color of \uvnir$= 3.75 \pm 0.62$~mag. Notably, this includes all three of the extreme BH$^{\star}$-like LRDs with \uvnir $\gtrsim 4$ from \citet{degraaff25}, \citet{naidu25}, and \citet{taylor25}. Consistent with their red optical colors, most also show strong Balmer breaks ($\gtrsim$3 mag) and pronounced optical curvature. These properties weaken toward the bluer end of the distribution: the three absorbers with \uvnir $< 3.25$ have shallower Balmer breaks ($\lesssim$2 mag), and one of them,  1286-159717 from \citet{deugenio25b}, has a clear power-law–like SED instead.

This strong correlation with optical color raises the question of whether all LRDs with \uvnir$\gtrsim 3$ exhibit such absorptions. However, most of these red LRDs are covered only by NIRSpec/PRISM spectroscopy, which lacks the spectral resolution required to detect narrow absorption features (e.g., all LRDs in UNCOVER and COSMOS). Of the 13 spectroscopic LRDs with \uvnir $\gtrsim 3$ not previously identified as absorbers, only five have higher-resolution grating spectra that cover one or more of the main broad lines (e.g., H$\alpha$, H$\beta$, or He~\textsc{i}): RUBIES-UDS-31749, -127928 (faint), RUBIES-EGS-42232, -926125, and JADES-GDN-1181-68797. All five show broad H$\alpha$ emission and are listed in the BLAGN catalogs from \citet{taylor24} and \citet{hviding25}, but none show reported narrow absorption features.


\subsection{Color Dependence of Emission-line Ratios}
\label{ss:color_lineratios}

Lastly, we examine the emission-line fluxes and equivalent widths of LRDs as a function of rest-frame \uvnir color, using line measurements from the DJA/NIRSpec v4.4 release. These line fluxes are derived using a combination of spline fits to the continuum and Gaussian profiles (convolved with the instrumental resolution) for the emission lines (see \citealt{valentino25} for details). We restrict our analysis to the high-continuum-S/N LRDs discussed throughout this section and consider only emission lines with robust detections (S/N~$>$5) measured in the NIRSpec/PRISM spectra.

Figure~\ref{fig:lineratios} shows the trends between key emission-line ratios and \uvnir color. Colored markers highlight well-studied LRDs from the literature, following the same color scheme as in Figure~\ref{fig:rest_slopes}. Black squares indicate the same line ratios measured on stacked spectra, binned by UV-to-optical color, from Pérez-González et al. (in prep.). The left panel shows the (blended) \Ha+\NII\ to \Hb\ ratio, with high-S/N detections available for nearly 86\% of the sample. The remaining sources are mostly at $z > 7$, where \Ha\ lies beyond the NIRSpec/PRISM spectral range. The line ratio shows a clear positive correlation: the reddest LRDs tend to show significantly stronger \Ha+\NII\ emission relative to \Hb. This trend is broadly consistent with values derived from the stacked spectra (Perez-Gonzalez et al., in prep.) and recent results from \citet{degraaff25b}. Some discrepancies arise in the reddest bins, where larger uncertainties and the stacking process likely dilute the contribution from the most extreme BH$^{\star}$ LRDs, which are rare and therefore underrepresented.

Extreme examples like ``The Cliff” and RUBIES-UDS-BLAGN1 lie at the upper right of the diagram, with the highest line ratios. Other prominent BH$^{\star}$ LRDs from \citet{naidu25} and \citet{taylor25} are absent due to the lack of \Ha\ coverage at their higher redshifts. The observed Balmer decrements span a broad range ($\sim$2–20), with most values exceeding the canonical Balmer decrement of \Ha/\Hb$=3.1$ expected for AGN narrow-line regions under Case B recombination \citep{osterbrock06}. Because LRDs generally exhibit broad Balmer lines, the \Ha+\NII\ to \Hb\ ratio measured at PRISM resolution largely probes the conditions of the broad-line region (BLR). In contrast, the narrow-line Balmer decrement measured in higher-resolution spectra typically yields much lower ratios, closer to the theoretical value \citep[e.g.,][]{killi23, akins24b, deugenio25, brooks25}.

The physical origin of the large Balmer decrement in LRDs remains poorly understood. Early interpretations invoking heavy obscuration (e.g., \citealt{kocevski23}, \citealt{greene23}) are disfavored in light of the BH$^{\star}$ models, which predict more complex radiative transfer effects. In this context, the elevated broad \Ha/\Hb\ ratios may be boosted by collisional excitation and resonant scattering in the dense, low-metallicity (and low-extinction) gas cocoon surrounding the accreting black hole \citep[e.g.,][]{chang25, torralba25b, yan25}. Supporting this scenario, the LRDs with the most extreme \Ha/\Hb\ ratios are also those that exhibit narrow self-absorption features superimposed on their broad Balmer lines.

The central panel of Figure~\ref{fig:lineratios} shows the \OIII/\Hb\ flux ratio as a function of \uvnir color. These lines are detected in nearly 90\% of the LRD sample. This line ratio exhibits a clear negative trend: redder LRDs exhibit weaker \OIII\ emission relative to \Hb, while the bluer LRDs show elevated \OIII/\Hb\ ratios. The three BH$^{\star}$ candidates (``The Cliff,” MoM-BH*1, and CAPERS-119334) stand out at the top-left of the diagram, with \OIII/\Hb\ $\ll$ 1 due to their strong broad \Hb\ emission and near-complete absence of \OIII. On the opposite end, the bluest LRDs feature some of the highest ratios with ``Virgil" \citep{rinaldi25} featuring one of the highest \OIII/\Hb$\sim6.5$ values. This trend is also supported by stacked spectra (Perez-Gonzalez et al., in prep.) and by \citet{degraaff25b}, who find similarly low \OIII/\Hb\ in the reddest LRDs.

\begin{figure}
\includegraphics[width=8.5cm,angle=0]{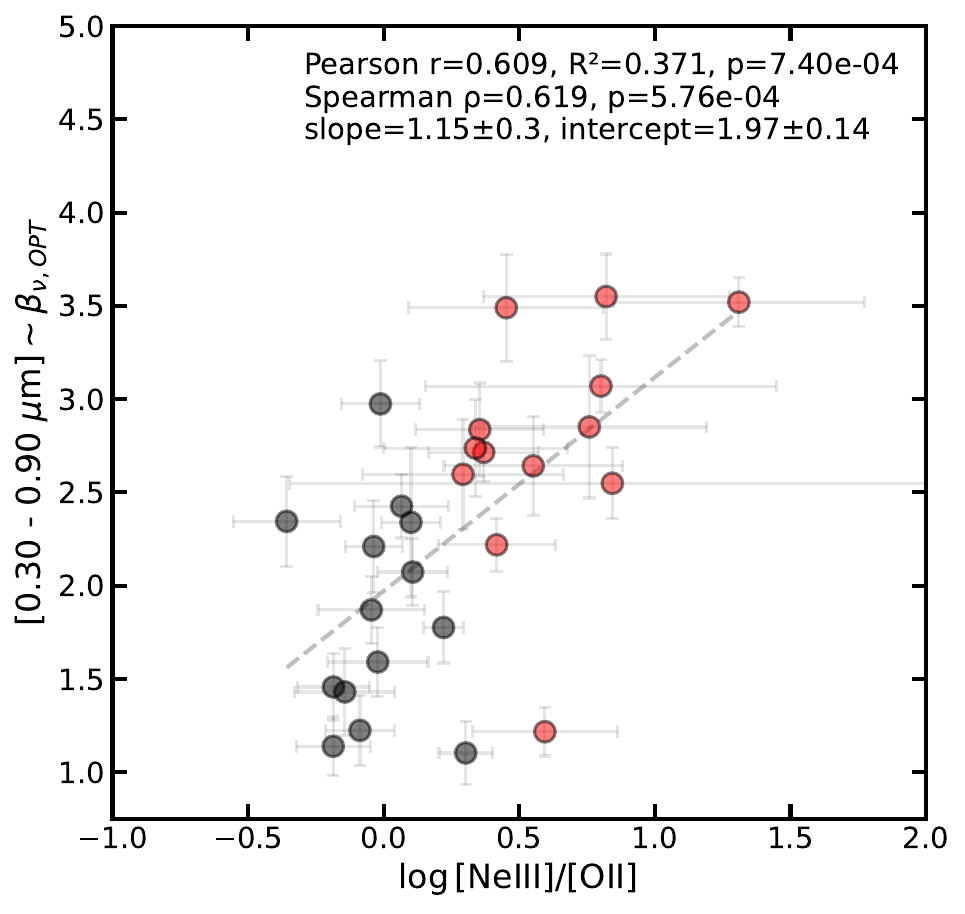}
\caption{\NeIII/\OII\ line ratio as a function of rest-frame \uvnir color for spectroscopic LRDs. Only 12\% of sources have both lines detected at S/N~$>$5 (gray), while an additional 10\% have high-S/N \NeIII\ and lower-S/N (2$<$S/N$<$5) \OII\ detections (red). These latter sources are predominantly redder in \uvnir and drive an overall positive correlation between color and line ratio. Since both lines are narrow and likely originate in the host galaxy, the trend may reflect changing ionization conditions or metallicity in the galaxy population.}
\label{fig:ne3o2}
\end{figure}

At first glance, the observed trend may seem counterintuitive, as classical excitation diagnostics (e.g., the BPT diagram; \citealt{bpt}) associate high \OIII/\Hb\ ratios with AGN activity in low-redshift galaxies. Yet, in LRDs, the reddest and most AGN-dominated systems, marked by broad lines and extreme SEDs, show the lowest \OIII/\Hb\ ratios. This can be understood if the broad \Hb\ originates in the BLR, while \OIII, typically emitted from the host galaxy or narrow-line region (NLR), is weak or absent. As \uvnir color becomes bluer, \OIII\ emission strengthens, likely tracing a nascent NLR or the galaxy host beyond the BH$^{\star}$ cocoon. In some LRD with high-resolution NIRSpec spectra, where broad and narrow components can be separated, the narrow-line ratio reaches (\OIII/\Hb)$_{\text{narrow}} \sim 10$, compared to the unresolved PRISM ratio of \OIII/\Hb\ $\sim$ 2 \citep[e.g.,][]{deugenio25c, jones25}. In even bluer LRDs, the broad \Hb\ component fades relative to the narrow, often becoming indistinguishable \citep{deugenio25, juodzbalis24b, akins24b, rinaldi25}. As a result, PRISM-based ratios more closely reflect the intrinsic narrow-line value, similar to those in normal galaxies \citep{robertsborsani24, backhaus24}, supporting a scenario in which narrow-line emission in blue LRDs arises primarily from the host.

\input{tab2.tex}

To further test whether narrow emission lines arise from the host galaxy while the optical continuum is AGN-dominated, we examine the correlation between equivalent width (EW) and \uvnir color for the most prominent narrow lines. The right panel of Figure~\ref{fig:lineratios} shows the EW of \OIII\ and the of blended \Hg+\OIIIs\, the latter serving as a proxy for the complex of lines near the Balmer-break (\OII, \NeIII, \Hd) commonly seen in high-EW galaxies. Among these, \Hg+\OIIIs\ is typically the strongest and most reliably detected, with $\sim$45\% of LRDs showing S/N~$>$5 in this feature, compared to 20\%, 18\%, and 1\% for \OIII, \NeIII, and \OII, respectively. Both EWs show a clear anti-correlation with \uvnir color: bluer LRDs exhibit stronger narrow-line emission, while redder LRDs show much weaker or undetected lines. This supports the picture that narrow-line features become more prominent toward bluer LRDs, where the host galaxy likely contributes more to the spectrum.

We caution that \Hg+\OIIIs\ may not be purely narrow, especially in the reddest LRDs where broad \Hg\ emission can dominate. Still, high-resolution NIRSpec spectra suggest that the narrow \Hg\ component becomes more prominent in bluer LRDs \citep[e.g.,][]{akins24b, juodzbalis24b, rinaldi25}, while the broad component dominates over the narrow in redder systems \citep[e.g.,][]{deugenio25, jones25, torralba25b} similarly to \Hb.

While other narrow lines in the Balmer-break complex are less frequently detected at high S/N, they still provide valuable diagnostics. In particular, the \NeIII/\OII\ ratio is a known AGN indicator at low redshift, where elevated values are linked to harder ionizing spectra \citep[e.g.,][]{trump23, backhaus22, backhaus24}. Figure~\ref{fig:ne3o2} shows \NeIII/\OII\ as a function of \uvnir color. We find a positive correlation, with redder LRDs tending to show stronger \NeIII\ relative to \OII. However, only about 12\% of LRDs have S/N$>$5 detections in both lines (grey circles), with a median color of \uvnir$\sim$1.95. Another 10\% (red circles) show weaker \OII\ (2$<$S/N$<5$) but measurable \NeIII, with a redder median color of \uvnir$\sim$2.88, driving the overall trend. 

Although elevated \NeIII/\OII\ typically signals AGN photoionization, if narrow lines in LRDs primarily trace the host galaxy, the trend could instead reflect changing host conditions, such as higher dust content or ionization parameter, in redder systems. We note, however, that \OII\ is difficult to measure in red LRDs, as it lies close to the Balmer break where the continuum drops steeply, making reliable fits challenging at PRISM resolution.

The UV emission lines in LRDs are generally faint, with few detections of classical AGN tracers such as \CIII, \CIV, or \MgII\ above S/N~$>$5. The exception is Ly$\alpha$, detected in $\sim$20\% of the sample. While Ly$\alpha$ equivalent widths show no clear trend with color, detections are concentrated in bluer LRDs, with a median \uvnir$\sim$2.0$^{+0.6}_{-0.7}$. Lowering the detection threshold to S/N~$>$2 boosts the \CIII\ detection rate to $\sim$22\%, consistent with it being the brightest UV line after Ly$\alpha$; \CIV\ and \MgII\ remain below 5\%. Like Ly$\alpha$, \CIII\ detections span a wide color range but are slightly redder on average, with \uvnir$\sim$2.4$^{+0.7}_{-0.6}$, consistent with previous reports of \CIII\ in some of the reddest LRDs \citep[e.g.,][]{labbe24b, deugenio25, torralba25b}.

Although \CIII\ is a common AGN tracer at low redshift, its interpretation in LRDs is less clear. Hard radiation fields in high-redshift star-forming galaxies can also power strong \CIII\ emission, complicating its association with BH$^{\star}$ activity \citep[e.g.,][]{nakajima18}. In fact, recent blind surveys for \CIII\ emitters and other high-EW UV-line sources at $z>4$ find that these systems tend to have much bluer \uvnir colors ($<$1), with only $\sim$10\% overlap with the LRD population (Arevalo-Gonzalez in prep.; Davis in prep.), suggesting they trace largely distinct populations.

\begin{figure*}
\centering
\includegraphics[width=18.2cm,angle=0]{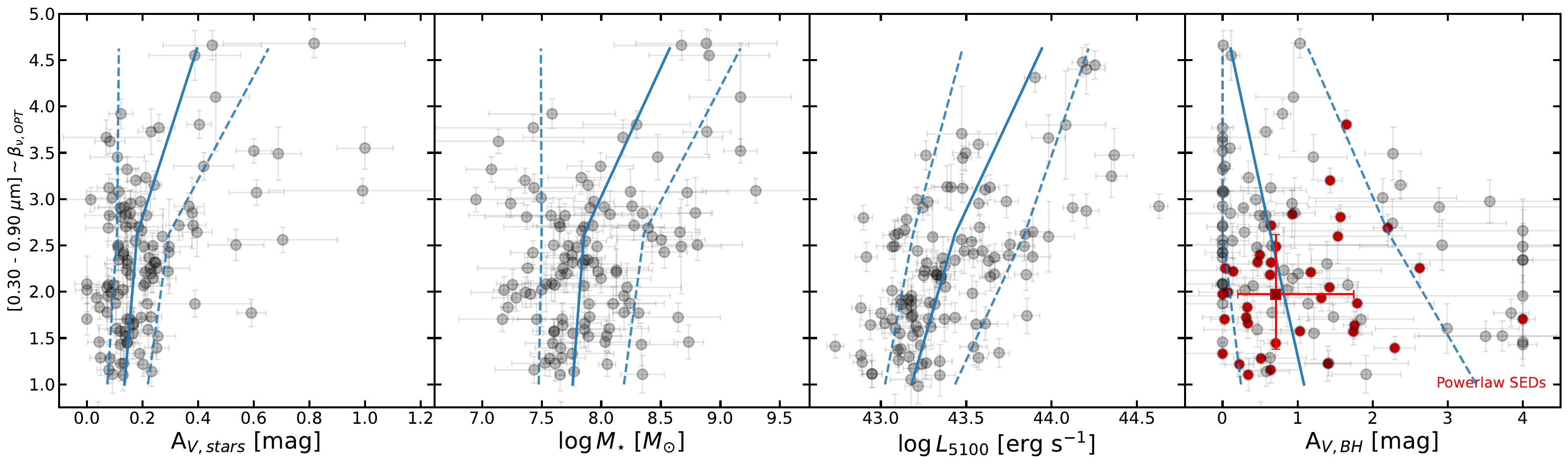}
\caption{Best-fit parameters of the two-component (galaxy plus BH$^{\star}$) model as a function of rest-frame \uvnir color for LRDs with high-S/N NIRSpec/PRISM spectra. The panels show stellar attenuation ($A_{V,\star}$), stellar mass, BH$^{\star}$ luminosity ($L_{5100}$), and BH$^{\star}$ attenuation ($A_{V,\mathrm{BH}}$). Stellar attenuation and mass remain approximately constant over \uvnir$\sim1$–3, but increase toward the reddest LRDs. Similarly, BH$^{\star}$ luminosity increases with color, while $A_{V,\mathrm{BH}}$ decreases, reaching its lowest values in the reddest LRDs. Red symbols highlight LRDs with power-law SEDs, which show systematically higher $A_{V,\mathrm{BH}}$.}
\label{fig:color_properties_trends}
\end{figure*}


\section{A Semi-Empirical Model for LRD SEDs: From ``The Cliff'' to ``Virgil"}
\label{s:model}

\subsection{A Two-Component model: galaxy plus BH$^{\star}$}

Initial efforts to model the peculiar blue UV and red optical SEDs of LRDs focused on multi-component fits combining stellar and AGN templates. These aimed to avoid the extreme stellar masses or non-standard attenuation laws required by single-component models \citep{kocevski23, barro24, wang24_lrd, labbe24b}. A key challenge in these models is reproducing the sharp inflection between the blue UV continuum and the red optical slope, which often aligns precisely with the Balmer limit \citep{setton24}. The recent confirmation of several LRDs with extreme Balmer breaks ($>$3 mag), inconsistent with any known stellar population \citep{degraaff24, naidu25, taylor25}, has prompted a shift toward models involving dense gas envelopes around accreting black holes. In these so-called BH$^{\star}$ scenarios, emission from hot ($T \sim 10^4$ K), dense gas, analogous to stellar atmospheres, can naturally produce Balmer breaks as strong or stronger than those from evolved stellar populations \citep{inayoshi24}. While early implementations cannot yet reproduce the full rest-optical to NIR spectrum probed by MIRI \citep{degraaff25, taylor25}, they provide a plausible mechanism for the sharp SED turnover at the Balmer limit, since the red BH$^{\star}$ component contributes little flux blueward of the break.

Building on this framework, \citet{naidu25} modeled the SED of the LRD from \citet{furtak24} (see also \citealt{ji25, deugenio25}) using a two-component model combining a UV-bright stellar population with a BH$^{\star}$ component dominating the optical continuum. This success suggests the model could be extended to the broader LRD population. However, to do so, it must account for: (1) the wide range in UV-to-optical colors, (2) the diversity in spectral shapes, including strong curvature and power-law types, and (3) the systematic shift in the SED inflection point from the Balmer limit to longer wavelengths in progressively bluer LRDs, as discussed in the previous section.

To address these constraints, we adopt a semi-empirical model combining a galaxy and a BH$^{\star}$ component. Instead of a complex parametric approach using photoionization codes like \texttt{CLOUDY} \citep{cloudy}, we use a simplified method where the BH$^{\star}$ component is modeled using the observed continuum of the extreme LRD ``The Cliff” \citep{degraaff25}. A similar strategy was recently adopted by \citet{degraaff25b} and \citet{umeda25}, who approximate the BH$^{\star}$ SED with a blackbody. Here, we construct the BH$^{\star}$ template by fitting a low-order polynomial to the emission-line–free regions of the ``The Cliff” spectrum, capturing its shape from the Balmer break to $\sim$1~$\mu$m rest-frame. We assume no UV emission from the BH$^{\star}$ component (see \S~\ref{s:discussion} for further discussion). Dust attenuation is applied using an SMC law \citep{pei92}, with $A_{\rm V}$ as the only free parameter modifying the BH$^{\star}$ shape. Even modest attenuation can transform the intrinsic curvature of the BH$^{\star}$ template into a redder optical continuum that more closely resembles a power law. Varying both the reddening and relative luminosity of the BH$^{\star}$ component naturally shifts the SED inflection point to longer wavelengths. Our model does not include an additional IR component to account for potential dust emission from the BH$^{\star}$, as in \citet{barro25}, since the current spectra do not probe the rest-frame NIR ($\lambda \sim 1$–3~$\mu$m). Incorporating MIRI data would be essential to assess this contribution and test for warm dust associated with the BH$^{\star}$ component.

The stellar component is modeled using \texttt{Prospector} \citep{leja19}, following the same approach as the two-component models in \citet{pgp23b} and \citet{barro25}, with a $\tau$-model star formation history and a reduced set of free parameters: age, $\tau$, stellar mass, and $A_V$, assuming fixed sub-solar metallicity. Best-fit properties for the 118 LRDs with high-S/N continuum detections are listed in Table~\ref{tab:lrd_spec_properties}.

\begin{figure*}
\centering
\includegraphics[width=18cm,angle=0]{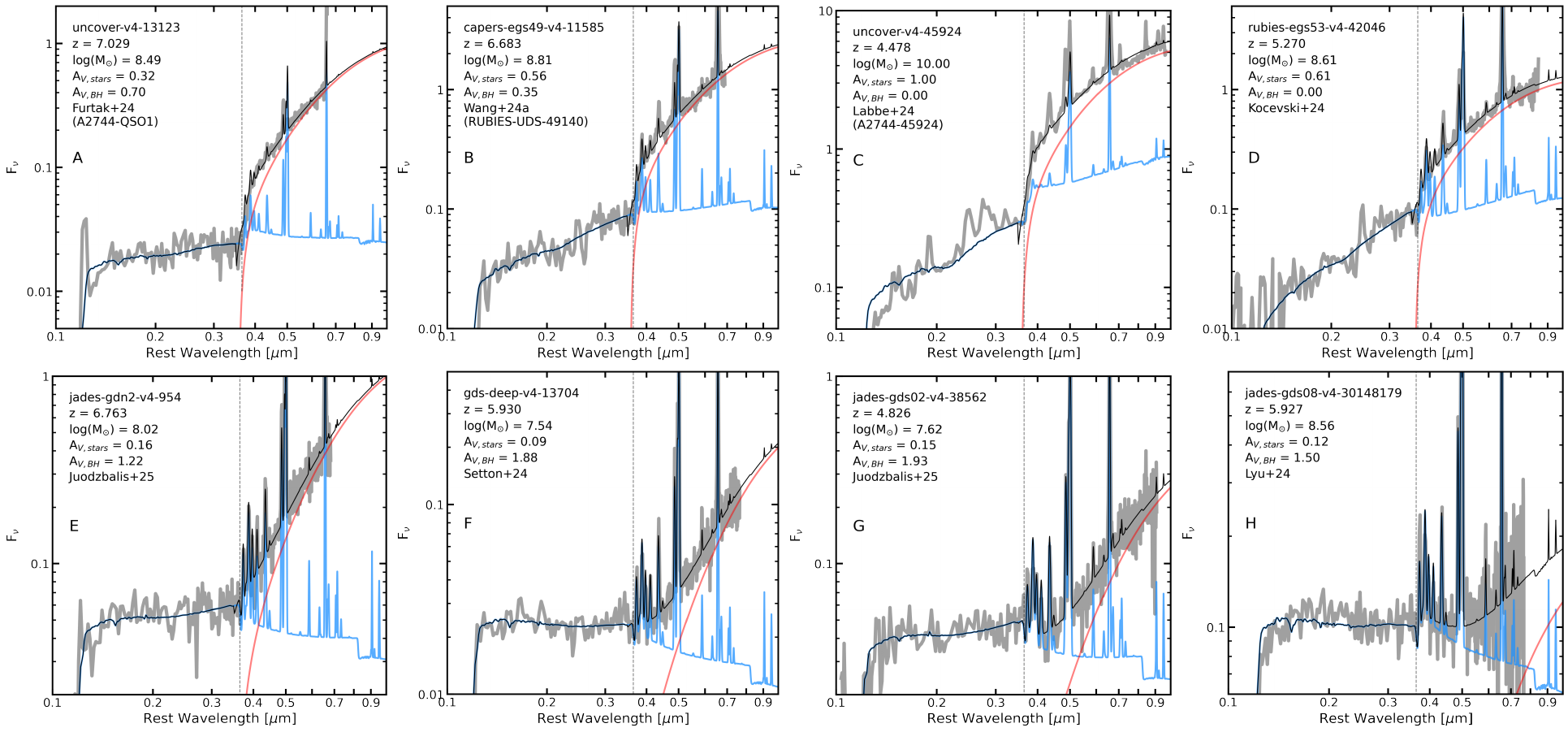}
\caption{Individual spectra (gray) for representative LRDs spanning the full range of \uvnir colors and SED shapes, from strong curvature to power-law–like continua. Each panel also shows the best-fit two-component model (black), with the galaxy and BH$^{\star}$ components plotted separately in blue and red. The dashed vertical line marks the Balmer limit (H${\infty}$). Best-fit stellar mass and dust attenuations for both components are indicated. \textit{Top:} Redder LRDs with prominent Balmer breaks and strong optical curvature are best-fit with low BH$^{\star}$ attenuation ($A_{\rm V, BH}$), but typically higher stellar attenuation ($A_{\rm V, stars}$) and stellar mass. \textit{Bottom:} Progressively bluer, power-law–like LRDs are best-fit with lower stellar attenuation and mass, but higher BH$^{\star}$ attenuation and lower BH$^{\star}$ luminosity. This combination shifts the crossover point between galaxy and BH$^{\star}$ components to longer wavelengths, shifting the inflection between the flat UV and steep optical continuum (see Figure~\ref{fig:inflection}).}
\label{fig:model_fits}
\end{figure*}

\subsection{Interpreting LRD Diversity Through the Two-Component Model}

Our qualitative expectations are confirmed by the distribution of best-fit parameters as a function of rest-frame color across the LRD population. Dividing the sample into three UV-to-optical color bins, \uvnir $>3$, $2<$\uvnir$<3$, and \uvnir $<2$, we find that the median stellar attenuation decreases steadily from $A_V = 0.39^{+0.26}_{-0.24}$ in the reddest bin to $0.18^{+0.16}_{-0.10}$ and $0.14^{+0.08}_{-0.10}$ in the intermediate and bluest bins, respectively. Similarly, the stellar mass declines from \lmass$=8.57^{+0.60}_{-1.14}$ to $7.85^{+0.49}_{-0.36}$ and $7.76^{+0.42}_{-0.29}$. In contrast, the attenuation of the BH$^{\star}$ component increases across the same bins, from $A_{V,\mathrm{BH}} = 0.10^{+1.00}_{-0.10}$ to $0.63^{+1.58}_{-0.63}$ and $1.05^{+2.23}_{-0.87}$. Meanwhile, the BH$^{\star}$ luminosity declines from $\log(\nu L_{\nu})[5100]\equiv \log (L_{5100}~[\mathrm{erg\,s}^{-1}]) = 43.93^{+0.45}_{-0.39}$ to $43.43^{+0.48}_{-0.15}$ and $43.18^{+0.33}_{-0.23}$. These trends suggest that redder LRDs are powered by more luminous, less obscured BH$^{\star}$ components embedded in more attenuated stellar hosts, while bluer LRDs tend to host fainter, more obscured BH$^{\star}$ components and nearly dust-free galaxies.

Figure~\ref{fig:color_properties_trends} illustrates these trends for the full parameter distributions. For the stellar host properties, both $A_V$ and stellar mass remain nearly constant across the bulk of the population (\uvnir$\sim1$–3), and then increase toward the reddest, strong-curvature LRDs, accompanied by a larger scatter and uncertainties that reflect the difficulty of constraining the galaxy component when the UV continuum is weak and increasingly dominated by the BH$^{\star}$ contribution.

The BH$^{\star}$ parameters show a more continuous variation with color. Both the BH$^{\star}$ luminosity and $A_{V,\mathrm{BH}}$ evolve smoothly across the full \uvnir range, but with substantially larger scatter than the stellar properties. In particular, $A_{V,\mathrm{BH}}$ spans a wide range ($A_V \sim 0$–4) and exhibits large uncertainties at high attenuation, reflecting the increasing difficulty of constraining the BH$^{\star}$ component as it becomes fainter relative to the galaxy in the composite SED. We also highlight LRDs with power-law SEDs (red circles), which preferentially show higher median $A_{V,\mathrm{BH}}\sim0.8$, consistent with the attenuation required to reproduce their steep optical continua.

Figure~\ref{fig:model_fits} shows the best-fit SEDs for representative LRDs with NIRSpec/PRISM spectra and high continuum S/N. The top row presents the reddest LRDs, characterized by prominent Balmer breaks and strong curvature. Our model reproduces the SED of the \citet{furtak23} LRD with a similar configuration to that in \citet{naidu25}: a minimally attenuated, low-mass stellar component dominating the UV, and an unreddened BH$^{\star}$ continuum dominating the optical. The model also fits other red LRDs with strong but less extreme breaks, such as those in \citet{labbe24b} and \citet{wang24_lrd}, with slightly higher BH$^{\star}$ attenuation ($A_{V,\mathrm{BH}} \lesssim 0.6$~mag) and more reddened stellar components, needed to reproduce their red UV slopes. A key reason this model outperforms previous multi-component fits is that the BH$^{\star}$ SED features an abrupt drop blueward of the Balmer break, allowing the composite SED to match the sharp inflections observed in these sources. Standard reddened AGN templates tend to produce smoother transitions that cannot reproduce the observed break amplitude \citep[e.g.,][]{ma25}.

The bottom row shows progressively bluer LRDs, with decreasing \uvnir colors and power-law–like SEDs. As discussed above, these objects typically have more attenuated and lower-luminosity BH$^{\star}$ components, along with nearly dust-free stellar hosts. The model reproduces this sequence by increasing $A_{V,\mathrm{BH}}$ to $\sim$1–2~mag while keeping $A_{V,\mathrm{stars}} \sim 0.2$~mag, consistent with their flat UV slopes ($\beta_{\nu,\mathrm{UV}} \sim 0$). Although the power-law shape extends out to $\lambda \sim 1~\mu$m, the model predicts a flattening in the rest-frame NIR, which should be testable with MIRI photometry.

Overall, the two-component model successfully fits LRDs across the full range of \uvnir colors and SED types, suggesting that the observed diversity is primarily driven by variations in the BH$^{\star}$-to-galaxy luminosity ratio (L$_{\rm BH^{\star}}$/L$_{\rm gal}$), together with increasing obscuration of the BH$^{\star}$ component. As L$_{\rm BH^{\star}}$/L$_{\rm gal}$ declines toward bluer colors, the crossover between the stellar (blue) and BH$^{\star}$ (red) continua shifts to longer wavelengths, from $\sim$0.4~$\mu$m to $\sim$0.7~$\mu$m, delaying the onset of the steep optical rise.

\begin{figure}
\includegraphics[width=8.5cm,angle=0]{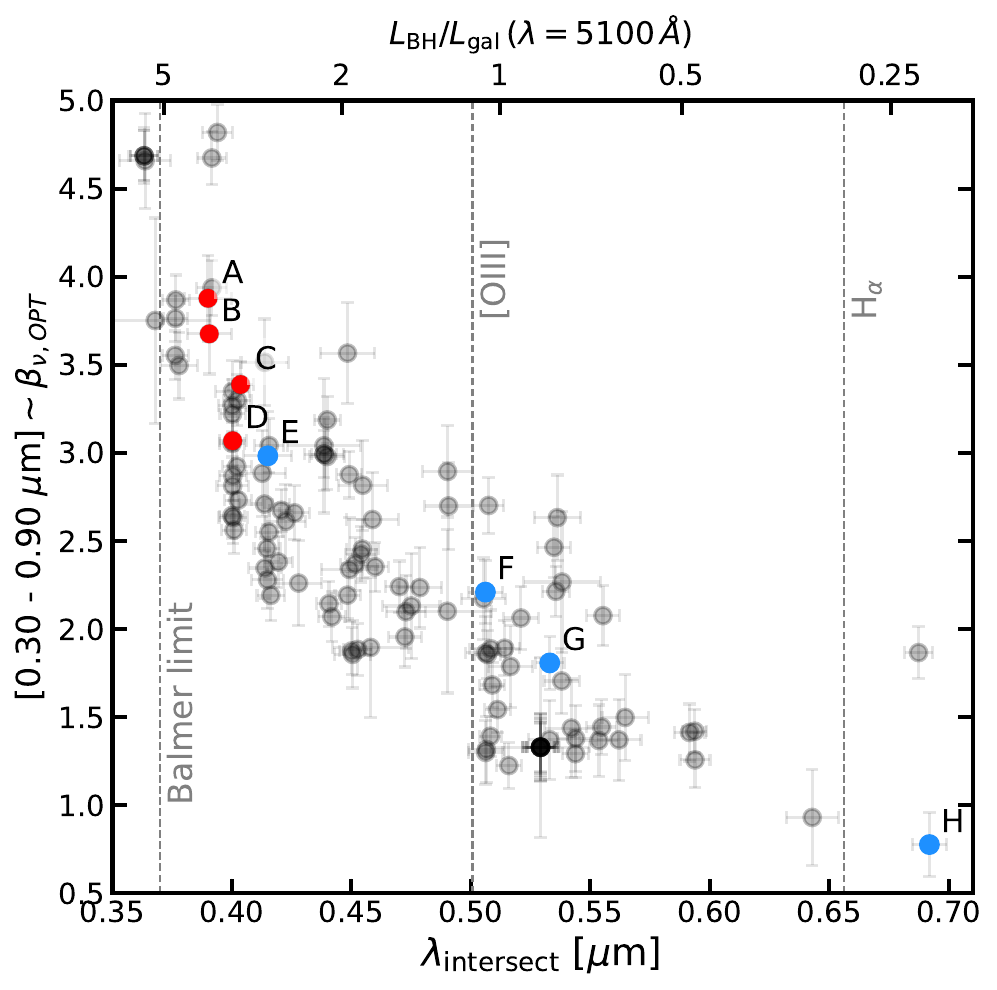}
\caption{Wavelength at which the BH$^{\star}$ and galaxy components have equal luminosity ($L_{\rm BH^{\star}}/L_{\rm gal}=1$), $\lambda_{\rm intersect}$, as a function of rest-frame \uvnir color for spectroscopic LRDs. This crossover serves as a proxy for the SED inflection, marking the transition from flat UV continua to steep optical slopes. Letters and colors correspond to the LRDs shown in Figure~\ref{fig:model_fits}, illustrating the shift from inflections near H$_{\infty}$ in redder LRDs to progressively longer wavelengths in bluer systems as \uvnir decreases.}
\label{fig:inflection}
\end{figure}

To quantify this behavior, Figure~\ref{fig:inflection} shows the wavelength at which the luminosities of the two components are equal, L$_{\rm BH^{\star}}$/L$_{\rm gal}=1$, denoted as $\lambda_{\rm intersect}$. While this crossover does not correspond exactly to the inflection point of the composite SED, the steep slope of the BH$^{\star}$ continuum ensures that the turnover typically occurs slightly blueward of $\lambda_{\rm intersect}$. The top y-axis also reports the luminosity ratio L$_{\rm BH^{\star}}$/L$_{\rm gal}$ at 5100~\AA, which provides a measure of the relative normalization of the two components and closely tracks $\lambda_{\rm intersect}$. We find a clear correlation with color: for the reddest LRDs (\uvnir$\gtrsim 3$), $\lambda_{\rm intersect}$ lies near the Balmer limit, consistent with \citet{setton24}. As \uvnir decreases and L$_{\rm BH^{\star}}$/L$_{\rm gal}$(5100) declines, $\lambda_{\rm intersect}$ shifts progressively redward, often approaching \OIII\ or longer wavelengths. Red and blue symbols mark the LRDs shown in Figure~\ref{fig:model_fits}, illustrating the continuous decline with color. For the bluer LRDs (letters E through H), L$_{\rm BH^{\star}}$/L$_{\rm gal}$(5100) decreases from $\sim$3 to $\sim$0.25, indicating that the BH$^{\star}$ becomes subdominant at 5100~\AA. The bluest example, JADES-30148179, exhibits the longest-wavelength crossover and one of the shallowest optical slopes. First reported by \citet{lyu24} as a MIRI-selected AGN, this source, like “Virgil” \citep{iani24}, exemplifies blue LRDs in which the steepening of the optical continuum becomes apparent only redward of \Ha, as traced by MIRI photometry.

\section{Discussion: Toward a Unified View of the LRD Population}
\label{s:discussion}

\begin{figure}
\includegraphics[width=8.7cm,angle=0]{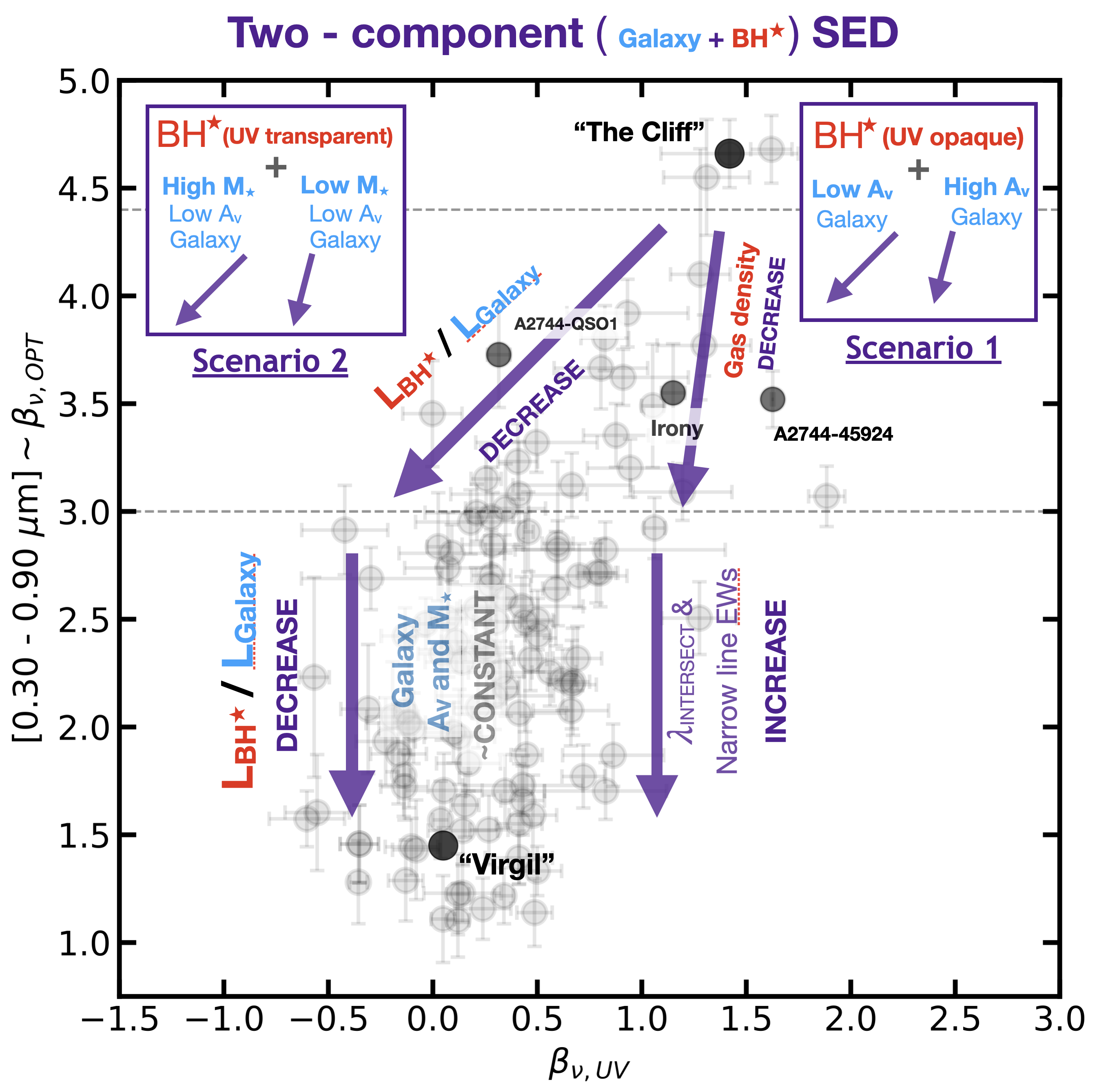}
\caption{Schematic diagram of two scenarios for interpreting trends in the LRD color–slope diagram based on the two-component galaxy (blue) plus BH$^{\star}$ (red) model. In scenario 1 the BH$^{\star}$ component never contributes to the UV (opaque). The UV slopes at the red end depend on the dust attenuation of the host. In scenario 2, the BH$^{\star}$ exhibits some reddened UV emission at lower gas densities. This UV emission, combined with blue galaxy hosts of different luminosities, can explain the observed range in $\beta_{\rm UV}$.  Labeled black points mark representative LRDs at key positions in the diagram. Overall, the large observed spread in \uvnir colors among LRDs is primarily driven by variations in the BH$^{\star}$–to–galaxy luminosity ratio, attenuation, and the presence or absence of UV emission from the BH$^{\star}$.}
\label{fig:LRD_evol}
\end{figure}

The semi-empirical two-component model discussed in \S~\ref{s:model}, in which the rest-frame UV continuum is dominated by stellar emission and the optical by a BH$^{\star}$, provides a useful framework for interpreting the diversity in color and SED shapes of LRD SEDs. Combined with the rest-frame color–slope diagram (Figure~\ref{fig:rest_slopes}), it serves as a diagnostic tool for exploring how variations in BH and host galaxy properties shape the observed continuum, and potentially tracing evolutionary paths within the LRD population. These ideas are illustrated schematically in Figure~\ref{fig:LRD_evol}.

As discussed throughout this work, three key trends emerge in the color–slope diagram when moving from the reddest to the bluest LRDs: (1) The reddest $\sim$20\% of LRDs (\uvnir$\gtrsim 3$) exhibit significantly redder UV slopes, while the majority of the population shows uniformly blue UV slopes with $\beta_{\nu,\mathrm{UV}} \sim 0.3$; (2) the optical continuum shape evolves from strong curvature with Balmer breaks and absorption features to smoother, power-law–like shapes; (3) the FWHM of broad Balmer lines become progressively narrower, while narrow metal lines grow stronger and more prominent in the bluest LRDs.

A simple schematic picture (downward arrows in Figure~\ref{fig:LRD_evol}) is that the spread in \uvnir color reflects variations in the relative luminosity of the BH$^{\star}$-to-host ($L_{\rm BH^{\star}}/L_{\rm gal}$). The reddest LRDs are dominated by luminous BH$^{\star}$ components, while the bluest ones are increasingly galaxy-dominated, not just in the UV but also into the optical (up to $\sim$5100~\AA), and show stronger narrow lines. As $L_{\rm BH^{\star}}/L_{\rm gal}$ declines (due to declining BH$^{\star}$ luminosity and/or increased attenuation), the SED inflection point shifts to longer wavelengths, reducing the curvature of the optical continuum and yielding more power-law–like SEDs.

\subsection{What is the Origin of the Red UV Continuum?}

While this picture explains the majority of LRDs with \uvnir$<3$, it raises questions for the reddest ones. What causes their red UV slopes? In our default two-component model, the galaxy dominates the UV (Scenario 1: BH$^{\star}$ is UV-opaque), so red UV slopes require higher stellar attenuation and masses ($A_V \sim 0.3$–0.5, $\log M_\star \sim 8$–9), in contrast to the uniformly lower values observed in most LRDs ($A_V \sim 0.15$, $\log M_\star \sim 7.5$–8). At face value, the higher stellar masses inferred for the reddest LRDs would be consistent with their larger inferred black hole masses, suggested by both higher luminosities and broader Balmer lines, and could help alleviate the tension with the large $M_{\rm BH}/M_\star \sim 0.01$–0.1 ratios in LRDs relative to local relations \citep[e.g.,][]{kocevski23, kocevski24, maiolino23, juodzbalis25, jones25}.

However, the preference for dustier hosts is unexpected in the context of dense, dust-poor gas envelopes favored in BH$^{\star}$ models for the reddest LRDs. Moreover, the unusual SED of LRDs and the presence of broad wings in their Balmer lines call into question standard bolometric corrections \citep{greene25} and virial linewidth calibrations \citep{rusakov25, kokorev25}, potentially lowering inferred M$_{\rm BH}$ by up to an order of magnitude depending on the accretion rates  \citep{deugenio25c, torralba25b}. Interpreted strictly within Scenario 1, the observed trends would imply an evolutionary sequence from bluer to redder LRDs, with both host galaxies and black holes growing in mass over time.

An alternative interpretation (Scenario 2: BH$^{\star}$ is UV-transparent) is that the BH$^{\star}$ component shows a faint, heavily reddened UV continuum. Although dense gas envelopes are generally expected to suppress emission blueward of the Balmer break, certain combinations of gas density and geometry may allow a small fraction of reddened UV light to escape \citep[e.g.,][]{inayoshi24, degraaff25, naidu25}. In this case, red UV slopes could arise naturally from the BH$^{\star}$ itself, without invoking dusty host galaxies. In bluer LRDs, any such BH$^{\star}$ UV emission would be hidden by the brighter galaxy UV continuum.

Within this scenario, shifts toward bluer \uvnir colors could occur through two channels: (1) decreasing $L_{\rm BH^{\star}}/L_{\rm galaxy}$ as the host becomes more luminous, or (2) declining gas density or covering fraction, increasing the UV continuum from the BH$^{\star}$. Together, these effects can explain the wide range of UV slopes observed at similar \uvnir colors. For example, A2744-45924 and A2744-QSO1 both have \uvnir$\sim3.5$ but very different UV slopes ($\beta_{\nu,\mathrm{UV}}\sim1.5$ and $\sim0$; A and C in Figure 9). A2744-45924 may represent a more transparent BH$^{\star}$ with AGN-dominated UV emission \citep{labbe24b}, whereas A2744-QSO1 resembles a “Cliff”-like system dominated in the UV by a bright, dust-free stellar host \citep{naidu25}. Other LRDs with similar colors but intermediate slopes (e.g., “Irony” or the Balmer absorbers in \citealt{kocevski24} and \citealt{wang24b}) likely reflect varying combinations of BH$^{\star}$ UV opacity and BH-to-galaxy luminosity ratios.

In this second scenario, the reddest LRDs do not require highly attenuated hosts and may instead have lower stellar masses, as indicated by their sub-dominant UV continua. This would imply an evolutionary sequence opposite to that in Scenario 1, with stellar mass increasing from red to blue LRDs. Under this interpretation, the reddest LRDs could correspond to an early phase of rapid, possibly super-Eddington black hole growth in relatively small galaxies. As LRDs evolve toward bluer \uvnir colors, their host galaxies would grow more massive while the BH$^{\star}$ component fades, either by exiting a high accretion phase or by dispersing the dense gas cocoon. A potential challenge to this picture is that black hole masses would need to remain similar or increase toward bluer LRDs, which appears at odds with their lower luminosities and narrower line widths. However, given the uncertainties in interpreting Balmer line profiles in LRDs, a systematic analysis of line shapes across the full population will be essential to robustly test this evolutionary sequence.

\subsection{Is there any Dust in LRDs?}

In the two-component model, the higher incidence of power-law–like optical SEDs among bluer LRDs is attributed to increased attenuation of the BH$^{\star}$ component, which is the primary parameter capable of modifying the continuum shape. In more physically motivated BH$^{\star}$ models, similar reddening could arise from higher gas densities or opacities, reducing the need to invoke dust explicitly. Distinguishing between these possibilities requires assessing whether independent evidence for dust is present.

Although a higher BH$^{\star}$ attenuation in bluer galaxies may seem counterintuitive, it is plausible if these systems host more evolved galaxies with higher metallicities and dust content, leading to stronger local attenuation near the BH$^{\star}$. Consistent with this picture, MIRI observations indicate that bluer LRDs exhibit redder rest-frame near-infrared colors \citep[e.g., 1–3~$\mu$m;][]{pg24a, barro25}, suggestive of warm dust emission. In contrast, the reddest LRDs typically show flatter near-infrared SEDs with little evidence for dust re-emission \citep[e.g.,][]{degraaff25, setton25}, supporting the view that they are comparatively dust-poor.

The rest-frame \nearmidir color provides a sensitive diagnostic of dust emission. Stellar populations and typical obscured AGNs without significant hot dust produce relatively mild colors (\nearmidir$\sim$0$–$0.5; Figure~8 of \citealt{barro25}), and even BH$^{\star}$ or blackbody-like photospheric emission would yield similarly flat or negative values. By contrast, the median \nearmidir color of MIRI-detected LRDs is closer to $\sim$1.5, implying the presence of warm dust emission or, alternatively, a more complex intrinsic continuum shape than captured by simple BH$^{\star}$ models.

Overall, the relatively simple semi-empirical model provides an effective framework for a unified scenario in which LRD diversity arises primarily from variations in the relative luminosities of galaxy and BH$^{\star}$ components. This involves some evolution in attenuation and possibly UV-opacity of the BH$^{\star}$, superimposed on a relatively homogeneous population of blue, low-mass, low-extinction galaxies with prominent (high-EW) emission lines. Testing these predictions will require further refinement of BH$^{\star}$ models to incorporate more realistic radiative transfer and geometric effects, particularly to explain the red UV slopes in the most extreme LRDs. Similarly, a more systematic MIRI follow-up is essential to confirm whether bluer LRDs indeed show more prominent dust re-emission, as suggested by their red rest-NIR colors. In parallel, deeper and higher-resolution spectroscopy will be crucial to assess the impact of line blending on the PRISM pseudo-continuum and to detect faint high-ionization lines that can help disentangle AGN- from star formation–driven emission.


\section{Summary}
\label{s:summary}

This paper presents a detailed analysis of a large, uniformly selected sample of Little Red Dots (LRDs) across six JWST deep fields. Using an exhaustive color selection, we identify nearly 1000 photometric LRDs, of which 191 have publicly available NIRSpec/PRISM spectra in the Dawn JWST Archive (DJA) compilation. Within this spectroscopic sample, 118 LRDs have high continuum S/N~$>2$ in the rest-UV, enabling robust measurements of UV slopes and rest-frame UV-to-optical colors. We use this dataset to quantify the prevalence of broad-line AGN features, characterize the continuum SED shapes among LRDs, and explore correlations between color, SED type, emission line widths, line ratios, and the presence of absorption features.

The main observational results are as follows:
\begin{itemize}
\item The fraction of broad-line AGNs (BLAGNs) satisfying the LRD photometric selection is high, ranging from $\sim$58–92\% across different fields. This fraction increases with a more inclusive selection that extends to bluer optical slopes.

\item Among photometric LRDs with NIRSpec grating spectra at $3.5 < z < 6.8$, $\sim$50\% to 90\% exhibit broad emission lines, depending on the field. These represent lower limits, as nearly all show clear emission lines, and the identification of broad components is limited by depth and sensitivity.

\item The overall LRD population spans nearly 4 magnitudes in \uvnir rest-frame color, indicating broad diversity in intrinsic properties. Most NIRSpec LRDs have blue UV slopes with a median $\beta_{\nu, \mathrm{UV}} = 0.3^{+0.5}_{-0.4}$ and red UV-to-optical colors with median \uvnir$ = 2.3^{+0.7}_{-0.7}$. However, the reddest 18\% (\uvnir$\gtrsim 3$) have significantly redder slopes, reaching $\beta_{\nu, \mathrm{UV}} = 1.1^{+0.5}_{-0.6}$.

\item A similar trend appears in the optical continuum: LRDs with \uvnir $\gtrsim 3$ show pronounced SED curvature, with Balmer breaks and flattening toward the NIR. All literature-reported LRDs with strong breaks fall in this regime. In contrast, bluer LRDs typically show power-law–like optical slopes, forming the canonical ``V-shaped” SED.

\item While curved SEDs dominate at redder colors and power-laws at the blue end, both types exist across the color range. This overlap reveals a degeneracy in UV-to-optical color: distinct SED shapes can yield similar \uvnir values. Like the age–dust degeneracy in stellar populations, this suggests the continuum shape reflects varying combinations of BH luminosity, attenuation, and host galaxy contribution.

\item Broad Balmer line widths decrease with color: median FWHM values drop from 2327~km~s$^{-1}$ in the reddest LRDs to 1677~km~s$^{-1}$ for those with \uvnir$< 2$.

\item Emission-line ratios and equivalent widths also correlate with [0.3$-$0.9 $\mu$m]: the \Ha+\NII/\Hb\ ratio increases with color (reaching $\sim$20 in the reddest), while \OIII/\Hb\ decreases. BH$^{\star}$ candidates show the lowest values ($\ll$1), and bluer LRDs reach up to $\sim$8. The equivalent widths of \OIII\ and \Hg+\OIIIs\ decline toward redder colors, suggesting narrow-line emission becomes more prominent in bluer LRDs, consistent with increased host galaxy contribution.
\end{itemize}

We interpret the range in rest-frame colors, SED types, and emission-line properties using a semi-empirical two-component model combining a galaxy and a BH$^{\star}$ template. Key results from the modeling include:

\begin{itemize}
\item The model successfully fits both curved and power-law SEDs. Curved LRDs require luminous, low-attenuation BH$^{\star}$ components and more obscured stellar hosts. Power-law LRDs are best fit with fainter, highly attenuated BH$^{\star}$ components and nearly dust-free galaxies.

\item The broad range in \uvnir color is primarily driven by the relative luminosity and attenuation of the BH$^{\star}$ and galaxy components.

\item The wavelength where the SED transitions from flat UV to steep optical shits with color: in redder LRDs (\uvnir$\gtrsim 3$), the inflection occurs near the Balmer limit; in bluer LRDs, it shifts progressively redward, beyond \OIII\ and toward \Ha.
\end{itemize}

\section*{Acknowledgments}
We thank G. Brammer and the Dawn JWST Archive (DJA) for making high-level data products publicly available. DJA is an initiative of the Cosmic Dawn Center, which is funded by the Danish National Research Foundation under grant No. 140. PGP-G acknowledges support from grant PID2022-139567NB-I00 funded by Spanish Ministerio de Ciencia, Innovaci\'on y Universidades MCIU/AEI/10.13039/501100011033, FEDER {\it Una manera de hacer Europa}. 

This work is based on observations made with the NASA/ESA/CSA \textit{James Webb Space Telescope}. 
The data were obtained from the Mikulski Archive for Space Telescopes (MAST) at the Space Telescope 
Science Institute (STScI), which is operated by the Association of Universities for Research in Astronomy, Inc., under NASA contract NAS~5-03127. These observations are associated with JWST programs 
GTO-1180, GTO-1181, GTO-1215, GTO-1286, GO-1345, GO-2198, GO-2561, DDT-2750, GO-3073, 
GO-4106, GO-4233, GO-5224, GO-6585, GO-6838, and DDT-6541.

JSD acknowledges the support of the Royal Society via the award of a Royal Society Research Professorship. This work has made use of the Rainbow Cosmological Surveys Database, which is operated by the Centro de Astrobiología (CAB), CSIC-INTA, partnered with the University of California Observatories at Santa Cruz (UCO/Lick, UCSC). This project has received funding from the European Union’s Horizon 2020 research and innovation programme under the Marie Skłodowska-Curie grant agreement No 101148925.

\software{Astropy \citep{astropy}, EAZY \citep{eazy}, GALFIT  \citep{galfit}, matplotlib \citep{matplotlib}, NumPy \citep{numpy}, PZETA (\citealt{pg08}), Prospector (\citealt{leja19} \citealt{johnson21}), Rainbow pipeline (\citealt{pg05,pg08}, \citealt{barro11b}), SExtractor \citep{sex}, Synthesizer (\citealt{pg05,pg08b})}

\bibliography{referencias}

\end{document}

%% file: tab2.tex
\begin{table*}
\caption{Properties of LRDs with NIRSpec PRISM spectra}
\centering
\footnotesize
\begin{tabular}{lcccccccc}
\hline
DJA Name & RA & Dec & $z_{\rm spec}$ & $\beta_{\nu, \rm UV}$ & [0.3$-$0.9 $\mu$m] & $\log(M_\star/M_\odot)$ & $A_{V,{\rm gal}}$ & $A_{V,{\rm BH}}$ \\
\hline
rubies-egs61-v4-69475 & 214.949188 & 52.964137 & 5.620 & $0.41 \pm 0.23$ & $2.06 \pm 0.17$ & $7.45 \pm 0.16$ & $0.15 \pm 0.06$ & $0.00 \pm 0.03$ \\
ceers-ddt-v4-1768 & 214.925759 & 52.945659 & 5.088 & $0.50 \pm 0.29$ & $3.32 \pm 0.14$ & $7.14 \pm 0.26$ & $0.20 \pm 0.05$ & $0.00 \pm 0.07$ \\
rubies-egs61-v4-61496 & 214.972440 & 52.962191 & 5.084 & $0.70 \pm 0.23$ & $2.70 \pm 0.15$ & $7.89 \pm 0.81$ & $0.25 \pm 0.14$ & $1.50 \pm 0.76$ \\
capers-egs53-v4-7761 & 214.840034 & 52.860648 & 3.859 & $0.22 \pm 0.12$ & $2.62 \pm 0.19$ & $7.92 \pm 0.35$ & $0.26 \pm 0.06$ & $0.00 \pm 0.20$ \\
capers-egs49-v4-8752 & 214.876137 & 52.880826 & 8.357 & $0.60 \pm 0.07$ & $2.85 \pm 0.38$ & $8.77 \pm 0.15$ & $0.32 \pm 0.04$ & $1.24 \pm 0.39$ \\
rubies-egs61-v4-55604 & 214.983030 & 52.956002 & 6.989 & $1.05 \pm 0.05$ & $3.49 \pm 0.29$ & $9.62 \pm 0.21$ & $0.59 \pm 0.07$ & $1.50 \pm 0.25$ \\
capers-egs47-v4-9244 & 214.985675 & 52.956230 & 5.200 & $0.12 \pm 0.05$ & $1.64 \pm 0.17$ & $7.58 \pm 0.04$ & $0.09 \pm 0.02$ & $0.08 \pm 0.20$ \\
capers-egs53-v4-10108 & 214.797532 & 52.818752 & 6.623 & $0.21 \pm 0.09$ & $1.98 \pm 0.19$ & $7.87 \pm 0.20$ & $0.17 \pm 0.04$ & $0.13 \pm 0.31$ \\
\hline
\end{tabular}
\tablecomments{
This table is available in its entirety at
\url{https://github.com/guillermobc/Barro25}.
}
\label{tab:lrd_spec_properties}
\end{table*}

%% file: referencias.bib
@Article{eazy,
  Author         = {{Brammer}, G.~B. and {van Dokkum}, P.~G. and {Coppi},
                   P.},
  Title          = "{EAZY: A Fast, Public Photometric Redshift Code}",
  Journal        = {\apj},
  Volume         = 686,
  Pages          = {1503-1513},
  adsurl         = {http://adsabs.harvard.edu/abs/2008ApJ...686.1503B},
  archiveprefix  = "arXiv",
  doi            = {10.1086/591786},
  eprint         = {0807.1533},
  month          = oct,
  year           = 2008
}

@Article{kron80,
  Author         = {{Kron}, R.~G.},
  Title          = "{Photometry of a complete sample of faint galaxies}",
  Journal        = {\apjs},
  Volume         = 43,
  Pages          = {305-325},
  adsurl         = {http://adsabs.harvard.edu/abs/1980ApJS...43..305K},
  doi            = {10.1086/190669},
  month          = jun,
  year           = 1980
}

@Article{sex,
  Author         = {{Bertin}, E. and {Arnouts}, S.},
  Title          = "{SExtractor: Software for source extraction.}",
  Journal        = {\aaps},
  Volume         = 117,
  Pages          = {393-404},
  adsurl         = {http://adsabs.harvard.edu/abs/1996A%26AS..117..393B},
  month          = jun,
  year           = 1996
}

@Article{pg05,
  Author         = {{P{\'e}rez-Gonz{\'a}lez}, P.~G. and {Rieke}, G.~H. and
                   {Egami}, E. and {Alonso-Herrero}, A. and {Dole}, H. and
                   {Papovich}, C. and {Blaylock}, M. and {Jones}, J. and
                   {Rieke}, M. and {Rigby}, J. and {Barmby}, P. and
                   {Fazio}, G.~G. and {Huang}, J. and {Martin}, C.},
  Title          = "{Spitzer View on the Evolution of Star-forming
                   Galaxies from z = 0 to z \~{} 3}",
  Journal        = {\apj},
  Volume         = 630,
  Pages          = {82-107},
  adsurl         = {http://adsabs.harvard.edu/abs/2005ApJ...630...82P},
  doi            = {10.1086/431894},
  eprint         = {arXiv:astro-ph/0505101},
  month          = sep,
  year           = 2005
}

@Article{cloudy,
  Author         = {{Ferland}, G.~J. and {Korista}, K.~T. and {Verner},
                   D.~A. and {Ferguson}, J.~W. and {Kingdon}, J.~B. and
                   {Verner}, E.~M.},
  Title          = "{CLOUDY 90: Numerical Simulation of Plasmas and Their
                   Spectra}",
  Journal        = {\pasp},
  Volume         = 110,
  Pages          = {761-778},
  adsurl         = {http://adsabs.harvard.edu/abs/1998PASP..110..761F},
  doi            = {10.1086/316190},
  month          = jul,
  year           = 1998
}

@Article{bpt,
  Author         = {{Baldwin}, J.~A. and {Phillips}, M.~M. and
                   {Terlevich}, R.},
  Title          = "{Classification parameters for the emission-line
                   spectra of extragalactic objects}",
  Journal        = {\pasp},
  Volume         = 93,
  Pages          = {5-19},
  adsurl         = {http://adsabs.harvard.edu/abs/1981PASP...93....5B},
  doi            = {10.1086/130766},
  month          = feb,
  year           = 1981
}

@Article{pg08b,
  Author         = {{P{\'e}rez-Gonz{\'a}lez}, P.~G. and {Trujillo}, I. and
                   {Barro}, G. and {Gallego}, J. and {Zamorano}, J. and
                   {Conselice}, C.~J.},
  Title          = "{Exploring the Evolutionary Paths of the Most Massive
                   Galaxies since z \~{} 2}",
  Journal        = {\apj},
  Volume         = 687,
  Pages          = {50-58},
  adsurl         = {http://adsabs.harvard.edu/abs/2008ApJ...687...50P},
  archiveprefix  = "arXiv",
  doi            = {10.1086/591843},
  eprint         = {0807.1069},
  month          = nov,
  year           = 2008
}

@Article{pg08,
  Author         = {{P{\'e}rez-Gonz{\'a}lez}, P.~G. and {Rieke}, G.~H. and
                   {Villar}, V. and {Barro}, G. and {Blaylock}, M. and
                   {Egami}, E. and {Gallego}, J. and {Gil de Paz}, A. and
                   {Pascual}, S. and {Zamorano}, J. and {Donley}, J.~L. },
  Title          = "{The Stellar Mass Assembly of Galaxies from z = 0 to z
                   = 4: Analysis of a Sample Selected in the Rest-Frame
                   Near-Infrared with Spitzer}",
  Journal        = {\apj},
  Volume         = 675,
  Pages          = {234-261},
  adsurl         = {http://adsabs.harvard.edu/abs/2008ApJ...675..234P},
  doi            = {10.1086/523690},
  eprint         = {arXiv:0709.1354},
  month          = mar,
  year           = 2008
}

@Article{wuyts07,
  Author         = {{Wuyts}, S. and {Labb{\'e}}, I. and {Franx}, M. and
                   {Rudnick}, G. and {van Dokkum}, P.~G. and {Fazio},
                   G.~G. and {F{\"o}rster Schreiber}, N.~M. and {Huang},
                   J. and {Moorwood}, A.~F.~M. and {Rix}, H.-W. and
                   {R{\"o}ttgering}, H. and {van der Werf}, P.},
  Title          = "{What Do We Learn from IRAC Observations of Galaxies
                   at 2 {\lt} z {\lt} 3.5?}",
  Journal        = {\apj},
  Volume         = 655,
  Pages          = {51-65},
  adsurl         = {http://adsabs.harvard.edu/abs/2007ApJ...655...51W},
  doi            = {10.1086/509708},
  eprint         = {arXiv:astro-ph/0609548},
  month          = jan,
  year           = 2007
}

@ARTICLE{williams10,
   author = {{Williams}, R.~J. and {Quadri}, R.~F. and {Franx}, M. and {van Dokkum}, P. and 
	{Toft}, S. and {Kriek}, M. and {Labb{\'e}}, I.},
    title = "{The Evolving Relations Between Size, Mass, Surface Density, and Star Formation in 3 {\times} 10$^{4}$ Galaxies Since z = 2}",
  journal = {\apj},
archivePrefix = "arXiv",
   eprint = {0906.4786},
 primaryClass = "astro-ph.CO",
     year = 2010,
    month = apr,
   volume = 713,
    pages = {738-750},
      doi = {10.1088/0004-637X/713/2/738},
   adsurl = {http://adsabs.harvard.edu/abs/2010ApJ...713..738W}
}

@ARTICLE{galfit,
   author = {{Peng}, C.~Y. and {Ho}, L.~C. and {Impey}, C.~D. and {Rix}, H.-W.
	},
    title = "{Detailed Structural Decomposition of Galaxy Images}",
  journal = {\aj},
   eprint = {arXiv:astro-ph/0204182},
     year = 2002,
    month = jul,
   volume = 124,
    pages = {266-293},
      doi = {10.1086/340952},
   adsurl = {http://adsabs.harvard.edu/abs/2002AJ....124..266P}
}

@ARTICLE{barro11b,
   author = {{Barro}, G. and {P{\'e}rez-Gonz{\'a}lez}, P.~G. and {Gallego}, J. and 
	{Ashby}, M.~L.~N. and {Kajisawa}, M. and {Miyazaki}, S. and 
	{Villar}, V. and {Yamada}, T. and {Zamorano}, J.},
    title = "{UV-to-FIR Analysis of Spitzer/IRAC Sources in the Extended Groth Strip. II. Photometric Redshifts, Stellar Masses, and Star Formation Rates}",
  journal = {\apjs},
archivePrefix = "arXiv",
   eprint = {1102.4335},
 primaryClass = "astro-ph.CO",
     year = 2011,
    month = apr,
   volume = 193,
      eid = {30},
    pages = {30},
      doi = {10.1088/0067-0049/193/2/30},
   adsurl = {http://adsabs.harvard.edu/abs/2011ApJS..193...30B}
}

@ARTICLE{labbe23,
       author = {{Labb{\'e}}, Ivo and {van Dokkum}, Pieter and {Nelson}, Erica and {Bezanson}, Rachel and {Suess}, Katherine A. and {Leja}, Joel and {Brammer}, Gabriel and {Whitaker}, Katherine and {Mathews}, Elijah and {Stefanon}, Mauro and {Wang}, Bingjie},
        title = "{A population of red candidate massive galaxies  600 Myr after the Big Bang}",
      journal = {\nat},
         year = 2023,
        month = apr,
       volume = {616},
       number = {7956},
        pages = {266-269},
          doi = {10.1038/s41586-023-05786-2},
archivePrefix = {arXiv},
       eprint = {2207.12446},
 primaryClass = {astro-ph.GA},
       adsurl = {https://ui.adsabs.harvard.edu/abs/2023Natur.616..266L}
}

@ARTICLE{labbe24,
       author = {{Labbe}, Ivo and {Greene}, Jenny E. and {Bezanson}, Rachel and {Fujimoto}, Seiji and {Furtak}, Lukas J. and {Goulding}, Andy D. and {Matthee}, Jorryt and {Naidu}, Rohan P. and {Oesch}, Pascal A. and {Atek}, Hakim and {Brammer}, Gabriel and {Chemerynska}, Iryna and {Coe}, Dan and {Cutler}, Sam E. and {Dayal}, Pratika and {Feldmann}, Robert and {Franx}, Marijn and {Glazebrook}, Karl and {Leja}, Joel and {Marchesini}, Danilo and {Maseda}, Michael and {Nanayakkara}, Themiya and {Nelson}, Erica J. and {Pan}, Richard and {Papovich}, Casey and {Price}, Sedona H. and {Suess}, Katherine A. and {Wang}, Bingjie and {Whitaker}, Katherine E. and {Williams}, Christina C. and {Zitrin}, Adi},
        title = "{UNCOVER: Candidate Red Active Galactic Nuclei at 3<z<7 with JWST and ALMA}",
      journal = {arXiv e-prints},
         year = 2023,
        month = jun,
          eid = {arXiv:2306.07320},
        pages = {arXiv:2306.07320},
          doi = {10.48550/arXiv.2306.07320},
archivePrefix = {arXiv},
       eprint = {2306.07320},
 primaryClass = {astro-ph.GA},
       adsurl = {https://ui.adsabs.harvard.edu/abs/2023arXiv230607320L}
}

@ARTICLE{kocevski23,
       author = {{Kocevski}, Dale D. and {Onoue}, Masafusa and {Inayoshi}, Kohei and {Trump}, Jonathan R. and {Arrabal Haro}, Pablo and {Grazian}, Andrea and {Dickinson}, Mark and {Finkelstein}, Steven L. and {Kartaltepe}, Jeyhan S. and {Hirschmann}, Michaela and {Fujimoto}, Seiji and {Juneau}, Stephanie and {Amorin}, Ricardo O. and {Bagley}, Micaela B. and {Barro}, Guillermo and {Bell}, Eric F. and {Bisigello}, Laura and {Calabro}, Antonello and {Cleri}, Nikko J. and {Cooper}, M.~C. and {Ding}, Xuheng and {Grogin}, Norman A. and {Ho}, Luis C. and {Inoue}, Akio K. and {Jiang}, Linhua and {Jones}, Brenda and {Koekemoer}, Anton M. and {Li}, Wenxiu and {Li}, Zhengrong and {McGrath}, Elizabeth J. and {Molina}, Juan and {Papovich}, Casey and {Perez-Gonzalez}, Pablo G. and {Pirzkal}, Nor and {Wilkins}, Stephen M. and {Yang}, Guang and {Yung}, L.~Y. Aaron},
        title = "{Hidden Little Monsters: Spectroscopic Identification of Low-Mass, Broad-Line AGN at $z>5$ with CEERS}",
      journal = {arXiv e-prints},
         year = 2023,
        month = jan,
          eid = {arXiv:2302.00012},
        pages = {arXiv:2302.00012},
          doi = {10.48550/arXiv.2302.00012},
archivePrefix = {arXiv},
       eprint = {2302.00012},
 primaryClass = {astro-ph.GA},
       adsurl = {https://ui.adsabs.harvard.edu/abs/2023arXiv230200012K}
}

@ARTICLE{johnson21,
       author = {{Johnson}, Benjamin D. and {Leja}, Joel and {Conroy}, Charlie and {Speagle}, Joshua S.},
        title = "{Stellar Population Inference with Prospector}",
      journal = {\apjs},
         year = 2021,
        month = jun,
       volume = {254},
       number = {2},
          eid = {22},
        pages = {22},
          doi = {10.3847/1538-4365/abef67},
archivePrefix = {arXiv},
       eprint = {2012.01426},
 primaryClass = {astro-ph.GA},
       adsurl = {https://ui.adsabs.harvard.edu/abs/2021ApJS..254...22J}
}

@ARTICLE{leja19,
       author = {{Leja}, Joel and {Carnall}, Adam C. and {Johnson}, Benjamin D. and {Conroy}, Charlie and {Speagle}, Joshua S.},
        title = "{How to Measure Galaxy Star Formation Histories. II. Nonparametric Models}",
      journal = {\apj},
         year = 2019,
        month = may,
       volume = {876},
       number = {1},
          eid = {3},
        pages = {3},
          doi = {10.3847/1538-4357/ab133c},
archivePrefix = {arXiv},
       eprint = {1811.03637},
 primaryClass = {astro-ph.GA},
       adsurl = {https://ui.adsabs.harvard.edu/abs/2019ApJ...876....3L}
}

@ARTICLE{polletta06,
       author = {{Polletta}, Maria del Carmen and {Wilkes}, Belinda J. and {Siana}, Brian and {Lonsdale}, Carol J. and {Kilgard}, Roy and {Smith}, Harding E. and {Kim}, Dong-Woo and {Owen}, Frazer and {Efstathiou}, Andreas and {Jarrett}, Tom and {Stacey}, Gordon and {Franceschini}, Alberto and {Rowan-Robinson}, Michael and {Babbedge}, Tom S.~R. and {Berta}, Stefano and {Fang}, Fan and {Farrah}, Duncan and {Gonz{\'a}lez-Solares}, Eduardo and {Morrison}, Glenn and {Surace}, Jason A. and {Shupe}, Dave L.},
        title = "{Chandra and Spitzer Unveil Heavily Obscured Quasars in the Chandra/SWIRE Survey}",
      journal = {\apj},
         year = 2006,
        month = may,
       volume = {642},
       number = {2},
        pages = {673-693},
          doi = {10.1086/500821},
archivePrefix = {arXiv},
       eprint = {astro-ph/0602228},
 primaryClass = {astro-ph},
       adsurl = {https://ui.adsabs.harvard.edu/abs/2006ApJ...642..673P}
}

@ARTICLE{pgp23b,
       author = {{P{\'e}rez-Gonz{\'a}lez}, Pablo G. and {Costantin}, Luca and {Langeroodi}, Danial and {Rinaldi}, Pierluigi and {Annunziatella}, Marianna and {Ilbert}, Olivier and {Colina}, Luis and {Noorgaard-Nielsen}, Hans Ulrik and {Greve}, Thomas and {Ostlin}, G{\"o}ran and {Wright}, Gillian and {Alonso-Herrero}, Almudena and {{\'A}lvarez-M{\'a}rquez}, Javier and {Caputi}, Karina I. and {Eckart}, Andreas and {Le F{\`e}vre}, Olivier and {Labiano}, {\'A}lvaro and {Garc{\'\i}a-Mar{\'\i}n}, Macarena and {Hjorth}, Jens and {Kendrew}, Sarah and {Pye}, John P. and {Tikkanen}, Tuomo and {van der Werf}, Paul and {Walter}, Fabian and {Ward}, Martin and {Bosman}, Sarah E.~I. and {Gillman}, Steven and {Garc{\'\i}a-Argum{\'a}nez}, {\'A}ngela and {Mar{\'\i}a M{\'e}rida}, Rosa},
        title = "{Life beyond 30: probing the $-20<M_\mathrm{UV}<-17$ luminosity function at $8<z<13$ with the NIRCam parallel field of the MIRI Deep Survey}",
      journal = {arXiv e-prints},
         year = 2023,
        month = feb,
          eid = {arXiv:2302.02429},
        pages = {arXiv:2302.02429},
          doi = {10.48550/arXiv.2302.02429},
archivePrefix = {arXiv},
       eprint = {2302.02429},
 primaryClass = {astro-ph.GA},
       adsurl = {https://ui.adsabs.harvard.edu/abs/2023arXiv230202429P}
}

@ARTICLE{pgp23a,
       author = {{P{\'e}rez-Gonz{\'a}lez}, Pablo G. and {Barro}, Guillermo and {Annunziatella}, Marianna and {Costantin}, Luca and {Garc{\'\i}a-Argum{\'a}nez}, {\'A}ngela and {McGrath}, Elizabeth J. and {M{\'e}rida}, Rosa M. and {Zavala}, Jorge A. and {Haro}, Pablo Arrabal and {Bagley}, Micaela B. and {Backhaus}, Bren E. and {Behroozi}, Peter and {Bell}, Eric F. and {Bisigello}, Laura and {Buat}, V{\'e}ronique and {Calabr{\`o}}, Antonello and {Casey}, Caitlin M. and {Cleri}, Nikko J. and {Coogan}, Rosemary T. and {Cooper}, M.~C. and {Cooray}, Asantha R. and {Dekel}, Avishai and {Dickinson}, Mark and {Elbaz}, David and {Ferguson}, Henry C. and {Finkelstein}, Steven L. and {Fontana}, Adriano and {Franco}, Maximilien and {Gardner}, Jonathan P. and {Giavalisco}, Mauro and {G{\'o}mez-Guijarro}, Carlos and {Grazian}, Andrea and {Grogin}, Norman A. and {Guo}, Yuchen and {Huertas-Company}, Marc and {Jogee}, Shardha and {Kartaltepe}, Jeyhan S. and {Kewley}, Lisa J. and {Kirkpatrick}, Allison and {Kocevski}, Dale D. and {Koekemoer}, Anton M. and {Long}, Arianna S. and {Lotz}, Jennifer M. and {Lucas}, Ray A. and {Papovich}, Casey and {Pirzkal}, Nor and {Ravindranath}, Swara and {Somerville}, Rachel S. and {Tacchella}, Sandro and {Trump}, Jonathan R. and {Wang}, Weichen and {Wilkins}, Stephen M. and {Wuyts}, Stijn and {Yang}, Guang and {Aaron Yung}, L.~Y.},
        title = "{CEERS Key Paper. IV. A Triality in the Nature of HST-dark Galaxies}",
      journal = {\apjl},
         year = 2023,
        month = mar,
       volume = {946},
       number = {1},
          eid = {L16},
        pages = {L16},
          doi = {10.3847/2041-8213/acb3a5},
archivePrefix = {arXiv},
       eprint = {2211.00045},
 primaryClass = {astro-ph.GA},
       adsurl = {https://ui.adsabs.harvard.edu/abs/2023ApJ...946L..16P}
}

@ARTICLE{furtak23,
       author = {{Furtak}, Lukas J. and {Zitrin}, Adi and {Plat}, Ad{\`e}le and {Fujimoto}, Seiji and {Wang}, Bingjie and {Nelson}, Erica J. and {Labb{\'e}}, Ivo and {Bezanson}, Rachel and {Brammer}, Gabriel B. and {van Dokkum}, Pieter and {Endsley}, Ryan and {Glazebrook}, Karl and {Greene}, Jenny E. and {Leja}, Joel and {Price}, Sedona H. and {Smit}, Renske and {Stark}, Daniel P. and {Weaver}, John R. and {Whitaker}, Katherine E. and {Atek}, Hakim and {Chevallard}, Jacopo and {Curtis-Lake}, Emma and {Dayal}, Pratika and {Feltre}, Anna and {Franx}, Marijn and {Fudamoto}, Yoshinobu and {Marchesini}, Danilo and {Mowla}, Lamiya A. and {Pan}, Richard and {Suess}, Katherine A. and {Vidal-Garc{\'\i}a}, Alba and {Williams}, Christina C.},
        title = "{JWST UNCOVER: Extremely Red and Compact Object at z $_{phot}$ ≃ 7.6 Triply Imaged by A2744}",
      journal = {\apj},
         year = 2023,
        month = aug,
       volume = {952},
       number = {2},
          eid = {142},
        pages = {142},
          doi = {10.3847/1538-4357/acdc9d},
archivePrefix = {arXiv},
       eprint = {2212.10531},
 primaryClass = {astro-ph.GA},
       adsurl = {https://ui.adsabs.harvard.edu/abs/2023ApJ...952..142F}
}

@ARTICLE{oke83,
       author = {{Oke}, J.~B. and {Gunn}, J.~E.},
        title = "{Secondary standard stars for absolute spectrophotometry.}",
      journal = {\apj},
         year = 1983,
        month = mar,
       volume = {266},
        pages = {713-717},
          doi = {10.1086/160817},
       adsurl = {https://ui.adsabs.harvard.edu/abs/1983ApJ...266..713O}
}

@ARTICLE{astropy,
       author = {{Astropy Collaboration} and {Price-Whelan}, Adrian M. and {Lim}, Pey Lian and {Earl}, Nicholas and {Starkman}, Nathaniel and {Bradley}, Larry and {Shupe}, David L. and {Patil}, Aarya A. and {Corrales}, Lia and {Brasseur}, C.~E. and {N{\"o}the}, Maximilian and {Donath}, Axel and {Tollerud}, Erik and {Morris}, Brett M. and {Ginsburg}, Adam and {Vaher}, Eero and {Weaver}, Benjamin A. and {Tocknell}, James and {Jamieson}, William and {van Kerkwijk}, Marten H. and {Robitaille}, Thomas P. and {Merry}, Bruce and {Bachetti}, Matteo and {G{\"u}nther}, H. Moritz and {Aldcroft}, Thomas L. and {Alvarado-Montes}, Jaime A. and {Archibald}, Anne M. and {B{\'o}di}, Attila and {Bapat}, Shreyas and {Barentsen}, Geert and {Baz{\'a}n}, Juanjo and {Biswas}, Manish and {Boquien}, M{\'e}d{\'e}ric and {Burke}, D.~J. and {Cara}, Daria and {Cara}, Mihai and {Conroy}, Kyle E. and {Conseil}, Simon and {Craig}, Matthew W. and {Cross}, Robert M. and {Cruz}, Kelle L. and {D'Eugenio}, Francesco and {Dencheva}, Nadia and {Devillepoix}, Hadrien A.~R. and {Dietrich}, J{\"o}rg P. and {Eigenbrot}, Arthur Davis and {Erben}, Thomas and {Ferreira}, Leonardo and {Foreman-Mackey}, Daniel and {Fox}, Ryan and {Freij}, Nabil and {Garg}, Suyog and {Geda}, Robel and {Glattly}, Lauren and {Gondhalekar}, Yash and {Gordon}, Karl D. and {Grant}, David and {Greenfield}, Perry and {Groener}, Austen M. and {Guest}, Steve and {Gurovich}, Sebastian and {Handberg}, Rasmus and {Hart}, Akeem and {Hatfield-Dodds}, Zac and {Homeier}, Derek and {Hosseinzadeh}, Griffin and {Jenness}, Tim and {Jones}, Craig K. and {Joseph}, Prajwel and {Kalmbach}, J. Bryce and {Karamehmetoglu}, Emir and {Ka{\l}uszy{\'n}ski}, Miko{\l}aj and {Kelley}, Michael S.~P. and {Kern}, Nicholas and {Kerzendorf}, Wolfgang E. and {Koch}, Eric W. and {Kulumani}, Shankar and {Lee}, Antony and {Ly}, Chun and {Ma}, Zhiyuan and {MacBride}, Conor and {Maljaars}, Jakob M. and {Muna}, Demitri and {Murphy}, N.~A. and {Norman}, Henrik and {O'Steen}, Richard and {Oman}, Kyle A. and {Pacifici}, Camilla and {Pascual}, Sergio and {Pascual-Granado}, J. and {Patil}, Rohit R. and {Perren}, Gabriel I. and {Pickering}, Timothy E. and {Rastogi}, Tanuj and {Roulston}, Benjamin R. and {Ryan}, Daniel F. and {Rykoff}, Eli S. and {Sabater}, Jose and {Sakurikar}, Parikshit and {Salgado}, Jes{\'u}s and {Sanghi}, Aniket and {Saunders}, Nicholas and {Savchenko}, Volodymyr and {Schwardt}, Ludwig and {Seifert-Eckert}, Michael and {Shih}, Albert Y. and {Jain}, Anany Shrey and {Shukla}, Gyanendra and {Sick}, Jonathan and {Simpson}, Chris and {Singanamalla}, Sudheesh and {Singer}, Leo P. and {Singhal}, Jaladh and {Sinha}, Manodeep and {Sip{\H{o}}cz}, Brigitta M. and {Spitler}, Lee R. and {Stansby}, David and {Streicher}, Ole and {{\v{S}}umak}, Jani and {Swinbank}, John D. and {Taranu}, Dan S. and {Tewary}, Nikita and {Tremblay}, Grant R. and {Val-Borro}, Miguel de and {Van Kooten}, Samuel J. and {Vasovi{\'c}}, Zlatan and {Verma}, Shresth and {de Miranda Cardoso}, Jos{\'e} Vin{\'\i}cius and {Williams}, Peter K.~G. and {Wilson}, Tom J. and {Winkel}, Benjamin and {Wood-Vasey}, W.~M. and {Xue}, Rui and {Yoachim}, Peter and {Zhang}, Chen and {Zonca}, Andrea and {Astropy Project Contributors}},
        title = "{The Astropy Project: Sustaining and Growing a Community-oriented Open-source Project and the Latest Major Release (v5.0) of the Core Package}",
      journal = {\apj},
         year = 2022,
        month = aug,
       volume = {935},
       number = {2},
          eid = {167},
        pages = {167},
          doi = {10.3847/1538-4357/ac7c74},
archivePrefix = {arXiv},
       eprint = {2206.14220},
 primaryClass = {astro-ph.IM},
       adsurl = {https://ui.adsabs.harvard.edu/abs/2022ApJ...935..167A}
}

@ARTICLE{matplotlib,
       author = {{Hunter}, John D.},
        title = "{Matplotlib: A 2D Graphics Environment}",
      journal = {Computing in Science and Engineering},
         year = 2007,
        month = may,
       volume = {9},
       number = {3},
        pages = {90-95},
          doi = {10.1109/MCSE.2007.55},
       adsurl = {https://ui.adsabs.harvard.edu/abs/2007CSE.....9...90H}
}

@ARTICLE{numpy,
       author = {{van der Walt}, St{\'e}fan and {Colbert}, S. Chris and {Varoquaux}, Ga{\"e}l},
        title = "{The NumPy Array: A Structure for Efficient Numerical Computation}",
      journal = {Computing in Science and Engineering},
         year = 2011,
        month = mar,
       volume = {13},
       number = {2},
        pages = {22-30},
          doi = {10.1109/MCSE.2011.37},
archivePrefix = {arXiv},
       eprint = {1102.1523},
 primaryClass = {cs.MS},
       adsurl = {https://ui.adsabs.harvard.edu/abs/2011CSE....13b..22V}
}

@ARTICLE{baggen23,
       author = {{Baggen}, Josephine F.~W. and {van Dokkum}, Pieter and {Labbe}, Ivo and {Brammer}, Gabriel and {Miller}, Tim B. and {Bezanson}, Rachel and {Leja}, Joel and {Wang}, Bingjie and {Whitaker}, Katherine E. and {Suess}, Katherine A. and {Nelson}, Erica J.},
        title = "{Sizes and mass profiles of candidate massive galaxies discovered by JWST at 7<z<9: evidence for very early formation of the central \raisebox{-0.5ex}\textasciitilde100 pc of present-day ellipticals}",
      journal = {arXiv e-prints},
         year = 2023,
        month = may,
          eid = {arXiv:2305.17162},
        pages = {arXiv:2305.17162},
          doi = {10.48550/arXiv.2305.17162},
archivePrefix = {arXiv},
       eprint = {2305.17162},
 primaryClass = {astro-ph.GA},
       adsurl = {https://ui.adsabs.harvard.edu/abs/2023arXiv230517162B}
}

@ARTICLE{kokorev24,
       author = {{Kokorev}, Vasily and {Caputi}, Karina I. and {Greene}, Jenny E. and {Dayal}, Pratika and {Trebitsch}, Maxime and {Cutler}, Sam E. and {Fujimoto}, Seiji and {Labb{\'e}}, Ivo and {Miller}, Tim B. and {Iani}, Edoardo and {Navarro-Carrera}, Rafael and {Rinaldi}, Pierluigi},
        title = "{A Census of Photometrically Selected Little Red Dots at 4 < z < 9 in JWST Blank Fields}",
      journal = {arXiv e-prints},
         year = 2024,
        month = jan,
          eid = {arXiv:2401.09981},
        pages = {arXiv:2401.09981},
          doi = {10.48550/arXiv.2401.09981},
archivePrefix = {arXiv},
       eprint = {2401.09981},
 primaryClass = {astro-ph.GA},
       adsurl = {https://ui.adsabs.harvard.edu/abs/2024arXiv240109981K}
}

@ARTICLE{pg24a,
       author = {{P{\'e}rez-Gonz{\'a}lez}, Pablo G. and {Barro}, Guillermo and {Rieke}, George H. and {Lyu}, Jianwei and {Rieke}, Marcia and {Alberts}, Stacey and {Williams}, Christina and {Hainline}, Kevin and {Sun}, Fengwu and {Puskas}, David and {Annunziatella}, Marianna and {Baker}, William M. and {Bunker}, Andrew J. and {Egami}, Eiichi and {Ji}, Zhiyuan and {Johnson}, Benjamin D. and {Robertson}, Brant and {Rodriguez Del Pino}, Bruno and {Rujopakarn}, Wiphu and {Shivaei}, Irene and {Tacchella}, Sandro and {Willmer}, Christopher N.~A. and {Willott}, Chris},
        title = "{What is the nature of Little Red Dots and what is not, MIRI SMILES edition}",
      journal = {arXiv e-prints},
         year = 2024,
        month = jan,
          eid = {arXiv:2401.08782},
        pages = {arXiv:2401.08782},
          doi = {10.48550/arXiv.2401.08782},
archivePrefix = {arXiv},
       eprint = {2401.08782},
 primaryClass = {astro-ph.GA},
       adsurl = {https://ui.adsabs.harvard.edu/abs/2024arXiv240108782P}
}

@ARTICLE{barro24,
       author = {{Barro}, Guillermo and {P{\'e}rez-Gonz{\'a}lez}, Pablo G. and {Kocevski}, Dale D. and {McGrath}, Elizabeth J. and {Trump}, Jonathan R. and {Simons}, Raymond C. and {Somerville}, Rachel S. and {Yung}, L.~Y. Aaron and {Arrabal Haro}, Pablo and {Akins}, Hollis B. and {Bagley}, Michaela B. and {Cleri}, Nikko J. and {Costantin}, Luca and {Davis}, Kelcey and {Dickinson}, Mark and {Finkelstein}, Steve L. and {Giavalisco}, Mauro and {G{\'o}mez-Guijarro}, Carlos and {Hathi}, Nimish P. and {Hirschmann}, Michaela and {Holwerda}, Benne W. and {Huertas-Company}, Marc and {Kartaltepe}, Jeyhan S. and {Koekemoer}, Anton M. and {Lucas}, Ray A. and {Papovich}, Casey and {Pirzkal}, Nor and {Seill{\'e}}, Lise-Marie and {Tacchella}, Sandro and {Wuyts}, Stijn and {Wilkins}, Stephen M. and {de la Vega}, Alexander and {Yang}, Guang and {Zavala}, Jorge A.},
        title = "{Extremely Red Galaxies at z = 5{\textendash}9 with MIRI and NIRSpec: Dusty Galaxies or Obscured Active Galactic Nuclei?}",
      journal = {\apj},
         year = 2024,
        month = mar,
       volume = {963},
       number = {2},
          eid = {128},
        pages = {128},
          doi = {10.3847/1538-4357/ad167e},
archivePrefix = {arXiv},
       eprint = {2305.14418},
 primaryClass = {astro-ph.GA},
       adsurl = {https://ui.adsabs.harvard.edu/abs/2024ApJ...963..128B}
}

@ARTICLE{greene23,
       author = {{Greene}, Jenny E. and {Labbe}, Ivo and {Goulding}, Andy D. and {Furtak}, Lukas J. and {Chemerynska}, Iryna and {Kokorev}, Vasily and {Dayal}, Pratika and {Volonteri}, Marta and {Williams}, Christina C. and {Wang}, Bingjie and {Setton}, David J. and {Burgasser}, Adam J. and {Bezanson}, Rachel and {Atek}, Hakim and {Brammer}, Gabriel and {Cutler}, Sam E. and {Feldmann}, Robert and {Fujimoto}, Seiji and {Glazebrook}, Karl and {de Graaff}, Anna and {Khullar}, Gourav and {Leja}, Joel and {Marchesini}, Danilo and {Maseda}, Michael V. and {Matthee}, Jorryt and {Miller}, Tim B. and {Naidu}, Rohan P. and {Nanayakkara}, Themiya and {Oesch}, Pascal A. and {Pan}, Richard and {Papovich}, Casey and {Price}, Sedona H. and {van Dokkum}, Pieter and {Weaver}, John R. and {Whitaker}, Katherine E. and {Zitrin}, Adi},
        title = "{UNCOVER Spectroscopy Confirms the Surprising Ubiquity of Active Galactic Nuclei in Red Sources at z > 5}",
      journal = {\apj},
         year = 2024,
        month = mar,
       volume = {964},
       number = {1},
          eid = {39},
        pages = {39},
          doi = {10.3847/1538-4357/ad1e5f},
archivePrefix = {arXiv},
       eprint = {2309.05714},
 primaryClass = {astro-ph.GA},
       adsurl = {https://ui.adsabs.harvard.edu/abs/2024ApJ...964...39G}
}

@ARTICLE{matthee24,
       author = {{Matthee}, Jorryt and {Naidu}, Rohan P. and {Brammer}, Gabriel and {Chisholm}, John and {Eilers}, Anna-Christina and {Goulding}, Andy and {Greene}, Jenny and {Kashino}, Daichi and {Labbe}, Ivo and {Lilly}, Simon J. and {Mackenzie}, Ruari and {Oesch}, Pascal A. and {Weibel}, Andrea and {Wuyts}, Stijn and {Xiao}, Mengyuan and {Bordoloi}, Rongmon and {Bouwens}, Rychard and {van Dokkum}, Pieter and {Illingworth}, Garth and {Kramarenko}, Ivan and {Maseda}, Michael V. and {Mason}, Charlotte and {Meyer}, Romain A. and {Nelson}, Erica J. and {Reddy}, Naveen A. and {Shivaei}, Irene and {Simcoe}, Robert A. and {Yue}, Minghao},
        title = "{Little Red Dots: An Abundant Population of Faint Active Galactic Nuclei at z {\ensuremath{\sim}} 5 Revealed by the EIGER and FRESCO JWST Surveys}",
      journal = {\apj},
         year = 2024,
        month = mar,
       volume = {963},
       number = {2},
          eid = {129},
        pages = {129},
          doi = {10.3847/1538-4357/ad2345},
       adsurl = {https://ui.adsabs.harvard.edu/abs/2024ApJ...963..129M}
}

@ARTICLE{wang24_lrd,
       author = {{Wang}, Bingjie and {de Graaff}, Anna and {Davies}, Rebecca L. and {Greene}, Jenny E. and {Leja}, Joel and {Goulding}, Andy D. and {Williams}, Christina C. and {Brammer}, Gabriel B. and {Suess}, Katherine A. and {Weibel}, Andrea and {Bezanson}, Rachel and {Boogaard}, Leindert A. and {Cleri}, Nikko J. and {Hirschmann}, Michaela and {Katz}, Harley and {Labbe}, Ivo and {Maseda}, Michael V. and {Matthee}, Jorryt and {McConachie}, Ian and {Naidu}, Rohan P. and {Oesch}, Pascal A. and {Rix}, Hans-Walter and {Setton}, David J. and {Whitaker}, Katherine E.},
        title = "{RUBIES: JWST/NIRSpec Confirmation of an Infrared-luminous, Broad-line Little Red Dot with an Ionized Outflow}",
      journal = {arXiv e-prints},
         year = 2024,
        month = mar,
          eid = {arXiv:2403.02304},
        pages = {arXiv:2403.02304},
          doi = {10.48550/arXiv.2403.02304},
archivePrefix = {arXiv},
       eprint = {2403.02304},
 primaryClass = {astro-ph.GA},
       adsurl = {https://ui.adsabs.harvard.edu/abs/2024arXiv240302304W}
}

@ARTICLE{williams24,
       author = {{Williams}, Christina C. and {Alberts}, Stacey and {Ji}, Zhiyuan and {Hainline}, Kevin N. and {Lyu}, Jianwei and {Rieke}, George and {Endsley}, Ryan and {Suess}, Katherine A. and {Sun}, Fengwu and {Johnson}, Benjamin D. and {Florian}, Michael and {Shivaei}, Irene and {Rujopakarn}, Wiphu and {Baker}, William M. and {Bhatawdekar}, Rachana and {Boyett}, Kristan and {Bunker}, Andrew J. and {Cameron}, Alex J. and {Carniani}, Stefano and {Charlot}, Stephane and {Curtis-Lake}, Emma and {DeCoursey}, Christa and {de Graaff}, Anna and {Egami}, Eiichi and {Eisenstein}, Daniel J. and {Gibson}, Justus L. and {Hausen}, Ryan and {Helton}, Jakob M. and {Maiolino}, Roberto and {Maseda}, Michael V. and {Nelson}, Erica J. and {P{\'e}rez-Gonz{\'a}lez}, Pablo G. and {Rieke}, Marcia J. and {Robertson}, Brant E. and {Saxena}, Aayush and {Tacchella}, Sandro and {Willmer}, Christopher N.~A. and {Willott}, Chris J.},
        title = "{The Galaxies Missed by Hubble and ALMA: The Contribution of Extremely Red Galaxies to the Cosmic Census at 3 < z < 8}",
      journal = {\apj},
         year = 2024,
        month = jun,
       volume = {968},
       number = {1},
          eid = {34},
        pages = {34},
          doi = {10.3847/1538-4357/ad3f17},
archivePrefix = {arXiv},
       eprint = {2311.07483},
 primaryClass = {astro-ph.GA},
       adsurl = {https://ui.adsabs.harvard.edu/abs/2024ApJ...968...34W}
}

@ARTICLE{kocevski24,
       author = {{Kocevski}, Dale D. and {Finkelstein}, Steven L. and {Barro}, Guillermo and {Taylor}, Anthony J. and {Calabr{\`o}}, Antonello and {Laloux}, Brivael and {Buchner}, Johannes and {Trump}, Jonathan R. and {Leung}, Gene C.~K. and {Yang}, Guang and {Dickinson}, Mark and {P{\'e}rez-Gonz{\'a}lez}, Pablo G. and {Pacucci}, Fabio and {Inayoshi}, Kohei and {Somerville}, Rachel S. and {McGrath}, Elizabeth J. and {Akins}, Hollis B. and {Bagley}, Micaela B. and {Bisigello}, Laura and {Bowler}, Rebecca A.~A. and {Carnall}, Adam and {Casey}, Caitlin M. and {Cheng}, Yingjie and {Cleri}, Nikko J. and {Costantin}, Luca and {Cullen}, Fergus and {Davis}, Kelcey and {Donnan}, Callum T. and {Dunlop}, James S. and {Ellis}, Richard S. and {Ferguson}, Henry C. and {Fujimoto}, Seiji and {Fontana}, Adriano and {Giavalisco}, Mauro and {Grazian}, Andrea and {Grogin}, Norman A. and {Hathi}, Nimish P. and {Hirschmann}, Michaela and {Huertas-Company}, Marc and {Holwerda}, Benne W. and {Illingworth}, Garth and {Juneau}, St{\'e}phanie and {Kartaltepe}, Jeyhan S. and {Koekemoer}, Anton M. and {Li}, Wenxiu and {Lucas}, Ray A. and {Magee}, Dan and {Mason}, Charlotte and {McLeod}, Derek J. and {McLure}, Ross J. and {Napolitano}, Lorenzo and {Papovich}, Casey and {Pirzkal}, Nor and {Rodighiero}, Giulia and {Santini}, Paola and {Wilkins}, Stephen M. and {Yung}, L.~Y. Aaron},
        title = "{The Rise of Faint, Red AGN at $z>4$: A Sample of Little Red Dots in the JWST Extragalactic Legacy Fields}",
      journal = {arXiv e-prints},
         year = 2024,
        month = apr,
          eid = {arXiv:2404.03576},
        pages = {arXiv:2404.03576},
          doi = {10.48550/arXiv.2404.03576},
archivePrefix = {arXiv},
       eprint = {2404.03576},
 primaryClass = {astro-ph.GA},
       adsurl = {https://ui.adsabs.harvard.edu/abs/2024arXiv240403576K}
}

@ARTICLE{akins24,
       author = {{Akins}, Hollis B. and {Casey}, Caitlin M. and {Lambrides}, Erini and {Allen}, Natalie and {Andika}, Irham T. and {Brinch}, Malte and {Champagne}, Jaclyn B. and {Cooper}, Olivia and {Ding}, Xuheng and {Drakos}, Nicole E. and {Faisst}, Andreas and {Finkelstein}, Steven L. and {Franco}, Maximilien and {Fujimoto}, Seiji and {Gentile}, Fabrizio and {Gillman}, Steven and {Gozaliasl}, Ghassem and {Harish}, Santosh and {Hayward}, Christopher C. and {Hirschmann}, Michaela and {Ilbert}, Olivier and {Kartaltepe}, Jeyhan S. and {Kocevski}, Dale D. and {Koekemoer}, Anton M. and {Kokorev}, Vasily and {Liu}, Daizhong and {Long}, Arianna S. and {McCracken}, Henry Joy and {McKinney}, Jed and {Onoue}, Masafusa and {Paquereau}, Louise and {Renzini}, Alvio and {Rhodes}, Jason and {Robertson}, Brant E. and {Shuntov}, Marko and {Silverman}, John D. and {Tanaka}, Takumi S. and {Toft}, Sune and {Trakhtenbrot}, Benny and {Valentino}, Francesco and {Zavala}, Jorge},
        title = "{COSMOS-Web: The over-abundance and physical nature of ``little red dots''--Implications for early galaxy and SMBH assembly}",
      journal = {arXiv e-prints},
         year = 2024,
        month = jun,
          eid = {arXiv:2406.10341},
        pages = {arXiv:2406.10341},
          doi = {10.48550/arXiv.2406.10341},
archivePrefix = {arXiv},
       eprint = {2406.10341},
 primaryClass = {astro-ph.GA},
       adsurl = {https://ui.adsabs.harvard.edu/abs/2024arXiv240610341A}
}

@ARTICLE{leung24,
       author = {{Leung}, Gene C.~K. and {Finkelstein}, Steven L. and {P{\'e}rez-Gonz{\'a}lez}, Pablo G. and {Morales}, Alexa M. and {Taylor}, Anthony J. and {Barro}, Guillermo and {Kocevski}, Dale D. and {Akins}, Hollis B. and {Carnall}, Adam C. and {Ch{\'a}vez Ortiz}, {\'O}scar A. and {Cleri}, Nikko J. and {Cullen}, Fergus and {Donnan}, Callum T. and {Dunlop}, James S. and {Ellis}, Richard S. and {Grogin}, Norman A. and {Hirschmann}, Michaela and {Koekemoer}, Anton M. and {Kokorev}, Vasily and {Lucas}, Ray A. and {McLeod}, Derek J. and {Papovich}, Casey and {Yung}, L.~Y. Aaron},
        title = "{Exploring the Nature of Little Red Dots: Constraints on AGN and Stellar Contributions from PRIMER MIRI Imaging}",
      journal = {arXiv e-prints},
         year = 2024,
        month = nov,
          eid = {arXiv:2411.12005},
        pages = {arXiv:2411.12005},
archivePrefix = {arXiv},
       eprint = {2411.12005},
 primaryClass = {astro-ph.GA},
       adsurl = {https://ui.adsabs.harvard.edu/abs/2024arXiv241112005L}
}

@ARTICLE{wang24b,
       author = {{Wang}, Bingjie and {Leja}, Joel and {de Graaff}, Anna and {Brammer}, Gabriel B. and {Weibel}, Andrea and {van Dokkum}, Pieter and {Baggen}, Josephine F.~W. and {Suess}, Katherine A. and {Greene}, Jenny E. and {Bezanson}, Rachel and {Cleri}, Nikko J. and {Hirschmann}, Michaela and {Labb{\'e}}, Ivo and {Matthee}, Jorryt and {McConachie}, Ian and {Naidu}, Rohan P. and {Nelson}, Erica and {Oesch}, Pascal A. and {Setton}, David J. and {Williams}, Christina C.},
        title = "{RUBIES: Evolved Stellar Populations with Extended Formation Histories at z {\ensuremath{\sim}} 7{\textendash}8 in Candidate Massive Galaxies Identified with JWST/NIRSpec}",
      journal = {\apjl},
         year = 2024,
        month = jul,
       volume = {969},
       number = {1},
          eid = {L13},
        pages = {L13},
          doi = {10.3847/2041-8213/ad55f7},
archivePrefix = {arXiv},
       eprint = {2405.01473},
 primaryClass = {astro-ph.GA},
       adsurl = {https://ui.adsabs.harvard.edu/abs/2024ApJ...969L..13W}
}

@ARTICLE{taylor24,
       author = {{Taylor}, Anthony J. and {Finkelstein}, Steven L. and {Kocevski}, Dale D. and {Jeon}, Junehyoung and {Bromm}, Volker and {Amorin}, Ricardo O. and {Arrabal Haro}, Pablo and {Backhaus}, Bren E. and {Bagley}, Micaela B. and {Ba{\~n}ados}, Eduardo and {Bhatawdekar}, Rachana and {Brooks}, Madisyn and {Calabro}, Antonello and {Chavez Ortiz}, Oscar A. and {Cheng}, Yingjie and {Cleri}, Nikko J. and {Cole}, Justin W. and {Davis}, Kelcey and {Dickinson}, Mark and {Donnan}, Callum and {Dunlop}, James S. and {Ellis}, Richard S. and {Fernandez}, Vital and {Fontana}, Adriano and {Fujimoto}, Seiji and {Giavalisco}, Mauro and {Grazian}, Andrea and {Guo}, Jingsong and {Hathi}, Nimish P. and {Holwerda}, Benne W. and {Hirschmann}, Michaela and {Inayoshi}, Kohei and {Kartaltepe}, Jeyhan S. and {Khusanova}, Yana and {Koekemoer}, Anton M. and {Kokorev}, Vasily and {Larson}, Rebecca L. and {Leung}, Gene C.~K. and {Lucas}, Ray A. and {McLeod}, Derek J. and {Napolitano}, Lorenzo and {Onoue}, Masafusa and {Pacucci}, Fabio and {Papovich}, Casey and {P{\'e}rez-Gonz{\'a}lez}, Pablo G. and {Pirzkal}, Nor and {Somerville}, Rachel S. and {Trump}, Jonathan R. and {Wilkins}, Stephen M. and {Yung}, L.~Y. Aaron and {Zhang}, Haowen},
        title = "{Broad-Line AGN at $3.5<z<6$: The Black Hole Mass Function and a Connection with Little Red Dots}",
      journal = {arXiv e-prints},
         year = 2024,
        month = sep,
          eid = {arXiv:2409.06772},
        pages = {arXiv:2409.06772},
          doi = {10.48550/arXiv.2409.06772},
archivePrefix = {arXiv},
       eprint = {2409.06772},
 primaryClass = {astro-ph.GA},
       adsurl = {https://ui.adsabs.harvard.edu/abs/2024arXiv240906772T}
}

@ARTICLE{inayoshi24,
       author = {{Inayoshi}, Kohei and {Maiolino}, Roberto},
        title = "{Extremely Dense Gas around Little Red Dots and High-redshift AGNs: A Non-stellar Origin of the Balmer Break and Absorption Features}",
      journal = {arXiv e-prints},
         year = 2024,
        month = sep,
          eid = {arXiv:2409.07805},
        pages = {arXiv:2409.07805},
          doi = {10.48550/arXiv.2409.07805},
archivePrefix = {arXiv},
       eprint = {2409.07805},
 primaryClass = {astro-ph.GA},
       adsurl = {https://ui.adsabs.harvard.edu/abs/2024arXiv240907805I}
}

@ARTICLE{kokorev24b,
       author = {{Kokorev}, Vasily and {Chisholm}, John and {Endsley}, Ryan and {Finkelstein}, Steven L. and {Greene}, Jenny E. and {Akins}, Hollis B. and {Bromm}, Volker and {Casey}, Caitlin M. and {Fujimoto}, Seiji and {Labb{\'e}}, Ivo and {Larson}, Rebecca L.},
        title = "{Silencing the Giant: Evidence of AGN Feedback and Quenching in a Little Red Dot at z = 4.13}",
      journal = {arXiv e-prints},
         year = 2024,
        month = jul,
          eid = {arXiv:2407.20320},
        pages = {arXiv:2407.20320},
          doi = {10.48550/arXiv.2407.20320},
archivePrefix = {arXiv},
       eprint = {2407.20320},
 primaryClass = {astro-ph.GA},
       adsurl = {https://ui.adsabs.harvard.edu/abs/2024arXiv240720320K}
}

@ARTICLE{casey24,
       author = {{Casey}, Caitlin M. and {Akins}, Hollis B. and {Kokorev}, Vasily and {McKinney}, Jed and {Cooper}, Olivia R. and {Long}, Arianna S. and {Franco}, Maximilien and {Manning}, Sinclaire M.},
        title = "{Dust in Little Red Dots}",
      journal = {arXiv e-prints},
         year = 2024,
        month = jul,
          eid = {arXiv:2407.05094},
        pages = {arXiv:2407.05094},
          doi = {10.48550/arXiv.2407.05094},
archivePrefix = {arXiv},
       eprint = {2407.05094},
 primaryClass = {astro-ph.GA},
       adsurl = {https://ui.adsabs.harvard.edu/abs/2024arXiv240705094C}
}

@ARTICLE{akins24b,
       author = {{Akins}, Hollis B. and {Casey}, Caitlin M. and {Berg}, Danielle A. and {Chisholm}, John and {Franco}, Maximilien and {Finkelstein}, Steven L. and {Fujimoto}, Seiji and {Kokorev}, Vasily and {Lambrides}, Erini and {Robertson}, Brant E. and {Taylor}, Anthony J. and {Coulter}, David A. and {Fox}, Ori and {Karmen}, Mitchell},
        title = "{Strong rest-UV emission lines in a ``little red dot'' AGN at $z=7$: Early SMBH growth alongside compact massive star formation?}",
      journal = {arXiv e-prints},
         year = 2024,
        month = oct,
          eid = {arXiv:2410.00949},
        pages = {arXiv:2410.00949},
          doi = {10.48550/arXiv.2410.00949},
archivePrefix = {arXiv},
       eprint = {2410.00949},
 primaryClass = {astro-ph.GA},
       adsurl = {https://ui.adsabs.harvard.edu/abs/2024arXiv241000949A}
}

@ARTICLE{killi23,
       author = {{Killi}, Meghana and {Watson}, Darach and {Brammer}, Gabriel and {McPartland}, Conor and {Antwi-Danso}, Jacqueline and {Newshore}, Rosa and {Coe}, Dan and {Allen}, Natalie and {Fynbo}, Johan P.~U. and {Gould}, Katriona and {Heintz}, Kasper E. and {Rusakov}, Vadim and {Vejlgaard}, Simone},
        title = "{Deciphering the JWST spectrum of a 'little red dot' at $z \sim 4.53$: An obscured AGN and its star-forming host}",
      journal = {arXiv e-prints},
         year = 2023,
        month = dec,
          eid = {arXiv:2312.03065},
        pages = {arXiv:2312.03065},
          doi = {10.48550/arXiv.2312.03065},
archivePrefix = {arXiv},
       eprint = {2312.03065},
 primaryClass = {astro-ph.GA},
       adsurl = {https://ui.adsabs.harvard.edu/abs/2023arXiv231203065K}
}

@software{brammer_grizli,
       author = {{Brammer}, Gabe},
        title = "{Grizli: Grism redshift and line analysis software}",
 howpublished = {Astrophysics Source Code Library, record ascl:1905.001},
         year = 2019,
        month = may,
          eid = {ascl:1905.001},
       adsurl = {https://ui.adsabs.harvard.edu/abs/2019ascl.soft05001B}
}

@software{SEP,
       author = {{Barbary}, Kyle and {Boone}, Kyle and {Deil}, Christoph},
        title = "{sep: v0.3.0}",
         year = 2015,
        month = feb,
          eid = {10.5281/zenodo.15669},
          doi = {10.5281/zenodo.15669},
      version = {v0.3.0},
    publisher = {Zenodo},
       adsurl = {https://ui.adsabs.harvard.edu/abs/2015zndo.....15669B}
}

@ARTICLE{eisenstein24,
       author = {{Eisenstein}, Daniel J. and {Johnson}, Benjamin D. and {Robertson}, Brant and {Tacchella}, Sandro and {Hainline}, Kevin and {Jakobsen}, Peter and {Maiolino}, Roberto and {Bonaventura}, Nina and {Bunker}, Andrew J. and {Cameron}, Alex J. and {Cargile}, Phillip A. and {Curtis-Lake}, Emma and {Hausen}, Ryan and {Pusk{\'a}s}, D{\'a}vid and {Rieke}, Marcia and {Sun}, Fengwu and {Willmer}, Christopher N.~A. and {Willott}, Chris and {Alberts}, Stacey and {Arribas}, Santiago and {Baker}, William M. and {Baum}, Stefi and {Bhatawdekar}, Rachana and {Carniani}, Stefano and {Charlot}, Stephane and {Chen}, Zuyi and {Chevallard}, Jacopo and {Curti}, Mirko and {DeCoursey}, Christa and {D'Eugenio}, Francesco and {de Graaff}, Anna and {Egami}, Eiichi and {Helton}, Jakob M. and {Ji}, Zhiyuan and {Jones}, Gareth C. and {Kumari}, Nimisha and {L{\"u}tzgendorf}, Nora and {Laseter}, Isaac and {Looser}, Tobias J. and {Lyu}, Jianwei and {Maseda}, Michael V. and {Nelson}, Erica and {Parlanti}, Eleonora and {Rauscher}, Bernard J. and {Rawle}, Tim and {Rieke}, George and {Rix}, Hans-Walter and {Rujopakarn}, Wiphu and {Sandles}, Lester and {Saxena}, Aayush and {Scholtz}, Jan and {Sharpe}, Katherine and {Shivaei}, Irene and {Simmonds}, Charlotte and {Smit}, Renske and {Topping}, Michael W. and {{\"U}bler}, Hannah and {Venturi}, Giacomo and {Williams}, Christina C. and {Witstok}, Joris and {Woodrum}, Charity},
        title = "{The JADES Origins Field: A New JWST Deep Field in the JADES Second NIRCam Data Release}",
      journal = {arXiv e-prints},
         year = 2023,
        month = oct,
          eid = {arXiv:2310.12340},
        pages = {arXiv:2310.12340},
          doi = {10.48550/arXiv.2310.12340},
archivePrefix = {arXiv},
       eprint = {2310.12340},
 primaryClass = {astro-ph.GA},
       adsurl = {https://ui.adsabs.harvard.edu/abs/2023arXiv231012340E}
}

@ARTICLE{valentino23,
       author = {{Valentino}, Francesco and {Brammer}, Gabriel and {Gould}, Katriona M.~L. and {Kokorev}, Vasily and {Fujimoto}, Seiji and {Jespersen}, Christian Kragh and {Vijayan}, Aswin P. and {Weaver}, John R. and {Ito}, Kei and {Tanaka}, Masayuki and {Ilbert}, Olivier and {Magdis}, Georgios E. and {Whitaker}, Katherine E. and {Faisst}, Andreas L. and {Gallazzi}, Anna and {Gillman}, Steven and {Gim{\'e}nez-Arteaga}, Clara and {G{\'o}mez-Guijarro}, Carlos and {Kubo}, Mariko and {Heintz}, Kasper E. and {Hirschmann}, Michaela and {Oesch}, Pascal and {Onodera}, Masato and {Rizzo}, Francesca and {Lee}, Minju and {Strait}, Victoria and {Toft}, Sune},
        title = "{An Atlas of Color-selected Quiescent Galaxies at z > 3 in Public JWST Fields}",
      journal = {\apj},
         year = 2023,
        month = apr,
       volume = {947},
       number = {1},
          eid = {20},
        pages = {20},
          doi = {10.3847/1538-4357/acbefa},
archivePrefix = {arXiv},
       eprint = {2302.10936},
 primaryClass = {astro-ph.GA},
       adsurl = {https://ui.adsabs.harvard.edu/abs/2023ApJ...947...20V}
}

@ARTICLE{lyu17,
       author = {{Lyu}, Jianwei and {Rieke}, G.~H. and {Shi}, Yong},
        title = "{Dust-deficient Palomar-Green Quasars and the Diversity of AGN Intrinsic IR Emission}",
      journal = {\apj},
         year = 2017,
        month = feb,
       volume = {835},
       number = {2},
          eid = {257},
        pages = {257},
          doi = {10.3847/1538-4357/835/2/257},
archivePrefix = {arXiv},
       eprint = {1612.06857},
 primaryClass = {astro-ph.GA},
       adsurl = {https://ui.adsabs.harvard.edu/abs/2017ApJ...835..257L}
}

@ARTICLE{lyu24,
       author = {{Lyu}, Jianwei and {Alberts}, Stacey and {Rieke}, George H. and {Shivaei}, Irene and {P{\'e}rez-Gonz{\'a}lez}, Pablo G. and {Sun}, Fengwu and {Hainline}, Kevin N. and {Baum}, Stefi and {Bonaventura}, Nina and {Bunker}, Andrew J. and {Egami}, Eiichi and {Eisenstein}, Daniel J. and {Florian}, Michael and {Ji}, Zhiyuan and {Johnson}, Benjamin D. and {Morrison}, Jane and {Rieke}, Marcia and {Robertson}, Brant and {Rujopakarn}, Wiphu and {Tacchella}, Sandro and {Scholtz}, Jan and {Willmer}, Christopher N.~A.},
        title = "{Active Galactic Nuclei Selection and Demographics: A New Age with JWST/MIRI}",
      journal = {\apj},
         year = 2024,
        month = may,
       volume = {966},
       number = {2},
          eid = {229},
        pages = {229},
          doi = {10.3847/1538-4357/ad3643},
archivePrefix = {arXiv},
       eprint = {2310.12330},
 primaryClass = {astro-ph.GA},
       adsurl = {https://ui.adsabs.harvard.edu/abs/2024ApJ...966..229L}
}

@ARTICLE{maiolino23,
       author = {{Maiolino}, Roberto and {Scholtz}, Jan and {Curtis-Lake}, Emma and {Carniani}, Stefano and {Baker}, William and {de Graaff}, Anna and {Tacchella}, Sandro and {{\"U}bler}, Hannah and {D'Eugenio}, Francesco and {Witstok}, Joris and {Curti}, Mirko and {Arribas}, Santiago and {Bunker}, Andrew J. and {Charlot}, St{\'e}phane and {Chevallard}, Jacopo and {Eisenstein}, Daniel J. and {Egami}, Eiichi and {Ji}, Zhiyuan and {Jones}, Gareth C. and {Lyu}, Jianwei and {Rawle}, Tim and {Robertson}, Brant and {Rujopakarn}, Wiphu and {Perna}, Michele and {Sun}, Fengwu and {Venturi}, Giacomo and {Williams}, Christina C. and {Willott}, Chris},
        title = "{JADES. The diverse population of infant Black Holes at 4<z<11: merging, tiny, poor, but mighty}",
      journal = {arXiv e-prints},
         year = 2023,
        month = aug,
          eid = {arXiv:2308.01230},
        pages = {arXiv:2308.01230},
          doi = {10.48550/arXiv.2308.01230},
archivePrefix = {arXiv},
       eprint = {2308.01230},
 primaryClass = {astro-ph.GA},
       adsurl = {https://ui.adsabs.harvard.edu/abs/2023arXiv230801230M}
}

@ARTICLE{hainline24,
       author = {{Hainline}, Kevin N. and {Maiolino}, Roberto and {Juodzbalis}, Ignas and {Scholtz}, Jan and {Ubler}, Hannah and {D'Eugenio}, Francesco and {Helton}, Jakob M. and {Sun}, Yang and {Sun}, Fengwu and {Robertson}, Brant and {Tacchella}, Sandro and {Bunker}, Andrew J. and {Carniani}, Stefano and {Charlot}, Stephane and {Curtis-Lake}, Emma and {Egami}, Eiichi and {Johnson}, Benjamin D. and {Lin}, Xiaojing and {Lyu}, Jianwei and {Perez-Gonzalez}, Pablo G. and {Rinaldi}, Pierluigi and {Silcock}, Maddie S. and {Williams}, Christina C. and {Willmer}, Christopher N.~A. and {Willott}, Chris and {Zhang}, Junyu and {Zhu}, Yongda},
        title = "{An Investigation Into The Selection and Colors of Little Red Dots and Active Galactic Nuclei}",
      journal = {arXiv e-prints},
         year = 2024,
        month = sep,
          eid = {arXiv:2410.00100},
        pages = {arXiv:2410.00100},
          doi = {10.48550/arXiv.2410.00100},
archivePrefix = {arXiv},
       eprint = {2410.00100},
 primaryClass = {astro-ph.GA},
       adsurl = {https://ui.adsabs.harvard.edu/abs/2024arXiv241000100H}
}

@ARTICLE{degraaff24,
       author = {{de Graaff}, Anna and {Brammer}, Gabriel and {Weibel}, Andrea and {Lewis}, Zach and {Maseda}, Michael V. and {Oesch}, Pascal A. and {Bezanson}, Rachel and {Boogaard}, Leindert A. and {Cleri}, Nikko J. and {Cooper}, Olivia R. and {Gottumukkala}, Rashmi and {Greene}, Jenny E. and {Hirschmann}, Michaela and {Hviding}, Raphael E. and {Katz}, Harley and {Labb{\'e}}, Ivo and {Leja}, Joel and {Matthee}, Jorryt and {McConachie}, Ian and {Miller}, Tim B. and {Naidu}, Rohan P. and {Price}, Sedona H. and {Rix}, Hans-Walter and {Setton}, David J. and {Suess}, Katherine A. and {Wang}, Bingjie and {Whitaker}, Katherine E. and {Williams}, Christina C.},
        title = "{RUBIES: a complete census of the bright and red distant Universe with JWST/NIRSpec}",
      journal = {arXiv e-prints},
         year = 2024,
        month = sep,
          eid = {arXiv:2409.05948},
        pages = {arXiv:2409.05948},
          doi = {10.48550/arXiv.2409.05948},
archivePrefix = {arXiv},
       eprint = {2409.05948},
 primaryClass = {astro-ph.GA},
       adsurl = {https://ui.adsabs.harvard.edu/abs/2024arXiv240905948D}
}

@ARTICLE{williams23,
       author = {{Williams}, Christina C. and {Tacchella}, Sandro and {Maseda}, Michael V. and {Robertson}, Brant E. and {Johnson}, Benjamin D. and {Willott}, Chris J. and {Eisenstein}, Daniel J. and {Willmer}, Christopher N.~A. and {Ji}, Zhiyuan and {Hainline}, Kevin N. and {Helton}, Jakob M. and {Alberts}, Stacey and {Baum}, Stefi and {Bhatawdekar}, Rachana and {Boyett}, Kristan and {Bunker}, Andrew J. and {Carniani}, Stefano and {Charlot}, Stephane and {Chevallard}, Jacopo and {Curtis-Lake}, Emma and {de Graaff}, Anna and {Egami}, Eiichi and {Franx}, Marijn and {Kumari}, Nimisha and {Maiolino}, Roberto and {Nelson}, Erica J. and {Rieke}, Marcia J. and {Sandles}, Lester and {Shivaei}, Irene and {Simmonds}, Charlotte and {Smit}, Renske and {Suess}, Katherine A. and {Sun}, Fengwu and {{\"U}bler}, Hannah and {Witstok}, Joris},
        title = "{JEMS: A Deep Medium-band Imaging Survey in the Hubble Ultra Deep Field with JWST NIRCam and NIRISS}",
      journal = {\apjs},
         year = 2023,
        month = oct,
       volume = {268},
       number = {2},
          eid = {64},
        pages = {64},
          doi = {10.3847/1538-4365/acf130},
archivePrefix = {arXiv},
       eprint = {2301.09780},
 primaryClass = {astro-ph.GA},
       adsurl = {https://ui.adsabs.harvard.edu/abs/2023ApJS..268...64W}
}

@ARTICLE{setton24,
       author = {{Setton}, David J. and {Greene}, Jenny E. and {de Graaff}, Anna and {Ma}, Yilun and {Leja}, Joel and {Matthee}, Jorryt and {Bezanson}, Rachel and {Boogaard}, Leindert A. and {Cleri}, Nikko J. and {Katz}, Harley and {Labbe}, Ivo and {Maseda}, Michael V. and {McConachie}, Ian and {Miller}, Tim B. and {Price}, Sedona H. and {Suess}, Katherine A. and {van Dokkum}, Pieter and {Wang}, Bingjie and {Weibel}, Andrea and {Whitaker}, Katherine E. and {Williams}, Christina C.},
        title = "{Little Red Dots at an Inflection Point: Ubiquitous ``V-Shaped'' Turnover Consistently Occurs at the Balmer Limit}",
      journal = {arXiv e-prints},
         year = 2024,
        month = nov,
          eid = {arXiv:2411.03424},
        pages = {arXiv:2411.03424},
          doi = {10.48550/arXiv.2411.03424},
archivePrefix = {arXiv},
       eprint = {2411.03424},
 primaryClass = {astro-ph.GA},
       adsurl = {https://ui.adsabs.harvard.edu/abs/2024arXiv241103424S}
}

@ARTICLE{setton25,
       author = {{Setton}, David J. and {Greene}, Jenny E. and {Spilker}, Justin S. and {Williams}, Christina C. and {Labb{\'e}}, Ivo and {Ma}, Yilun and {Wang}, Bingjie and {Whitaker}, Katherine E. and {Leja}, Joel and {de Graaff}, Anna and {Alberts}, Stacey and {Bezanson}, Rachel and {Boogaard}, Leindert A. and {Brammer}, Gabriel and {Cutler}, Sam E. and {Cleri}, Nikko J. and {Cooper}, Olivia R. and {Dayal}, Pratika and {Fujimoto}, Seiji and {Furtak}, Lukas J. and {Goulding}, Andy D. and {Hirschmann}, Michaela and {Kokorev}, Vasily and {Maseda}, Michael V. and {McConachie}, Ian and {Matthee}, Jorryt and {Miller}, Tim B. and {Naidu}, Rohan P. and {Oesch}, Pascal A. and {Pan}, Richard and {Price}, Sedona H. and {Suess}, Katherine A. and {Weaver}, John R. and {Xiao}, Mengyuan and {Zhang}, Yunchong and {Zitrin}, Adi},
        title = "{A Confirmed Deficit of Hot and Cold Dust Emission in the Most Luminous Little Red Dots}",
      journal = {\apjl},
         year = 2025,
        month = sep,
       volume = {991},
       number = {1},
          eid = {L10},
        pages = {L10},
          doi = {10.3847/2041-8213/ade78b},
archivePrefix = {arXiv},
       eprint = {2503.02059},
 primaryClass = {astro-ph.GA},
       adsurl = {https://ui.adsabs.harvard.edu/abs/2025ApJ...991L..10S}
}

@ARTICLE{labbe24b,
       author = {{Labbe}, Ivo and {Greene}, Jenny E. and {Matthee}, Jorryt and {Treiber}, Helena and {Kokorev}, Vasily and {Miller}, Tim B. and {Kramarenko}, Ivan and {Setton}, David J. and {Ma}, Yilun and {Goulding}, Andy D. and {Bezanson}, Rachel and {Naidu}, Rohan P. and {Williams}, Christina C. and {Atek}, Hakim and {Brammer}, Gabriel and {Cutler}, Sam E. and {Chemerynska}, Iryna and {Cloonan}, Aidan P. and {Dayal}, Pratika and {de Graaff}, Anna and {Fudamoto}, Yoshinobu and {Fujimoto}, Seiji and {Furtak}, Lukas J. and {Glazebrook}, Karl and {Heintz}, Kasper E. and {Leja}, Joel and {Marchesini}, Danilo and {Nanayakkara}, Themiya and {Nelson}, Erica J. and {Oesch}, Pascal A. and {Pan}, Richard and {Price}, Sedona H. and {Shivaei}, Irene and {Sobral}, David and {Suess}, Katherine A. and {van Dokkum}, Pieter and {Wang}, Bingjie and {Weaver}, John R. and {Whitaker}, Katherine E. and {Zitrin}, Adi},
        title = "{An unambiguous AGN and a Balmer break in an Ultraluminous Little Red Dot at z=4.47 from Ultradeep UNCOVER and All the Little Things Spectroscopy}",
      journal = {arXiv e-prints},
         year = 2024,
        month = dec,
          eid = {arXiv:2412.04557},
        pages = {arXiv:2412.04557},
          doi = {10.48550/arXiv.2412.04557},
archivePrefix = {arXiv},
       eprint = {2412.04557},
 primaryClass = {astro-ph.GA},
       adsurl = {https://ui.adsabs.harvard.edu/abs/2024arXiv241204557L}
}

@ARTICLE{hainline25,
       author = {{Hainline}, Kevin N. and {Maiolino}, Roberto and {Juod{\v{z}}balis}, Ignas and {Scholtz}, Jan and {{\"U}bler}, Hannah and {D'Eugenio}, Francesco and {Helton}, Jakob M. and {Sun}, Yang and {Sun}, Fengwu and {Robertson}, Brant and {Tacchella}, Sandro and {Bunker}, Andrew J. and {Carniani}, Stefano and {Charlot}, Stephane and {Curtis-Lake}, Emma and {Egami}, Eiichi and {Johnson}, Benjamin D. and {Lin}, Xiaojing and {Lyu}, Jianwei and {P{\'e}rez-Gonz{\'a}lez}, Pablo G. and {Rinaldi}, Pierluigi and {Silcock}, Maddie S. and {Venturi}, Giacomo and {Williams}, Christina C. and {Willmer}, Christopher N.~A. and {Willott}, Chris and {Zhang}, Junyu and {Zhu}, Yongda},
        title = "{An Investigation into the Selection and Colors of Little Red Dots and Active Galactic Nuclei}",
      journal = {\apj},
         year = 2025,
        month = feb,
       volume = {979},
       number = {2},
          eid = {138},
        pages = {138},
          doi = {10.3847/1538-4357/ad9920},
archivePrefix = {arXiv},
       eprint = {2410.00100},
 primaryClass = {astro-ph.GA},
       adsurl = {https://ui.adsabs.harvard.edu/abs/2025ApJ...979..138H}
}

@ARTICLE{weibel24,
       author = {{Weibel}, Andrea and {de Graaff}, Anna and {Setton}, David J. and {Miller}, Tim B. and {Oesch}, Pascal A. and {Brammer}, Gabriel and {Lagos}, Claudia D.~P. and {Whitaker}, Katherine E. and {Williams}, Christina C. and {Baggen}, Josephine F.~W. and {Bezanson}, Rachel and {Boogaard}, Leindert A. and {Cleri}, Nikko J. and {Greene}, Jenny E. and {Hirschmann}, Michaela and {Hviding}, Raphael E. and {Kuruvanthodi}, Adarsh and {Labb{\'e}}, Ivo and {Leja}, Joel and {Maseda}, Michael V. and {Matthee}, Jorryt and {McConachie}, Ian and {Naidu}, Rohan P. and {Roberts-Borsani}, Guido and {Schaerer}, Daniel and {Suess}, Katherine A. and {Valentino}, Francesco and {van Dokkum}, Pieter and {Wang}, Bingjie},
        title = "{RUBIES Reveals a Massive Quiescent Galaxy at z=7.3}",
      journal = {arXiv e-prints},
         year = 2024,
        month = sep,
          eid = {arXiv:2409.03829},
        pages = {arXiv:2409.03829},
          doi = {10.48550/arXiv.2409.03829},
archivePrefix = {arXiv},
       eprint = {2409.03829},
 primaryClass = {astro-ph.GA},
       adsurl = {https://ui.adsabs.harvard.edu/abs/2024arXiv240903829W}
}

@software{brammer23,
       author = {{Brammer}, Gabriel},
        title = "{msaexp: NIRSpec analyis tools}",
         year = 2023,
        month = sep,
          eid = {10.5281/zenodo.7299500},
          doi = {10.5281/zenodo.7299500},
      version = {0.6.17},
    publisher = {Zenodo},
       adsurl = {https://ui.adsabs.harvard.edu/abs/2022zndo...7299500B}
}

@ARTICLE{degraaff25,
       author = {{de Graaff}, Anna and {Rix}, Hans-Walter and {Naidu}, Rohan P. and {Labbe}, Ivo and {Wang}, Bingjie and {Leja}, Joel and {Matthee}, Jorryt and {Katz}, Harley and {Greene}, Jenny E. and {Hviding}, Raphael E. and {Baggen}, Josephine and {Bezanson}, Rachel and {Boogaard}, Leindert A. and {Brammer}, Gabriel and {Dayal}, Pratika and {van Dokkum}, Pieter and {Goulding}, Andy D. and {Hirschmann}, Michaela and {Maseda}, Michael V. and {McConachie}, Ian and {Miller}, Tim B. and {Nelson}, Erica and {Oesch}, Pascal A. and {Setton}, David J. and {Shivaei}, Irene and {Weibel}, Andrea and {Whitaker}, Katherine E. and {Williams}, Christina C.},
        title = "{A remarkable Ruby: Absorption in dense gas, rather than evolved stars, drives the extreme Balmer break of a Little Red Dot at $z=3.5$}",
      journal = {arXiv e-prints},
         year = 2025,
        month = mar,
          eid = {arXiv:2503.16600},
        pages = {arXiv:2503.16600},
          doi = {10.48550/arXiv.2503.16600},
archivePrefix = {arXiv},
       eprint = {2503.16600},
 primaryClass = {astro-ph.GA},
       adsurl = {https://ui.adsabs.harvard.edu/abs/2025arXiv250316600D}
}

@ARTICLE{naidu25,
       author = {{Naidu}, Rohan P. and {Matthee}, Jorryt and {Katz}, Harley and {de Graaff}, Anna and {Oesch}, Pascal and {Smith}, Aaron and {Greene}, Jenny E. and {Brammer}, Gabriel and {Weibel}, Andrea and {Hviding}, Raphael and {Chisholm}, John and {Labb\textbackslash'e}, Ivo and {Simcoe}, Robert A. and {Witten}, Callum and {Atek}, Hakim and {Baggen}, Josephine F.~W. and {Belli}, Sirio and {Bezanson}, Rachel and {Boogaard}, Leindert A. and {Bose}, Sownak and {Covelo-Paz}, Alba and {Dayal}, Pratika and {Fudamoto}, Yoshinobu and {Furtak}, Lukas J. and {Giovinazzo}, Emma and {Goulding}, Andy and {Gronke}, Max and {Heintz}, Kasper E. and {Hirschmann}, Michaela and {Illingworth}, Garth and {Inoue}, Akio K. and {Johnson}, Benjamin D. and {Leja}, Joel and {Leonova}, Ecaterina and {McConachie}, Ian and {Maseda}, Michael V. and {Natarajan}, Priyamvada and {Nelson}, Erica and {Setton}, David J. and {Shivaei}, Irene and {Sobral}, David and {Stefanon}, Mauro and {Tacchella}, Sandro and {Toft}, Sune and {Torralba}, Alberto and {van Dokkum}, Pieter and {van der Wel}, Arjen and {Volonteri}, Marta and {Walter}, Fabian and {Wang}, Bingjie and {Watson}, Darach},
        title = "{A ``Black Hole Star'' Reveals the Remarkable Gas-Enshrouded Hearts of the Little Red Dots}",
      journal = {arXiv e-prints},
         year = 2025,
        month = mar,
          eid = {arXiv:2503.16596},
        pages = {arXiv:2503.16596},
          doi = {10.48550/arXiv.2503.16596},
archivePrefix = {arXiv},
       eprint = {2503.16596},
 primaryClass = {astro-ph.GA},
       adsurl = {https://ui.adsabs.harvard.edu/abs/2025arXiv250316596N}
}

@ARTICLE{juodzbalis25,
       author = {{Juod{\v{z}}balis}, Ignas and {Maiolino}, Roberto and {Baker}, William M. and {Lake}, Emma Curtis and {Scholtz}, Jan and {D'Eugenio}, Francesco and {Trefoloni}, Bartolomeo and {Isobe}, Yuki and {Tacchella}, Sandro and {Bunker}, Andrew J. and {Carniani}, Stefano and {Charlot}, St{\'e}phane and {Jones}, Gareth C. and {Parlanti}, Eleonora and {Perna}, Michele and {Rinaldi}, Pierluigi and {Robertson}, Brant and {{\"U}bler}, Hannah and {Venturi}, Giacomo and {Willott}, Chris},
        title = "{JADES: comprehensive census of broad-line AGN from Reionization to Cosmic Noon revealed by JWST}",
      journal = {arXiv e-prints},
         year = 2025,
        month = apr,
          eid = {arXiv:2504.03551},
        pages = {arXiv:2504.03551},
          doi = {10.48550/arXiv.2504.03551},
archivePrefix = {arXiv},
       eprint = {2504.03551},
 primaryClass = {astro-ph.GA},
       adsurl = {https://ui.adsabs.harvard.edu/abs/2025arXiv250403551J}
}

@ARTICLE{juodzbalis24b,
       author = {{Juod{\v{z}}balis}, Ignas and {Maiolino}, Roberto and {Baker}, William M. and {Tacchella}, Sandro and {Scholtz}, Jan and {D'Eugenio}, Francesco and {Witstok}, Joris and {Schneider}, Raffaella and {Trinca}, Alessandro and {Valiante}, Rosa and {DeCoursey}, Christa and {Curti}, Mirko and {Carniani}, Stefano and {Chevallard}, Jacopo and {de Graaff}, Anna and {Arribas}, Santiago and {Bennett}, Jake S. and {Bourne}, Martin A. and {Bunker}, Andrew J. and {Charlot}, St{\'e}phane and {Jiang}, Brian and {Koudmani}, Sophie and {Perna}, Michele and {Robertson}, Brant and {Sijacki}, Debora and {{\"U}bler}, Hannah and {Williams}, Christina C. and {Willott}, Chris},
        title = "{A dormant overmassive black hole in the early Universe}",
      journal = {\nat},
         year = 2024,
        month = dec,
       volume = {636},
       number = {8043},
        pages = {594-597},
          doi = {10.1038/s41586-024-08210-5},
archivePrefix = {arXiv},
       eprint = {2403.03872},
 primaryClass = {astro-ph.GA},
       adsurl = {https://ui.adsabs.harvard.edu/abs/2024Natur.636..594J}
}

@ARTICLE{juodzbalis24a,
       author = {{Juod{\v{z}}balis}, Ignas and {Ji}, Xihan and {Maiolino}, Roberto and {D'Eugenio}, Francesco and {Scholtz}, Jan and {Risaliti}, Guido and {Fabian}, Andrew C. and {Mazzolari}, Giovanni and {Gilli}, Roberto and {Prandoni}, Isabella and {Arribas}, Santiago and {Bunker}, Andrew J. and {Carniani}, Stefano and {Charlot}, St{\'e}phane and {Curtis-Lake}, Emma and {de Graaff}, Anna and {Hainline}, Kevin and {Parlanti}, Eleonora and {Perna}, Michele and {P{\'e}rez-Gonz{\'a}lez}, Pablo G. and {Robertson}, Brant and {Tacchella}, Sandro and {{\"U}bler}, Hannah and {Williams}, Christina C. and {Willott}, Chris and {Witstok}, Joris},
        title = "{JADES - the Rosetta stone of JWST-discovered AGN: deciphering the intriguing nature of early AGN}",
      journal = {\mnras},
         year = 2024,
        month = nov,
       volume = {535},
       number = {1},
        pages = {853-873},
          doi = {10.1093/mnras/stae2367},
archivePrefix = {arXiv},
       eprint = {2407.08643},
 primaryClass = {astro-ph.GA},
       adsurl = {https://ui.adsabs.harvard.edu/abs/2024MNRAS.535..853J}
}

@ARTICLE{iani24,
       author = {{Iani}, Edoardo and {Rinaldi}, Pierluigi and {Caputi}, Karina I. and {Annunziatella}, Marianna and {Langeroodi}, Danial and {Melinder}, Jens and {P{\'e}rez-Gonz{\'a}lez}, Pablo G. and {{\'A}lvarez-M{\'a}rquez}, Javier and {Boogaard}, Leindert A. and {Bosman}, Sarah E.~I. and {Costantin}, Luca and {Moutard}, Thibaud and {Colina}, Luis and {{\"O}stlin}, G{\"o}ran and {Greve}, Thomas R. and {Wright}, Gillian and {Alonso-Herrero}, Almudena and {Bik}, Arjan and {Gillman}, Steven and {Crespo G{\'o}mez}, Alejandro and {Hjorth}, Jens and {Kendrew}, Sarah and {Labiano}, Alvaro and {Pye}, John P. and {Tikkanen}, Tuomo V. and {Walter}, Fabian and {van der Werf}, Paul P.},
        title = "{MIDIS: MIRI uncovers Virgil, the first Little Red Dot with clear detection of its host galaxy at z \raisebox{-0.5ex}\textasciitilde 6.6}",
      journal = {arXiv e-prints},
         year = 2024,
        month = jun,
          eid = {arXiv:2406.18207},
        pages = {arXiv:2406.18207},
          doi = {10.48550/arXiv.2406.18207},
archivePrefix = {arXiv},
       eprint = {2406.18207},
 primaryClass = {astro-ph.GA},
       adsurl = {https://ui.adsabs.harvard.edu/abs/2024arXiv240618207I}
}

@ARTICLE{rinaldi25,
       author = {{Rinaldi}, Pierluigi and {P{\'e}rez-Gonz{\'a}lez}, Pablo G. and {Rieke}, George H. and {Lyu}, Jianwei and {D'Eugenio}, Francesco and {Wu}, Zihao and {Carniani}, Stefano and {Looser}, Tobias J. and {Shivaei}, Irene and {Boogaard}, Leindert A. and {Diaz-Santos}, Tanio and {Colina}, Luis and {{\"O}stlin}, G{\"o}ran and {Alberts}, Stacey and {{\'A}lvarez-M{\'a}rquez}, Javier and {Annuziatella}, Marianna and {Aravena}, Manuel and {Bhatawdekar}, Rachana and {Bunker}, Andrew J. and {Caputi}, Karina I. and {Charlot}, St{\'e}phane and {Crespo G{\'o}mez}, Alejandro and {Curti}, Mirko and {Eckart}, Andreas and {Gillman}, Steven and {Hainline}, Kevin and {Kumari}, Nimisha and {Hjorth}, Jens and {Iani}, Edoardo and {Inami}, Hanae and {Ji}, Zhiyuan and {Johnson}, Benjamin D. and {Jones}, Gareth C. and {Labiano}, {\'A}lvaro and {Maiolino}, Roberto and {Melinder}, Jens and {Moutard}, Thibaud and {Pei{\ss}ker}, Florian and {Rieke}, Marcia and {Robertson}, Brant and {Scholtz}, Jan and {Tacchella}, Sandro and {van der Werf}, Paul P. and {Walter}, Fabian and {Williams}, Christina C. and {Willott}, Chris and {Witstok}, Joris and {{\"U}bler}, Hannah and {Zhu}, Yongda},
        title = "{Deciphering the Nature of Virgil: An Obscured AGN Lurking Within an Apparently Normal Lyman-{\ensuremath{\alpha}} Emitter During Cosmic Reionization}",
      journal = {arXiv e-prints},
         year = 2025,
        month = apr,
          eid = {arXiv:2504.01852},
        pages = {arXiv:2504.01852},
          doi = {10.48550/arXiv.2504.01852},
archivePrefix = {arXiv},
       eprint = {2504.01852},
 primaryClass = {astro-ph.GA},
       adsurl = {https://ui.adsabs.harvard.edu/abs/2025arXiv250401852R}
}

@ARTICLE{ma25,
       author = {{Ma}, Yilun and {Greene}, Jenny E. and {Setton}, David J. and {Volonteri}, Marta and {Leja}, Joel and {Wang}, Bingjie and {Bezanson}, Rachel and {Brammer}, Gabriel and {Cutler}, Sam E. and {Dayal}, Pratika and {van Dokkum}, Pieter and {Furtak}, Lukas J. and {Glazebrook}, Karl and {Goulding}, Andy D. and {de Graaff}, Anna and {Kokorev}, Vasily and {Labbe}, Ivo and {Pan}, Richard and {Price}, Sedona H. and {Weaver}, John R. and {Williams}, Christina C. and {Whitaker}, Katherine E. and {Zitrin}, Adi},
        title = "{UNCOVER: 404 Error{\textemdash}Models Not Found for the Triply Imaged Little Red Dot A2744-QSO1}",
      journal = {\apj},
         year = 2025,
        month = mar,
       volume = {981},
       number = {2},
          eid = {191},
        pages = {191},
          doi = {10.3847/1538-4357/ada613},
archivePrefix = {arXiv},
       eprint = {2410.06257},
 primaryClass = {astro-ph.GA},
       adsurl = {https://ui.adsabs.harvard.edu/abs/2025ApJ...981..191M}
}

@ARTICLE{harikane23,
       author = {{Harikane}, Yuichi and {Zhang}, Yechi and {Nakajima}, Kimihiko and {Ouchi}, Masami and {Isobe}, Yuki and {Ono}, Yoshiaki and {Hatano}, Shun and {Xu}, Yi and {Umeda}, Hiroya},
        title = "{A JWST/NIRSpec First Census of Broad-line AGNs at z = 4-7: Detection of 10 Faint AGNs with M $_{BH}$ {}10$^{6}$-{}10$^{8}$ M $_{{\ensuremath{\odot}}}$ and Their Host Galaxy Properties}",
      journal = {\apj},
         year = 2023,
        month = dec,
       volume = {959},
       number = {1},
          eid = {39},
        pages = {39},
          doi = {10.3847/1538-4357/ad029e},
archivePrefix = {arXiv},
       eprint = {2303.11946},
 primaryClass = {astro-ph.GA},
       adsurl = {https://ui.adsabs.harvard.edu/abs/2023ApJ...959...39H}
}

@ARTICLE{taylor25,
       author = {{Taylor}, Anthony J. and {Kokorev}, Vasily and {Kocevski}, Dale D. and {Akins}, Hollis B. and {Cullen}, Fergus and {Dickinson}, Mark and {Finkelstein}, Steven L. and {Arrabal Haro}, Pablo and {Bromm}, Volker and {Giavalisco}, Mauro and {Inayoshi}, Kohei and {Juneau}, Stephanie and {Leung}, Gene C.~K. and {Perez-Gonzalez}, Pablo G. and {Somerville}, Rachel S. and {Trump}, Jonathan R. and {Amorin}, Ricardo O. and {Barro}, Guillermo and {Burgarella}, Denis and {Brooks}, Madisyn and {Carnall}, Adam and {Casey}, Caitlin M. and {Cheng}, Yingjie and {Chisholm}, John and {Chworowsky}, Katherine and {Davis}, Kelcey and {Donnan}, Callum T. and {Dunlop}, James S. and {Ellis}, Richard S. and {Fernandez}, Vital and {Fujimoto}, Seiji and {Grogin}, Norman A. and {Gupta}, Ansh R. and {Hathi}, Nimish P. and {Jung}, Intae and {Hirschmann}, Michaela and {Kartaltepe}, Jeyhan S. and {Koekemoer}, Anton M. and {Larson}, Rebecca L. and {Leung}, Ho-Hin and {Llerena}, Mario and {Lucas}, Ray A. and {McLeod}, Derek J. and {McLure}, Ross and {Napolitano}, Lorenzo and {Papovich}, Casey and {Stanton}, Thomas M. and {Tripodi}, Roberta and {Wang}, Xin and {Wilkins}, Stephen M. and {Yung}, L.~Y. Aaron and {Zavala}, Jorge A.},
        title = "{CAPERS-LRD-z9: A Gas Enshrouded Little Red Dot Hosting a Broad-line AGN at z=9.288}",
      journal = {arXiv e-prints},
         year = 2025,
        month = may,
          eid = {arXiv:2505.04609},
        pages = {arXiv:2505.04609},
          doi = {10.48550/arXiv.2505.04609},
archivePrefix = {arXiv},
       eprint = {2505.04609},
 primaryClass = {astro-ph.GA},
       adsurl = {https://ui.adsabs.harvard.edu/abs/2025arXiv250504609T}
}

@ARTICLE{degraaff24b,
       author = {{de Graaff}, Anna and {Setton}, David J. and {Brammer}, Gabriel and {Cutler}, Sam and {Suess}, Katherine A. and {Labb{\'e}}, Ivo and {Leja}, Joel and {Weibel}, Andrea and {Maseda}, Michael V. and {Whitaker}, Katherine E. and {Bezanson}, Rachel and {Boogaard}, Leindert A. and {Cleri}, Nikko J. and {De Lucia}, Gabriella and {Franx}, Marijn and {Greene}, Jenny E. and {Hirschmann}, Michaela and {Matthee}, Jorryt and {McConachie}, Ian and {Naidu}, Rohan P. and {Oesch}, Pascal A. and {Price}, Sedona H. and {Rix}, Hans-Walter and {Valentino}, Francesco and {Wang}, Bingjie and {Williams}, Christina C.},
        title = "{Efficient formation of a massive quiescent galaxy at redshift 4.9}",
      journal = {Nature Astronomy},
         year = 2025,
        month = feb,
       volume = {9},
        pages = {280-292},
          doi = {10.1038/s41550-024-02424-3},
archivePrefix = {arXiv},
       eprint = {2404.05683},
 primaryClass = {astro-ph.GA},
       adsurl = {https://ui.adsabs.harvard.edu/abs/2025NatAs...9..280D}
}

@ARTICLE{zhang25,
       author = {{Zhang}, Junyu and {Egami}, Eiichi and {Sun}, Fengwu and {Lin}, Xiaojing and {Lyu}, Jianwei and {Zhu}, Yongda and {Rinaldi}, Pierluigi and {Sun}, Yang and {Bunker}, Andrew J. and {Bhatawdekar}, Rachana and {Helton}, Jakob M. and {Maiolino}, Roberto and {Ma}, Zheng and {Robertson}, Brant and {Tacchella}, Sandro and {Venturi}, Giacomo and {Williams}, Christina C. and {Willott}, Chris},
        title = "{Abundant Population of Broad H$α$ Emitters in the GOODS-N Field Revealed by CONGRESS, FRESCO, and JADES}",
      journal = {arXiv e-prints},
         year = 2025,
        month = may,
          eid = {arXiv:2505.02895},
        pages = {arXiv:2505.02895},
          doi = {10.48550/arXiv.2505.02895},
archivePrefix = {arXiv},
       eprint = {2505.02895},
 primaryClass = {astro-ph.GA},
       adsurl = {https://ui.adsabs.harvard.edu/abs/2025arXiv250502895Z}
}

@ARTICLE{ji25,
       author = {{Ji}, Xihan and {Maiolino}, Roberto and {{\"U}bler}, Hannah and {Scholtz}, Jan and {D'Eugenio}, Francesco and {Sun}, Fengwu and {Perna}, Michele and {Turner}, Hannah and {Arribas}, Santiago and {Bennett}, Jake S. and {Bunker}, Andrew and {Carniani}, Stefano and {Charlot}, St{\'e}phane and {Cresci}, Giovanni and {Curti}, Mirko and {Egami}, Eiichi and {Fabian}, Andy and {Inayoshi}, Kohei and {Isobe}, Yuki and {Jones}, Gareth and {Juod{\v{z}}balis}, Ignas and {Kumari}, Nimisha and {Lyu}, Jianwei and {Mazzolari}, Giovanni and {Parlanti}, Eleonora and {Robertson}, Brant and {Rodr{\'\i}guez Del Pino}, Bruno and {Schneider}, Raffaella and {Sijacki}, Debora and {Tacchella}, Sandro and {Trinca}, Alessandro and {Valiante}, Rosa and {Venturi}, Giacomo and {Volonteri}, Marta and {Willott}, Chris and {Witten}, Callum and {Witstok}, Joris},
        title = "{BlackTHUNDER -- A non-stellar Balmer break in a black hole-dominated little red dot at $z=7.04$}",
      journal = {arXiv e-prints},
         year = 2025,
        month = jan,
          eid = {arXiv:2501.13082},
        pages = {arXiv:2501.13082},
          doi = {10.48550/arXiv.2501.13082},
archivePrefix = {arXiv},
       eprint = {2501.13082},
 primaryClass = {astro-ph.GA},
       adsurl = {https://ui.adsabs.harvard.edu/abs/2025arXiv250113082J}
}

@ARTICLE{deugenio25,
       author = {{D'Eugenio}, Francesco and {Maiolino}, Roberto and {Perna}, Michele and {Uebler}, Hannah and {Ji}, Xihan and {McClymont}, William and {Koudmani}, Sophie and {Sijacki}, Debora and {Juod{\v{z}}balis}, Ignas and {Scholtz}, Jan and {Bennett}, Jake and {Bunker}, Andrew J. and {Carniani}, Stefano and {Charlot}, St{\'e}phane and {Cresci}, Giovanni and {Curtis-Lake}, Emma and {Dalla Bont{\`a}}, Elena and {Jones}, Gareth C. and {Lyu}, Jianwei and {Marconi}, Alessandro and {Mazzolari}, Giovanni and {Nelson}, Erica J. and {Parlanti}, Eleonora and {Robertson}, Brant E. and {Schneider}, Raffaella and {Simmonds}, Charlotte and {Tacchella}, Sandro and {Venturi}, Giacomo and {Willott}, Chris and {Witstok}, Joris and {Witten}, Callum},
        title = "{BlackTHUNDER strikes twice: rest-frame Balmer-line absorption and high Eddington accretion rate in a Little Red Dot at $z=7.04$}",
      journal = {arXiv e-prints},
         year = 2025,
        month = mar,
          eid = {arXiv:2503.11752},
        pages = {arXiv:2503.11752},
          doi = {10.48550/arXiv.2503.11752},
archivePrefix = {arXiv},
       eprint = {2503.11752},
 primaryClass = {astro-ph.GA},
       adsurl = {https://ui.adsabs.harvard.edu/abs/2025arXiv250311752D}
}

@ARTICLE{hviding25,
       author = {{Hviding}, Raphael E. and {de Graaff}, Anna and {Miller}, Tim B. and {Setton}, David J. and {Greene}, Jenny E. and {Labb{\'e}}, Ivo and {Brammer}, Gabriel and {Bezanson}, Rachel and {Boogaard}, Leindert A. and {Cleri}, Nikko J. and {Leja}, Joel and {Maseda}, Michael V. and {McConachie}, Ian and {Matthee}, Jorryt and {Naidu}, Rohan P. and {Oesch}, Pascal A. and {Wang}, Bingjie and {Whitaker}, Katherine E. and {Williams}, Christina C.},
        title = "{RUBIES: A spectroscopic census of little red dots: All point sources with v-shaped continua have broad lines}",
      journal = {\aap},
         year = 2025,
        month = oct,
       volume = {702},
          eid = {A57},
        pages = {A57},
          doi = {10.1051/0004-6361/202555816},
archivePrefix = {arXiv},
       eprint = {2506.05459},
 primaryClass = {astro-ph.GA},
       adsurl = {https://ui.adsabs.harvard.edu/abs/2025A&A...702A..57H}
}

@ARTICLE{deugenio25b,
       author = {{D'Eugenio}, Francesco and {Juod{\v{z}}balis}, Ignas and {Ji}, Xihan and {Scholtz}, Jan and {Maiolino}, Roberto and {Carniani}, Stefano and {Perna}, Michele and {Mazzolari}, Giovanni and {{\"U}bler}, Hannah and {Arribas}, Santiago and {Bhatawdekar}, Rachana and {Bunker}, Andrew J. and {Cresci}, Giovanni and {Curtis-Lake}, Emma and {Hainline}, Kevin and {Inayoshi}, Kohei and {Isobe}, Yuki and {Johnson}, Benjamin D. and {Jones}, Gareth C. and {Looser}, Tobias J. and {Nelson}, Erica J. and {Parlanti}, Eleonora and {Pusk{\'a}s}, D{\'a}vid and {Rinaldi}, Pierluigi and {Robertson}, Brant and {Rodr{\'\i}guez Del Pino}, Bruno and {Shivaei}, Irene and {Sun}, Fengwu and {Tacchella}, Sandro and {Venturi}, Giacomo and {Volonteri}, Marta and {Williams}, Christina C. and {Willmer}, Christopher N.~A. and {Willott}, Chris and {Witstok}, Joris},
        title = "{JADES and BlackTHUNDER: rest-frame Balmer-line absorption and the local environment in a Little Red Dot at z = 5}",
      journal = {arXiv e-prints},
         year = 2025,
        month = jun,
          eid = {arXiv:2506.14870},
        pages = {arXiv:2506.14870},
          doi = {10.48550/arXiv.2506.14870},
archivePrefix = {arXiv},
       eprint = {2506.14870},
 primaryClass = {astro-ph.GA},
       adsurl = {https://ui.adsabs.harvard.edu/abs/2025arXiv250614870D}
}

@ARTICLE{deugenio25c,
       author = {{D'Eugenio}, Francesco and {Nelson}, Erica and {Ji}, Xihan and {Baggen}, Josephine and {Greene}, Jenny and {Labb{\'e}}, Ivo and {Pezzulli}, Gabriele and {Brown}, Vanessa and {Maiolino}, Roberto and {Matthee}, Jorryt and {Terlevich}, Elena and {Terlevich}, Roberto and {Torralba}, Alberto and {Carniani}, Stefano},
        title = "{Irony at z=6.68: a bright AGN with forbidden Fe emission and multi-component Balmer absorption}",
      journal = {arXiv e-prints},
         year = 2025,
        month = sep,
          eid = {arXiv:2510.00101},
        pages = {arXiv:2510.00101},
          doi = {10.48550/arXiv.2510.00101},
archivePrefix = {arXiv},
       eprint = {2510.00101},
 primaryClass = {astro-ph.GA},
       adsurl = {https://ui.adsabs.harvard.edu/abs/2025arXiv251000101D}
}

@ARTICLE{torralba25b,
       author = {{Torralba}, Alberto and {Matthee}, Jorryt and {Pezzulli}, Gabriele and {Naidu}, Rohan P. and {Ishikawa}, Yuzo and {Brammer}, Gabriel B. and {Chang}, Seok-Jun and {Chisholm}, John and {de Graaff}, Anna and {D'Eugenio}, Francesco and {Di Cesare}, Claudia and {Eilers}, Anna-Christina and {Greene}, Jenny E. and {Gronke}, Max and {Iani}, Edoardo and {Kokorev}, Vasily and {Kotiwale}, Gauri and {Kramarenko}, Ivan and {Ma}, Yilun and {Mascia}, Sara and {Navarrete}, Benjam{\'\i}n and {Nelson}, Erica and {Oesch}, Pascal and {Simcoe}, Robert A. and {Wuyts}, Stijn},
        title = "{The warm outer layer of a Little Red Dot as the source of [Fe II] and collisional Balmer lines with scattering wings}",
      journal = {arXiv e-prints},
         year = 2025,
        month = sep,
          eid = {arXiv:2510.00103},
        pages = {arXiv:2510.00103},
          doi = {10.48550/arXiv.2510.00103},
archivePrefix = {arXiv},
       eprint = {2510.00103},
 primaryClass = {astro-ph.GA},
       adsurl = {https://ui.adsabs.harvard.edu/abs/2025arXiv251000103T}
}

@ARTICLE{valentino25,
       author = {{Valentino}, F. and {Heintz}, K.~E. and {Brammer}, G. and {Ito}, K. and {Kokorev}, V. and {Whitaker}, K.~E. and {Gallazzi}, A. and {de Graaff}, A. and {Weibel}, A. and {Frye}, B.~L. and {Kamieneski}, P.~S. and {Jin}, S. and {Ceverino}, D. and {Faisst}, A. and {Farcy}, M. and {Fujimoto}, S. and {Gillman}, S. and {Gottumukkala}, R. and {Hamadouche}, M. and {Harrington}, K.~C. and {Hirschmann}, M. and {Jespersen}, C.~K. and {Kakimoto}, T. and {Kubo}, M. and {Lagos}, C. d. P. and {Lee}, M. and {Magdis}, G.~E. and {Man}, A.~W.~S. and {Onodera}, M. and {Rizzo}, F. and {Shimakawa}, R. and {Setton}, D.~J. and {Tanaka}, M. and {Toft}, S. and {Wu}, P.-F. and {Zhu}, P.},
        title = "{Gas outflows in two recently quenched galaxies at z = 4 and 7}",
      journal = {\aap},
         year = 2025,
        month = jul,
       volume = {699},
          eid = {A358},
        pages = {A358},
          doi = {10.1051/0004-6361/202553908},
archivePrefix = {arXiv},
       eprint = {2503.01990},
 primaryClass = {astro-ph.GA},
       adsurl = {https://ui.adsabs.harvard.edu/abs/2025A&A...699A.358V}
}

@BOOK{osterbrock06,
       author = {{Osterbrock}, Donald E. and {Ferland}, Gary J.},
        title = "{Astrophysics of gaseous nebulae and active galactic nuclei}",
         year = 2006,
       adsurl = {https://ui.adsabs.harvard.edu/abs/2006agna.book.....O}
}

@ARTICLE{chang25,
       author = {{Chang}, Seok-Jun and {Gronke}, Max and {Matthee}, Jorryt and {Mason}, Charlotte},
        title = "{Impact of Resonance, Raman, and Thomson Scattering on Hydrogen Line Formation in Little Red Dots}",
      journal = {arXiv e-prints},
         year = 2025,
        month = aug,
          eid = {arXiv:2508.08768},
        pages = {arXiv:2508.08768},
          doi = {10.48550/arXiv.2508.08768},
archivePrefix = {arXiv},
       eprint = {2508.08768},
 primaryClass = {astro-ph.GA},
       adsurl = {https://ui.adsabs.harvard.edu/abs/2025arXiv250808768C}
}

@ARTICLE{jones25,
       author = {{Jones}, Gareth C. and {{\"U}bler}, Hannah and {Maiolino}, Roberto and {Ji}, Xihan and {Marconi}, Alessandro and {D'Eugenio}, Francesco and {Arribas}, Santiago and {Bunker}, Andrew J. and {Carniani}, Stefano and {Charlot}, St{\'e}phane and {Cresci}, Giovanni and {Inayoshi}, Kohei and {Isobe}, Yuki and {Juod{\v{z}}balis}, Ignas and {Mazzolari}, Giovanni and {P{\'e}rez-Gonz{\'a}lez}, Pablo G. and {Perna}, Michele and {Schneider}, Raffaella and {Scholtz}, Jan and {Tacchella}, Sandro},
        title = "{BlackTHUNDER: Shedding light on a dormant and extreme little red dot at z=8.50}",
      journal = {arXiv e-prints},
         year = 2025,
        month = sep,
          eid = {arXiv:2509.20455},
        pages = {arXiv:2509.20455},
          doi = {10.48550/arXiv.2509.20455},
archivePrefix = {arXiv},
       eprint = {2509.20455},
 primaryClass = {astro-ph.GA},
       adsurl = {https://ui.adsabs.harvard.edu/abs/2025arXiv250920455J}
}

@ARTICLE{trump23,
       author = {{Trump}, Jonathan R. and {Arrabal Haro}, Pablo and {Simons}, Raymond C. and {Backhaus}, Bren E. and {Amor{\'\i}n}, Ricardo O. and {Dickinson}, Mark and {Fern{\'a}ndez}, Vital and {Papovich}, Casey and {Nicholls}, David C. and {Kewley}, Lisa J. and {Brunker}, Samantha W. and {Salzer}, John J. and {Wilkins}, Stephen M. and {Almaini}, Omar and {Bagley}, Micaela B. and {Berg}, Danielle A. and {Bhatawdekar}, Rachana and {Bisigello}, Laura and {Buat}, V{\'e}ronique and {Burgarella}, Denis and {Calabr{\`o}}, Antonello and {Casey}, Caitlin M. and {Ciesla}, Laure and {Cleri}, Nikko J. and {Cole}, Justin W. and {Cooper}, M.~C. and {Cooray}, Asantha R. and {Costantin}, Luca and {Croton}, Darren and {Ferguson}, Henry C. and {Finkelstein}, Steven L. and {Fujimoto}, Seiji and {Gardner}, Jonathan P. and {Gawiser}, Eric and {Giavalisco}, Mauro and {Grazian}, Andrea and {Grogin}, Norman A. and {Hathi}, Nimish P. and {Hirschmann}, Michaela and {Holwerda}, Benne W. and {Huertas-Company}, Marc and {Hutchison}, Taylor A. and {Jogee}, Shardha and {Juneau}, St{\'e}phanie and {Jung}, Intae and {Kartaltepe}, Jeyhan S. and {Kirkpatrick}, Allison and {Kocevski}, Dale D. and {Koekemoer}, Anton M. and {Lotz}, Jennifer M. and {Lucas}, Ray A. and {Magnelli}, Benjamin and {Matharu}, Jasleen and {P{\'e}rez-Gonz{\'a}lez}, Pablo G. and {Pirzkal}, Nor and {Rafelski}, Marc and {Rose}, Caitlin and {Seill{\'e}}, Lise-Marie and {Somerville}, Rachel S. and {Straughn}, Amber N. and {Tacchella}, Sandro and {Vanderhoof}, Brittany N. and {Weiner}, Benjamin J. and {Wuyts}, Stijn and {Yung}, L.~Y. Aaron and {Zavala}, Jorge A.},
        title = "{The Physical Conditions of Emission-line Galaxies at Cosmic Dawn from JWST/NIRSpec Spectroscopy in the SMACS 0723 Early Release Observations}",
      journal = {\apj},
         year = 2023,
        month = mar,
       volume = {945},
       number = {1},
          eid = {35},
        pages = {35},
          doi = {10.3847/1538-4357/acba8a},
archivePrefix = {arXiv},
       eprint = {2207.12388},
 primaryClass = {astro-ph.GA},
       adsurl = {https://ui.adsabs.harvard.edu/abs/2023ApJ...945...35T}
}

@ARTICLE{backhaus22,
       author = {{Backhaus}, Bren E. and {Trump}, Jonathan R. and {Cleri}, Nikko J. and {Simons}, Raymond and {Momcheva}, Ivelina and {Papovich}, Casey and {Estrada-Carpenter}, Vicente and {Finkelstein}, Steven L. and {Matharu}, Jasleen and {Ji}, Zhiyuan and {Weiner}, Benjamin and {Giavalisco}, Mauro and {Jung}, Intae},
        title = "{CLEAR: Emission-line Ratios at Cosmic High Noon}",
      journal = {\apj},
         year = 2022,
        month = feb,
       volume = {926},
       number = {2},
          eid = {161},
        pages = {161},
          doi = {10.3847/1538-4357/ac3919},
archivePrefix = {arXiv},
       eprint = {2109.08147},
 primaryClass = {astro-ph.GA},
       adsurl = {https://ui.adsabs.harvard.edu/abs/2022ApJ...926..161B}
}

@ARTICLE{backhaus24,
       author = {{Backhaus}, Bren E. and {Trump}, Jonathan R. and {Pirzkal}, Nor and {Barro}, Guillermo and {Finkelstein}, Steven L. and {Arrabal Haro}, Pablo and {Simons}, Raymond C. and {Wessner}, Jessica and {Cleri}, Nikko J. and {Bagley}, Micaela B. and {Hirschmann}, Michaela and {Nicholls}, David C. and {Dickinson}, Mark and {Kartaltepe}, Jeyhan S. and {Papovich}, Casey and {Kocevski}, Dale D. and {Koekemoer}, Anton M. and {Bisigello}, Laura and {Jaskot}, Anne E. and {Lucas}, Ray A. and {Jung}, Intae and {Wilkins}, Stephen M. and {Yung}, L.~Y. Aaron and {Ferguson}, Henry C. and {Fontana}, Adriano and {Grazian}, Andrea and {Grogin}, Norman A. and {Kewley}, Lisa J. and {Kirkpatrick}, Allison and {Lotz}, Jennifer M. and {Pentericci}, Laura and {P{\'e}rez-Gonz{\'a}lez}, Pablo G. and {Ravindranath}, Swara and {Somerville}, Rachel S. and {Yang}, Guang and {Holwerda}, Benne W. and {Kurczynski}, Peter and {Hathi}, Nimish P. and {Rose}, Caitlin and {Davis}, Kelcey},
        title = "{CEERS Key Paper. VIII. Emission-line Ratios from NIRSpec and NIRCam Wide-Field Slitless Spectroscopy at z > 2}",
      journal = {\apj},
         year = 2024,
        month = feb,
       volume = {962},
       number = {2},
          eid = {195},
        pages = {195},
          doi = {10.3847/1538-4357/ad1520},
archivePrefix = {arXiv},
       eprint = {2307.09503},
 primaryClass = {astro-ph.GA},
       adsurl = {https://ui.adsabs.harvard.edu/abs/2024ApJ...962..195B}
}

@ARTICLE{barro25,
       author = {{Barro}, Guillermo and {Perez-Gonzalez}, Pablo G. and {Kocevski}, Dale D. and {McGrath}, Elizabeth J. and {Leung}, Gene C.~K. and {Cullen}, Fergus and {Dunlop}, James S. and {Ellis}, Richard S. and {Finkelstein}, Steven L. and {Grogin}, Norman A. and {Illingworth}, Garth and {Kartaltepe}, Jeyhan S. and {Koekemoer}, Anton M. and {Lucas}, Ray A. and {McLure}, Ross J. and {Yang}, Guang},
        title = "{A Comprehensive Photometric Selection of `Little Red Dots' in MIRI Fields: An IR-Bright LRD at $z=3.1386$ with Warm Dust Emission}",
      journal = {arXiv e-prints},
         year = 2024,
        month = dec,
          eid = {arXiv:2412.01887},
        pages = {arXiv:2412.01887},
          doi = {10.48550/arXiv.2412.01887},
archivePrefix = {arXiv},
       eprint = {2412.01887},
 primaryClass = {astro-ph.GA},
       adsurl = {https://ui.adsabs.harvard.edu/abs/2024arXiv241201887B}
}

@ARTICLE{brooks25,
       author = {{Brooks}, Madisyn and {Simons}, Raymond C. and {Trump}, Jonathan R. and {Taylor}, Anthony J. and {Bagley}, Micaela B. and {Backhaus}, Bren and {Davis}, Kelcey and {Buat}, V{\'e}ronique and {Cleri}, Nikko J. and {de la Vega}, Alexander and {Finkelstein}, Steven L. and {Hirschmann}, Michaela and {Holwerda}, Benne W. and {Kocevski}, Dale D. and {Koekemoer}, Anton M. and {Lucas}, Ray A. and {Pacucci}, Fabio and {Seill{\'e}}, Lise-Marie},
        title = "{Here There Be (Dusty) Monsters: High-redshift Active Galactic Nuclei Are Dustier than Their Hosts}",
      journal = {\apj},
         year = 2025,
        month = jun,
       volume = {986},
       number = {2},
          eid = {177},
        pages = {177},
          doi = {10.3847/1538-4357/addac4},
archivePrefix = {arXiv},
       eprint = {2410.07340},
 primaryClass = {astro-ph.GA},
       adsurl = {https://ui.adsabs.harvard.edu/abs/2025ApJ...986..177B}
}

@ARTICLE{degraaff25b,
       author = {{de Graaff}, Anna and {Hviding}, Raphael E. and {Naidu}, Rohan P. and {Greene}, Jenny E. and {Miller}, Tim B. and {Leja}, Joel and {Matthee}, Jorryt and {Brammer}, Gabriel and {Katz}, Harley and {Bezanson}, Rachel and {Boogaard}, Leindert A. and {Bose}, Sownak and {Chisholm}, John and {Cleri}, Nikko J. and {Dayal}, Pratika and {Feldmann}, Robert and {Fudamoto}, Yoshinobu and {Fujimoto}, Seiji and {Furtak}, Lukas J. and {Glazebrook}, Karl and {Gottumukkala}, Rashmi and {Heintz}, Kasper E. and {Kokorev}, Vasily and {Labbe}, Ivo and {Maseda}, Michael V. and {McConachie}, Ian and {Nanayakkara}, Themiya and {Nelson}, Erica and {Nowaczyk}, Przemys{\l}aw and {Oesch}, Pascal A. and {Rix}, Hans-Walter and {Setton}, David J. and {Torralba}, Alberto and {Walter}, Fabian and {Wang}, Bingjie and {Weibel}, Andrea and {van der Wel}, Arjen},
        title = "{Little Red Dots host Black Hole Stars: A unified family of gas-reddened AGN revealed by JWST/NIRSpec spectroscopy}",
      journal = {arXiv e-prints},
         year = 2025,
        month = nov,
          eid = {arXiv:2511.21820},
        pages = {arXiv:2511.21820},
          doi = {10.48550/arXiv.2511.21820},
archivePrefix = {arXiv},
       eprint = {2511.21820},
 primaryClass = {astro-ph.GA},
       adsurl = {https://ui.adsabs.harvard.edu/abs/2025arXiv251121820D}
}

@ARTICLE{robertsborsani24,
       author = {{Roberts-Borsani}, Guido and {Treu}, Tommaso and {Shapley}, Alice and {Fontana}, Adriano and {Pentericci}, Laura and {Castellano}, Marco and {Morishita}, Takahiro and {Bergamini}, Pietro and {Rosati}, Piero},
        title = "{Between the Extremes: A JWST Spectroscopic Benchmark for High-redshift Galaxies Using {\ensuremath{\sim}}500 Confirmed Sources at z {\ensuremath{\geq}} 5}",
      journal = {\apj},
         year = 2024,
        month = dec,
       volume = {976},
       number = {2},
          eid = {193},
        pages = {193},
          doi = {10.3847/1538-4357/ad85d3},
archivePrefix = {arXiv},
       eprint = {2403.07103},
 primaryClass = {astro-ph.GA},
       adsurl = {https://ui.adsabs.harvard.edu/abs/2024ApJ...976..193R}
}

@ARTICLE{nakajima18,
       author = {{Nakajima}, K. and {Schaerer}, D. and {Le F{\`e}vre}, O. and {Amor{\'\i}n}, R. and {Talia}, M. and {Lemaux}, B.~C. and {Tasca}, L.~A.~M. and {Vanzella}, E. and {Zamorani}, G. and {Bardelli}, S. and {Grazian}, A. and {Guaita}, L. and {Hathi}, N.~P. and {Pentericci}, L. and {Zucca}, E.},
        title = "{The VIMOS Ultra Deep Survey: Nature, ISM properties, and ionizing spectra of CIII]{\ensuremath{\lambda}}1909 emitters at z = 2-4}",
      journal = {\aap},
         year = 2018,
        month = may,
       volume = {612},
          eid = {A94},
        pages = {A94},
          doi = {10.1051/0004-6361/201731935},
archivePrefix = {arXiv},
       eprint = {1709.03990},
 primaryClass = {astro-ph.GA},
       adsurl = {https://ui.adsabs.harvard.edu/abs/2018A&A...612A..94N}
}

@ARTICLE{furtak24,
       author = {{Furtak}, Lukas J. and {Labb{\'e}}, Ivo and {Zitrin}, Adi and {Greene}, Jenny E. and {Dayal}, Pratika and {Chemerynska}, Iryna and {Kokorev}, Vasily and {Miller}, Tim B. and {Goulding}, Andy D. and {de Graaff}, Anna and {Bezanson}, Rachel and {Brammer}, Gabriel B. and {Cutler}, Sam E. and {Leja}, Joel and {Pan}, Richard and {Price}, Sedona H. and {Wang}, Bingjie and {Weaver}, John R. and {Whitaker}, Katherine E. and {Atek}, Hakim and {Bogd{\'a}n}, {\'A}kos and {Charlot}, St{\'e}phane and {Curtis-Lake}, Emma and {van Dokkum}, Pieter and {Endsley}, Ryan and {Feldmann}, Robert and {Fudamoto}, Yoshinobu and {Fujimoto}, Seiji and {Glazebrook}, Karl and {Juneau}, St{\'e}phanie and {Marchesini}, Danilo and {Maseda}, Micheal V. and {Nelson}, Erica and {Oesch}, Pascal A. and {Plat}, Ad{\`e}le and {Setton}, David J. and {Stark}, Daniel P. and {Williams}, Christina C.},
        title = "{A high black-hole-to-host mass ratio in a lensed AGN in the early Universe}",
      journal = {\nat},
         year = 2024,
        month = apr,
       volume = {628},
       number = {8006},
        pages = {57-61},
          doi = {10.1038/s41586-024-07184-8},
archivePrefix = {arXiv},
       eprint = {2308.05735},
 primaryClass = {astro-ph.GA},
       adsurl = {https://ui.adsabs.harvard.edu/abs/2024Natur.628...57F}
}

@ARTICLE{pei92,
       author = {{Pei}, Yichuan C.},
        title = "{Interstellar Dust from the Milky Way to the Magellanic Clouds}",
      journal = {\apj},
         year = 1992,
        month = aug,
       volume = {395},
        pages = {130},
          doi = {10.1086/171637},
       adsurl = {https://ui.adsabs.harvard.edu/abs/1992ApJ...395..130P}
}

@ARTICLE{umeda25,
       author = {{Umeda}, Hiroya and {Inayoshi}, Kohei and {Harikane}, Yuichi and {Murase}, Kohta},
        title = "{A Black-Hole Envelope Interpretation for Cosmological Demographics of Little Red Dots}",
      journal = {arXiv e-prints},
         year = 2025,
        month = dec,
          eid = {arXiv:2512.04208},
        pages = {arXiv:2512.04208},
          doi = {10.48550/arXiv.2512.04208},
archivePrefix = {arXiv},
       eprint = {2512.04208},
 primaryClass = {astro-ph.GA},
       adsurl = {https://ui.adsabs.harvard.edu/abs/2025arXiv251204208U}
}

@ARTICLE{heintz25,
       author = {{Heintz}, K.~E. and {Brammer}, G.~B. and {Watson}, D. and {Oesch}, P.~A. and {Keating}, L.~C. and {Hayes}, M.~J. and {Abdurro'uf} and {Arellano-C{\'o}rdova}, K.~Z. and {Carnall}, A.~C. and {Christiansen}, C.~R. and {Cullen}, F. and {Dav{\'e}}, R. and {Dayal}, P. and {Ferrara}, A. and {Finlator}, K. and {Fynbo}, J.~P.~U. and {Flury}, S.~R. and {Gelli}, V. and {Gillman}, S. and {Gottumukkala}, R. and {Gould}, K. and {Greve}, T.~R. and {Hardin}, S.~E. and {Hsiao}, T.~Y.-Y. and {Hutter}, A. and {Jakobsson}, P. and {Killi}, M. and {Khosravaninezhad}, N. and {Laursen}, P. and {Lee}, M.~M. and {Magdis}, G.~E. and {Matthee}, J. and {Naidu}, R.~P. and {Narayanan}, D. and {Pollock}, C. and {Prescott}, M.~K.~M. and {Rusakov}, V. and {Shuntov}, M. and {Sneppen}, A. and {Smit}, R. and {Tanvir}, N.~R. and {Terp}, C. and {Toft}, S. and {Valentino}, F. and {Vijayan}, A.~P. and {Weaver}, J.~R. and {Wise}, J.~H. and {Witstok}, J.},
        title = "{The JWST-PRIMAL archival survey: A JWST/NIRSpec reference sample for the physical properties and Lyman-{\ensuremath{\alpha}} absorption and emission of {\ensuremath{\sim}}600 galaxies at z = 5.0 ‑ 13.4}",
      journal = {\aap},
         year = 2025,
        month = jan,
       volume = {693},
          eid = {A60},
        pages = {A60},
          doi = {10.1051/0004-6361/202450243},
archivePrefix = {arXiv},
       eprint = {2404.02211},
 primaryClass = {astro-ph.GA},
       adsurl = {https://ui.adsabs.harvard.edu/abs/2025A&A...693A..60H}
}

@ARTICLE{curtislake25,
       author = {{Curtis-Lake}, Emma and {Cameron}, Alex J. and {Bunker}, Andrew J. and {Scholtz}, Jan and {Carniani}, Stefano and {Parlanti}, Eleonora and {D'Eugenio}, Francesco and {Jakobsen}, Peter and {Willmer}, Christopher N.~A. and {Arribas}, Santiago and {Baker}, William M. and {Charlot}, St{\'e}phane and {Chevallard}, Jacopo and {Circosta}, Chiara and {Curti}, Mirko and {Eisenstein}, Daniel J. and {Hainline}, Kevin and {Ji}, Zhiyuan and {Johnson}, Benjamin D. and {Jones}, Gareth C. and {Maiolino}, Roberto and {Maseda}, Michael V. and {P{\'e}rez-Gonz{\'a}lez}, Pablo G. and {Rawle}, Tim and {Rieke}, Marcia and {Rinaldi}, Pierluigi and {Robertson}, Brant and {Rodr{\'\i}gez Del Pino}, Bruno and {Saxena}, Aayush and {Shivaei}, Irene and {Smit}, Renske and {Tacchella}, Sandro and {{\"U}bler}, Hannah and {Venturi}, Giacomo and {Williams}, Christina C. and {Willott}, Chris and {Duan}, Qiao},
        title = "{JADES Data Release 4 Paper I: Sample Selection, Observing Strategy and Redshifts of the complete spectroscopic sample}",
      journal = {arXiv e-prints},
         year = 2025,
        month = oct,
          eid = {arXiv:2510.01033},
        pages = {arXiv:2510.01033},
          doi = {10.48550/arXiv.2510.01033},
archivePrefix = {arXiv},
       eprint = {2510.01033},
 primaryClass = {astro-ph.GA},
       adsurl = {https://ui.adsabs.harvard.edu/abs/2025arXiv251001033C}
}

@ARTICLE{bezanson24,
       author = {{Bezanson}, Rachel and {Labbe}, Ivo and {Whitaker}, Katherine E. and {Leja}, Joel and {Price}, Sedona H. and {Franx}, Marijn and {Brammer}, Gabriel and {Marchesini}, Danilo and {Zitrin}, Adi and {Wang}, Bingjie and {Weaver}, John R. and {Furtak}, Lukas J. and {Atek}, Hakim and {Coe}, Dan and {Cutler}, Sam E. and {Dayal}, Pratika and {van Dokkum}, Pieter and {Feldmann}, Robert and {F{\"o}rster Schreiber}, Natascha M. and {Fujimoto}, Seiji and {Geha}, Marla and {Glazebrook}, Karl and {de Graaff}, Anna and {Greene}, Jenny E. and {Juneau}, St{\'e}phanie and {Kassin}, Susan and {Kriek}, Mariska and {Khullar}, Gourav and {Maseda}, Michael and {Mowla}, Lamiya A. and {Muzzin}, Adam and {Nanayakkara}, Themiya and {Nelson}, Erica J. and {Oesch}, Pascal A. and {Pacifici}, Camilla and {Pan}, Richard and {Papovich}, Casey and {Setton}, David J. and {Shapley}, Alice E. and {Smit}, Renske and {Stefanon}, Mauro and {Taylor}, Edward N. and {Williams}, Christina C.},
        title = "{The JWST UNCOVER Treasury Survey: Ultradeep NIRSpec and NIRCam Observations before the Epoch of Reionization}",
      journal = {\apj},
         year = 2024,
        month = oct,
       volume = {974},
       number = {1},
          eid = {92},
        pages = {92},
          doi = {10.3847/1538-4357/ad66cf},
archivePrefix = {arXiv},
       eprint = {2212.04026},
 primaryClass = {astro-ph.GA},
       adsurl = {https://ui.adsabs.harvard.edu/abs/2024ApJ...974...92B}
}

@ARTICLE{finkelstein25,
       author = {{Finkelstein}, Steven L. and {Bagley}, Micaela B. and {Arrabal Haro}, Pablo and {Dickinson}, Mark and {Ferguson}, Henry C. and {Kartaltepe}, Jeyhan S. and {Kocevski}, Dale D. and {Koekemoer}, Anton M. and {Lotz}, Jennifer M. and {Papovich}, Casey and {P{\'e}rez-Gonz{\'a}lez}, Pablo G. and {Pirzkal}, Nor and {Somerville}, Rachel S. and {Trump}, Jonathan R. and {Yang}, Guang and {Yung}, L.~Y. Aaron and {Fontana}, Adriano and {Grazian}, Andrea and {Grogin}, Norman A. and {Kewley}, Lisa J. and {Kirkpatrick}, Allison and {Larson}, Rebecca L. and {Pentericci}, Laura and {Ravindranath}, Swara and {Wilkins}, Stephen M. and {Almaini}, Omar and {Amor{\'\i}n}, Ricardo O. and {Barro}, Guillermo and {Bhatawdekar}, Rachana and {Bisigello}, Laura and {Brooks}, Madisyn and {Buat}, V{\'e}ronique and {Buitrago}, Fernando and {Burgarella}, Denis and {Calabr{\`o}}, Antonello and {Castellano}, Marco and {Cheng}, Yingjie and {Cleri}, Nikko J. and {Cole}, Justin W. and {Cooper}, M.~C. and {Cooper}, Olivia R. and {Costantin}, Luca and {Cox}, Isa G. and {Croton}, Darren and {Daddi}, Emanuele and {Davis}, Kelcey and {Dekel}, Avishai and {Elbaz}, David and {Fern{\'a}ndez}, Vital and {Fujimoto}, Seiji and {Gandolfi}, Giovanni and {Gardner}, Jonathan P. and {Gawiser}, Eric and {Giavalisco}, Mauro and {G{\'o}mez-Guijarro}, Carlos and {Guo}, Yuchen and {Gupta}, Ansh R. and {Hathi}, Nimish P. and {Harish}, Santosh and {Henry}, Aur{\'e}lien and {Hirschmann}, Michaela and {Hu}, Weida and {Hutchison}, Taylor A. and {Iyer}, Kartheik G. and {Jaskot}, Anne E. and {Jha}, Saurabh W. and {Jung}, Intae and {Kassin}, Susan A. and {Kokorev}, Vasily and {Kurczynski}, Peter and {Leung}, Gene C.~K. and {Llerena}, Mario and {Long}, Arianna S. and {Lucas}, Ray A. and {Lu}, Shiying and {McGrath}, Elizabeth J. and {McIntosh}, Daniel H. and {Merlin}, Emiliano and {Mobasher}, Bahram and {Morales}, Alexa M. and {Napolitano}, Lorenzo and {Pacucci}, Fabio and {Pandya}, Viraj and {Rafelski}, Marc and {Rodighiero}, Giulia and {Rose}, Caitlin and {Santini}, Paola and {Seill{\'e}}, Lise-Marie and {Simons}, Raymond C. and {Shen}, Lu and {Straughn}, Amber N. and {Tacchella}, Sandro and {Taylor}, Anthony J. and {Vanderhoof}, Brittany N. and {Vega-Ferrero}, Jes{\'u}s and {Weiner}, Benjamin J. and {Willmer}, Christopher N.~A. and {Zhu}, Peixin and {Bell}, Eric F. and {Wuyts}, Stijn and {Holwerda}, Benne W. and {Wang}, Xin and {Wang}, Weichen and {Zavala}, Jorge A. and {CEERS Collaboration}},
        title = "{The Cosmic Evolution Early Release Science Survey (CEERS)}",
      journal = {\apjl},
         year = 2025,
        month = apr,
       volume = {983},
       number = {1},
          eid = {L4},
        pages = {L4},
          doi = {10.3847/2041-8213/adbbd3},
archivePrefix = {arXiv},
       eprint = {2501.04085},
 primaryClass = {astro-ph.GA},
       adsurl = {https://ui.adsabs.harvard.edu/abs/2025ApJ...983L...4F}
}

@ARTICLE{maseda24,
       author = {{Maseda}, Michael V. and {de Graaff}, Anna and {Franx}, Marijn and {Rix}, Hans-Walter and {Carniani}, Stefano and {Laseter}, Isaac and {Dudzevi{\v{c}}i{\={u}}t{\.{e}}}, Ugn{\.{e}} and {Rawle}, Tim and {Parlanti}, Eleonora and {Arribas}, Santiago and {Bunker}, Andrew J. and {Cameron}, Alex J. and {Charlot}, Stephane and {Curti}, Mirko and {D'Eugenio}, Francesco and {Jones}, Gareth C. and {Kumari}, Nimisha and {Maiolino}, Roberto and {{\"U}bler}, Hannah and {Saxena}, Aayush and {Smit}, Renske and {Willott}, Chris and {Witstok}, Joris},
        title = "{The NIRSpec Wide GTO Survey}",
      journal = {\aap},
         year = 2024,
        month = sep,
       volume = {689},
          eid = {A73},
        pages = {A73},
          doi = {10.1051/0004-6361/202449914},
archivePrefix = {arXiv},
       eprint = {2403.05506},
 primaryClass = {astro-ph.GA},
       adsurl = {https://ui.adsabs.harvard.edu/abs/2024A&A...689A..73M}
}

@ARTICLE{scholtz25,
       author = {{Scholtz}, J. and {Carniani}, S. and {Parlanti}, E. and {D'Eugenio}, F. and {Curtis-Lake}, E. and {Jakobsen}, P. and {Bunker}, A.~J. and {Cameron}, A.~J. and {Arribas}, S. and {Baker}, W.~M. and {Charlot}, S. and {Chevellard}, J. and {Circosta}, C. and {Curti}, M. and {Duan}, Q. and {Eisenstein}, D.~J. and {Hainline}, K. and {Ji}, Z. and {Johnson}, B.~D. and {Jones}, G.~C. and {Kumari}, N. and {Maiolino}, R. and {Maseda}, M.~V. and {Perna}, M. and {P{\'e}rez-Gonz{\'a}lez}, P.~G. and {Rawle}, T. and {Rieke}, M. and {Rinaldi}, P. and {Robertson}, B. and {Saxena}, A. and {Shivaei}, I. and {Silcock}, M.~S. and {Sun}, Y. and {Rodr{\'\i}guez Del Pino}, B. and {Tacchella}, S. and {{\"U}bler}, H. and {Venturi}, G. and {Williams}, C.~C. and {Willmer}, C.~N.~A. and {Willott}, C. and {Witstok}, J.},
        title = "{JADES Data Release 4 -- Paper II: Data reduction, analysis and emission-line fluxes of the complete spectroscopic sample}",
      journal = {arXiv e-prints},
         year = 2025,
        month = oct,
          eid = {arXiv:2510.01034},
        pages = {arXiv:2510.01034},
          doi = {10.48550/arXiv.2510.01034},
archivePrefix = {arXiv},
       eprint = {2510.01034},
 primaryClass = {astro-ph.GA},
       adsurl = {https://ui.adsabs.harvard.edu/abs/2025arXiv251001034S}
}

@ARTICLE{ronayne25,
       author = {{Ronayne}, Kaila and {Papovich}, Casey and {Kirkpatrick}, Allison and {Backhaus}, Bren E. and {Cullen}, Fergus and {Shen}, Lu and {Bagley}, Micaela B. and {Barro}, Guillermo and {Finkelstein}, Steven L. and {Hamblin}, Kurt and {Kartaltepe}, Jeyhan S. and {Kocevski}, Dale D. and {Koekemoer}, Anton M. and {Lambrides}, Erini and {Pacucci}, Fabio and {Yang}, Guang},
        title = "{MEGA: Spectrophotometric SED Fitting of Little Red Dots Detected in JWST MIRI}",
      journal = {arXiv e-prints},
         year = 2025,
        month = aug,
          eid = {arXiv:2508.20177},
        pages = {arXiv:2508.20177},
          doi = {10.48550/arXiv.2508.20177},
archivePrefix = {arXiv},
       eprint = {2508.20177},
 primaryClass = {astro-ph.GA},
       adsurl = {https://ui.adsabs.harvard.edu/abs/2025arXiv250820177R}
}

@ARTICLE{begelman25,
       author = {{Begelman}, Mitchell C. and {Dexter}, Jason},
        title = "{Little Red Dots As Late-stage Quasi-stars}",
      journal = {arXiv e-prints},
         year = 2025,
        month = jul,
          eid = {arXiv:2507.09085},
        pages = {arXiv:2507.09085},
          doi = {10.48550/arXiv.2507.09085},
archivePrefix = {arXiv},
       eprint = {2507.09085},
 primaryClass = {astro-ph.GA},
       adsurl = {https://ui.adsabs.harvard.edu/abs/2025arXiv250709085B}
}

@ARTICLE{rusakov25,
       author = {{Rusakov}, V. and {Watson}, D. and {Nikopoulos}, G.~P. and {Brammer}, G. and {Gottumukkala}, R. and {Harvey}, T. and {Heintz}, K.~E. and {Nielsen}, R.~D. and {Sim}, S.~A. and {Sneppen}, A. and {Vijayan}, A.~P. and {Adams}, N. and {Austin}, D. and {Conselice}, C.~J. and {Goolsby}, C.~M. and {Toft}, S. and {Witstok}, J.},
        title = "{JWST's little red dots: an emerging population of young, low-mass AGN cocooned in dense ionized gas}",
      journal = {arXiv e-prints},
         year = 2025,
        month = mar,
          eid = {arXiv:2503.16595},
        pages = {arXiv:2503.16595},
          doi = {10.48550/arXiv.2503.16595},
archivePrefix = {arXiv},
       eprint = {2503.16595},
 primaryClass = {astro-ph.GA},
       adsurl = {https://ui.adsabs.harvard.edu/abs/2025arXiv250316595R}
}

@ARTICLE{yan25,
       author = {{Yan}, Zu and {Inayoshi}, Kohei and {Chen}, Kejian and {Guo}, Jingsong},
        title = "{Balmer Transition Signatures from Gas-Enshrouded, Dust-Poor Active Galactic Nuclei}",
      journal = {arXiv e-prints},
         year = 2025,
        month = dec,
          eid = {arXiv:2512.11050},
        pages = {arXiv:2512.11050},
          doi = {10.48550/arXiv.2512.11050},
archivePrefix = {arXiv},
       eprint = {2512.11050},
 primaryClass = {astro-ph.GA},
       adsurl = {https://ui.adsabs.harvard.edu/abs/2025arXiv251211050Y}
}

@ARTICLE{kocevski23b,
       author = {{Kocevski}, Dale D. and {Barro}, Guillermo and {McGrath}, Elizabeth J. and {Finkelstein}, Steven L. and {Bagley}, Micaela B. and {Ferguson}, Henry C. and {Jogee}, Shardha and {Yang}, Guang and {Dickinson}, Mark and {Hathi}, Nimish P. and {Backhaus}, Bren E. and {Bell}, Eric F. and {Bisigello}, Laura and {Buat}, V{\'e}ronique and {Burgarella}, Denis and {Casey}, Caitlin M. and {Cleri}, Nikko J. and {Cooper}, M.~C. and {Costantin}, Luca and {Croton}, Darren and {Daddi}, Emanuele and {Fontana}, Adriano and {Fujimoto}, Seiji and {Gardner}, Jonathan P. and {Gawiser}, Eric and {Giavalisco}, Mauro and {Grazian}, Andrea and {Grogin}, Norman A. and {Guo}, Yuchen and {Arrabal Haro}, Pablo and {Hirschmann}, Michaela and {Holwerda}, Benne W. and {Huertas-Company}, Marc and {Hutchison}, Taylor A. and {Iyer}, Kartheik G. and {Jones}, Brenda and {Juneau}, St{\'e}phanie and {Kartaltepe}, Jeyhan S. and {Kewley}, Lisa J. and {Kirkpatrick}, Allison and {Koekemoer}, Anton M. and {Kurczynski}, Peter and {Le Bail}, Aur{\'e}lien and {Long}, Arianna S. and {Lotz}, Jennifer M. and {Lucas}, Ray A. and {Papovich}, Casey and {Pentericci}, Laura and {P{\'e}rez-Gonz{\'a}lez}, Pablo G. and {Pirzkal}, Nor and {Rafelski}, Marc and {Ravindranath}, Swara and {Somerville}, Rachel S. and {Straughn}, Amber N. and {Tacchella}, Sandro and {Trump}, Jonathan R. and {Wilkins}, Stephen M. and {Wuyts}, Stijn and {Yung}, L.~Y. Aaron and {Zavala}, Jorge A.},
        title = "{CEERS Key Paper. II. A First Look at the Resolved Host Properties of AGN at 3 < z < 5 with JWST}",
      journal = {\apjl},
         year = 2023,
        month = mar,
       volume = {946},
       number = {1},
          eid = {L14},
        pages = {L14},
          doi = {10.3847/2041-8213/acad00},
archivePrefix = {arXiv},
       eprint = {2208.14480},
 primaryClass = {astro-ph.GA},
       adsurl = {https://ui.adsabs.harvard.edu/abs/2023ApJ...946L..14K}
}

@ARTICLE{deugeniojades25,
       author = {{D'Eugenio}, Francesco and {Cameron}, Alex J. and {Scholtz}, Jan and {Carniani}, Stefano and {Willott}, Chris J. and {Curtis-Lake}, Emma and {Bunker}, Andrew J. and {Parlanti}, Eleonora and {Maiolino}, Roberto and {Willmer}, Christopher N.~A. and {Jakobsen}, Peter and {Robertson}, Brant E. and {Johnson}, Benjamin D. and {Tacchella}, Sandro and {Cargile}, Phillip A. and {Rawle}, Tim and {Arribas}, Santiago and {Chevallard}, Jacopo and {Curti}, Mirko and {Egami}, Eiichi and {Eisenstein}, Daniel J. and {Kumari}, Nimisha and {Looser}, Tobias J. and {Rieke}, Marcia J. and {Rodr{\'\i}guez Del Pino}, Bruno and {Saxena}, Aayush and {{\"U}bler}, Hannah and {Venturi}, Giacomo and {Witstok}, Joris and {Baker}, William M. and {Bhatawdekar}, Rachana and {Bonaventura}, Nina and {Boyett}, Kristan and {Charlot}, Stephane and {Danhaive}, A. Lola and {Hainline}, Kevin N. and {Hausen}, Ryan and {Helton}, Jakob M. and {Ji}, Xihan and {Ji}, Zhiyuan and {Jones}, Gareth C. and {Juod{\v{z}}balis}, Ignas and {Maseda}, Michael V. and {P{\'e}rez-Gonz{\'a}lez}, Pablo G. and {Perna}, Michele and {Pusk{\'a}s}, D{\'a}vid and {Shivaei}, Irene and {Silcock}, Maddie S. and {Simmonds}, Charlotte and {Smit}, Renske and {Sun}, Fengwu and {Villanueva}, Natalia C. and {Williams}, Christina C. and {Zhu}, Yongda},
        title = "{JADES Data Release 3: NIRSpec/Microshutter Assembly Spectroscopy for 4000 Galaxies in the GOODS Fields}",
      journal = {\apjs},
         year = 2025,
        month = mar,
       volume = {277},
       number = {1},
          eid = {4},
        pages = {4},
          doi = {10.3847/1538-4365/ada148},
archivePrefix = {arXiv},
       eprint = {2404.06531},
 primaryClass = {astro-ph.GA},
       adsurl = {https://ui.adsabs.harvard.edu/abs/2025ApJS..277....4D}
}
